\documentclass[%
reprint,
superscriptaddress,
nofootinbib,
 amsmath,amssymb,
 aps,
 prd,
twocolumn,
]{revtex4-2}

\usepackage[colorlinks,urlcolor=blue,citecolor=blue,linkcolor=blue]{hyperref}
\usepackage[hyphenbreaks]{breakurl}
\usepackage{color,graphicx,graphics,amsfonts,mathrsfs,amsmath,bm,psfrag,natbib,dcolumn,textcase}
\usepackage{enumerate}
\allowdisplaybreaks
\usepackage{epsfig, soul, latexsym, multirow}
\usepackage{comment}
\usepackage{subfigure}
\usepackage{mathtools,nccmath}
\usepackage{bigints}
\usepackage{float}
\usepackage{footmisc}
\usepackage{placeins}
\usepackage[normalem]{ulem}
\usepackage{physics}
\usepackage{setspace}
\usepackage[usenames,dvipsnames]{xcolor}

\begin{document}
\title{Parametrized tests of general relativity using eccentric compact binaries}
\author{Sajad A. Bhat}
\email{sajad.bhat@iucaa.in}
\affiliation{Chennai Mathematical Institute, Plot H1 SIPCOT IT Park, Siruseri 603103, India.}
\affiliation{Inter-University Centre for Astronomy and Astrophysics, Post Bag 4, Ganeshkhind, Pune - 411007, India}
\author{Pankaj Saini}
\email{pankajsaini@cmi.ac.in}
\affiliation{Chennai Mathematical Institute, Plot H1 SIPCOT IT Park, Siruseri 603103, India.}
\author{Marc Favata}
\email{marc.favata@montclair.edu}
\affiliation{Department of Physics \& Astronomy, Montclair State University,
1 Normal Avenue, Montclair, New Jersey 07043, USA}
\author{Chinmay Gandevikar}
\affiliation{Department of Physics, Indian Institute of Technology Madras, Chennai 600036, India}
\affiliation{Centre for Strings, Gravitation and Cosmology, Department of Physics, Indian Institute of Technology Madras, Chennai 600036, India}
\author{Chandra Kant Mishra}
\affiliation{Department of Physics, Indian Institute of Technology Madras, Chennai 600036, India}
\affiliation{Centre for Strings, Gravitation and Cosmology, Department of Physics, Indian Institute of Technology Madras, Chennai 600036, India}
\author{K. G. Arun}
\email{kgarun@cmi.ac.in}
\affiliation{Chennai Mathematical Institute, Plot H1 SIPCOT IT Park, Siruseri 603103, India.}
\date{\today}
\begin{abstract}
 Astrophysical population simulations predict that a subset of dynamically formed binary black holes (BBHs) may possess eccentricity $\gtrsim 0.1$ at a gravitational wave (GW) frequency of $10 \,\text{Hz}$. Presently, the LIGO-Virgo-KAGRA (LVK) Collaboration tests general relativity (GR) assuming that the binary eccentricity has decayed well before it enters the detector's frequency band. The detection of an eccentric binary could allow us to test GR in a regime inaccessible with binaries in circular orbits. Previous works have also shown that binary eccentricity can bias GR tests if unaccounted for. Here we develop two methods to extend parametrized tests of GR to eccentric binaries. The first method extends the standard null parametrized test for quasicircular binaries by adding fractional deviations at each post-Newtonian (PN) order in the eccentric part of the GW phasing (assuming the small-eccentricity limit). Simultaneous measurement of the circular and eccentric deviation parameters ($\delta\hat{\varphi}, \delta\hat{\varphi}^e$) allows us to constrain deviations from GR for eccentric binaries. While strong constraints on the deviation parameters are not achievable with LIGO's projected sensitivity, Cosmic Explorer (CE) can constrain 1PN deviations to $|\delta\hat{\varphi}_2| \lesssim 2\times 10^{-2}$ and $|\delta\hat{\varphi}^e_2|\lesssim10^{-1}$ for spinning BBHs. The multibanding of LISA and CE observations can constrain these deviations to $|\delta\hat{\varphi}_2| \lesssim 3 \times 10^{-3}$ and $|\delta\hat{\varphi}^e_2|\lesssim 2\times 10^{-2}$. The second method looks for GR deviations in the rate of periastron advance. 
We modify the relation between the radial and azimuthal frequencies by modifying the periastron advance per orbit $\Delta \Phi \equiv 2\pi k$ via $k\rightarrow k(1+\Delta \alpha)$. We derive a parametrized phase expansion for nonspinning eccentric binaries in terms of a non-GR deviation parameter $\Delta\alpha$ that captures deviations in the periastron advance relative to the GR prediction ($\Delta\alpha^{\rm GR} \to 0$). The parameter $\Delta\alpha$ can be constrained with LIGO to $|\Delta\alpha|\lesssim 4 \times 10^{-2}$ (with $1 \sigma$ confidence). For a typical binary neutron star (BNS) system, $\Delta\alpha$ can be measured with $|\Delta\alpha|\lesssim 8 \times 10^{-3}$ in LIGO. With CE sensitivity, the bounds on $\Delta\alpha$ improve by $\sim 1$ order of magnitude for BBHs and $\sim 2$ orders of magnitude for BNS. Multiband sources observed by LISA and CE provide an improved constraint of $|\Delta\alpha|\lesssim 3\times 10^{-5}$. The space-based detector DECIGO provides the best constraint on $\Delta\alpha$ with $|\Delta\alpha|\lesssim 8 \times 10^{-6}$.
\end{abstract}

\maketitle

\section{Introduction}
Gravitational waves (GWs) from the mergers of compact binaries allow us to test general relativity (GR) in the highly dynamical and strong-field regime~\cite{Will:2005va, Sathyaprakash:2009xs, Yunes:2013dva, Berti:2015itd, Krishnendu:2021fga}. With close to $90$ detections of compact binary coalescence~\cite{LIGOScientific:2021djp} by Advanced LIGO (LIGO)~\cite{LIGOScientific:2014pky} and Advanced Virgo (Virgo)~\cite{VIRGO:2014yos}, GR has been tested extensively~\cite{LIGOScientific:2016lio,  LIGOScientific:2018dkp, LIGOScientific:2019fpa, LIGOScientific:2020tif, GWTC3:2021sio}. To date, no statistically significant deviation from GR has been found~\cite{GWTC3:2021sio}. As the sensitivity of current GW detectors improves~\cite{LIGOScientific:2014pky,VIRGO:2014yos,KAGRA:2013rdx,KAGRA:2020tym, LIGO:2020xsf} and new detectors such as LIGO-India come online~\cite{Saleem:2021iwi}, more stringent tests of GR can be performed by combining information from multiple events. With $\sim\mathcal{O}(10)$ times improved sensitivity and enhanced low frequency performance, the third-generation (3G) ground-based detectors Cosmic Explorer (CE)~\cite{LIGOScientific:2016wof,Reitze:2019iox} and Einstein Telescope (ET)~\cite{Hild:2010id} are expected to detect $\sim 10^5$ binary black hole (BBH) mergers per year~\cite{Baibhav:2019gxm, Evans:2021gyd, Gupta:2023lga}, facilitating tests of GR with exquisite precision. The LIGO-Virgo-KAGRA (LVK) Collaboration currently employs quasicircular waveform models for testing GR. This assumes that GW emission circularizes the binary's orbit~\cite{PetersMathews:1963ux,Peters:1964zz} well before it enters the LVK frequency band ($\gtrsim 20 \, {\rm Hz}$).  

 To be measurable by LIGO, a binary typically requires an eccentricity of $\gtrsim 0.008  \;(0.05)$ for low (high)-mass systems at a GW frequency of $10 \, \text{Hz}$~\cite{Lower:2018seu,Moore:2019vjj,Favata:2021vhw}. However, note that this lower limit depends on the waveform models used~\cite{Shaikh:2023ypz}. Since binaries are more eccentric earlier in their inspiral, the better low-frequency sensitivity of CE and ET will allow eccentricities as small as $\gtrsim 10^{-3}$ (at $10 \, {\rm Hz}$)~\cite{Saini:2023wdk} to be resolved. The Laser
Interferometer Space Antenna (LISA)~\cite{Babak:2017tow,Yagi:2013du,Kawamura:2020pcg} can measure eccentricities $\gtrsim 10^{-2.5}$ (at one year before the merger) for supermassive BBHs~\cite{Garg:2023lfg}. Apart from the mergers of supermassive BHs, LISA will also see the inspiral of stellar-mass/intermediate-mass BBHs that will merge in the low-frequency band of ground-based 3G detectors. LISA will be able to constrain the eccentricity of stellar-mass BBHs with high precision~\cite{Nishizawa:2016jji, 2019arXiv190702283R}. The multiband observations of stellar-mass/intermediate-mass BBHs in LISA and CE will further improve eccentricity measurements~\cite{Klein:2022rbf}. The potential of multiband observation of stellar-mass black hole binaries to constrain various alternative theories of gravity has been studied by many authors~\cite{Barausse:2016eii,Nair:2015bga,Gnocchi:2019jzp,Chang_Liu:2020nwz,Toubiana:2020vtf, Carson:2019rda,Toubiana:2020vtf,Gupta:2020lxa,Datta:2020vcj,Klein:2022rbf,Wang:2023tle}.

There has been no confident detection of an eccentric binary during the first three LVK observing runs; and current LVK constraints on the eccentric binary merger rates based on model-independent searches are pessimistic~\cite{LIGOScientific:2019dag,LIGOScientific:2023lpe}. However, studies based on N-body simulations have shown that $\sim 5\%$ of dynamically-formed binaries in dense stellar environments such as globular clusters and nuclear star clusters can retain a non-negligible eccentricity ($\gtrsim 0.1$ at $10$ Hz)~\cite{Wen:2002km,OLeary:2008myb,Antonini:2012ad,Antonini:2013tea, Antonini:2015zsa, Samsing:2017xmd, Samsing:2018isx,  Rodriguez:2018pss, Zevin:2018kzq, Zevin:2020gbd, Gondan:2020svr, Zevin:2021rtf,DallAmico:2023neb}. Different formation channels leave unique imprints on binary properties, providing valuable information about their formation history. Multiple formation mechanisms have been proposed that can produce highly eccentric binaries. These include the secular evolution of stellar-mass binaries near a supermassive black hole~\cite{2014ApJ...794..122M, 2017ApJ...841...77A, Hoang:2017fvh, Hamers:2021kby}, gravitational interaction of a binary with a third body~\cite{Silsbee:2016djf,Liu:2018nrf,2018ApJ...863....7R,Michaely:2019aet,2023ApJ...955..134R}, and mergers in active-galactic nuclei discs or gaseous environments~\cite{Samsing:2020tda}. Recent studies have reported signatures of the presence of eccentricity in a subset of GWTC-3 events~\cite{Romero-Shaw:2020thy, Gayathri:2020coq, Romero-Shaw:2021ual, OShea:2021ugg, Gupte:2024jfe}. However, it is hard to distinguish between eccentricity and spin-precession effects, especially for short-duration signals like GW190521~\cite{CalderonBustillo:2020xms,Romero-Shaw:2022fbf}. The detection of eccentric binaries will provide a unique opportunity to probe certain physical effects associated with eccentric binaries that are absent for binaries in circular orbits. 

The LVK Collaboration presently uses quasicircular waveform models for parameter estimation and testing GR. If a quasicircular waveform model is used for analyzing a GW signal from an eccentric binary, this will introduce a systematic bias in the estimated parameters~\cite{Favata:2013rwa,  OShea:2021ugg, Favata:2021vhw, Narikawa:2016uwr, Divyajyoti:2023rht, DuttaRoy:2024aew}. For the standard parametrized test of GR, these systematic biases will shift the values of the ``circular'' deviation parameters from their true value. These biases, if statistically significant, will appear as a false violation of GR~\cite{Saini:2022igm,Bhat:2022amc,Narayan:2023vhm,Shaikh:2024wyn}. While small systematic biases for individual events may be hidden in the statistical errors, even small systematic errors can become more severe when performing tests of GR from a catalog of events~\cite{Moore:2021eok,Saini:2023rto}. (See the recent review in ~\cite{Gupta:2024gun} for a broader discussion of possible causes of false GR violations.) This necessitates the inclusion of the orbital eccentricity parameter in the waveform models employed for testing GR. Therefore, it is crucial to develop parametrized tests of GR for eccentric binaries. Here, we explore two approaches for including eccentricity in GR tests based on GWs.

\subsection{Parametrized tests of GR for circular binaries}
 When performing tests of GR using GWs, we have no prior knowledge about the most likely types of GR violations that may be present in a modified theory of gravity (see, for example,~\cite{Wolf:2019hun,AbhishekChowdhuri:2023gvu,Ghosh:2023xes,Kumar:2024utz,Nair:2024xdb}). Even for well-studied modified theories like scalar-tensor theories~\cite{Damour:1992we, Bernard:2018ivi, Bernard:2018hta, Tahura:2018zuq, Khalil:2018aaj,Okounkova:2019zjf,Okounkova:2020rqw, Cayuso:2020lca, East:2020hgw, East:2021bqk, Shiralilou:2021mfl, Julie:2024fwy}, it is hard to quantify the accuracy of these beyond-GR waveform models due to lack of long and accurate NR simulations for these theories.( However, see~\cite{Ma:2023sok,Mirshekari:2013vb,Lang:2013fna,Lang:2014osa,Corman:2024vlk} for some theoretical and numerical efforts in this direction.) Therefore, theory-agnostic parametrized tests that search for generic departures from GR~\cite{BSat95,Arun:2004hn,Arun:2006hn,Yunes:2009ke,PPE:2011ys,TIGER:2013upa,Mehta:2022pcn} provide a very useful alternative.

For circular orbits the stationary phase approximation (SPA)~\cite{Cutler:1994ys, Droz:1999qx} allows the frequency-domain phase $\Psi(f)$ of the gravitational waveform to be expanded as a post-Newtonian (PN)~\cite{Blanchet:2002av} series in the relative orbital speed:
\begin{equation}
    \Psi(f) \sim \frac{3}{128\eta v^5}\bigg(\varphi_0 + \varphi_1 v + \varphi_2 v^2 + \varphi_3 v^3 + \cdots \bigg) \,.
\end{equation}
Here $\eta=m_1m_2/M^2$ is the symmetric mass ratio, $v=(\pi M f)^{1/3}$ is the relative orbital speed, and $M= m_1+m_2$ is the total binary mass. The $\varphi_i$ are the PN coefficients and are functions of the intrinsic binary properties such as the component masses, spins, tidal deformabilities, etc. In a PN series, after factoring out the leading-order $v$ dependence, a PN coefficient multiplying $v^{2n}$ is called the $n$PN order coefficient.

 In parametrized tests with circular binaries~\cite{BSat95,Arun:2004hn,Arun:2006hn,Yunes:2009ke,PPE:2011ys,TIGER:2013upa,Mehta:2022pcn}, the GR waveform is modified by adding phenomenological deviation parameters at each PN order that capture potential deviations from GR: $\varphi_{i} \xrightarrow{} \varphi_{i} (1 + \delta \hat{\varphi}_{i})$.
The terms $\delta\hat{\varphi}_i$ are the fractional deviation parameters; they reduce to $\delta\hat{\varphi}_i\to 0$ in GR.\footnote{This relation is true for the case where the PN coefficient is non-zero in GR. However, for PN orders where the coefficients are zero [e.g., at -1PN and 0.5PN orders corresponding to $\mathcal{O}(v^{-2})$ and $\mathcal{O}(v)$ corrections], the appropriate relation is $\varphi_{i} \xrightarrow{} \delta \hat{\varphi}_{i}$.} In most applications of this formalism, only one $\delta\hat{\varphi}_i$ parameter at a time is allowed to vary along with the other GR parameters, given the limited sensitivity of current GW detectors. Allowing multiple deviation parameters to vary at once severely degrades parameter extraction accuracy~\cite{Gupta:2020lxa}. 

\subsection{Extending the standard parametrized tests for eccentric binaries}
The standard parametrized test can be extended to eccentric binaries. For generic eccentric orbits, a simple PN series for the GW phase is not available. However, in the small eccentricity limit, the SPA phase can be written as the circular phase $\Psi^{\rm circ.}$ plus leading-order corrections due to eccentricity $\Delta\Psi^{\rm ecc.}$ as
\begin{equation}\label{eq:spa_intro}
    \Psi(f) \sim \Psi^{\rm circ.} + \Delta\Psi^{\rm ecc.} \,.
\end{equation}
The eccentric corrections $\Delta\Psi^{\rm ecc.}$ also have a PN structure that can be written as~\cite{Moore:2016qxz}:
\begin{align}
    \label{eccentric phase introduction}
    \Delta\Psi^{\rm ecc.}_{\rm 3PN} 
    &= -\frac{2355}{1462} e_{0}^{2} \Big(\frac{v_{0}}{v} \nonumber \Big)^{19/3} \bigg(\varphi_0^e + \varphi_1^{e} v + \varphi_1^{e,0} v_0  \\
    &+ \varphi_2^{e} v^2 + \varphi_2^{e,0} v^2_0 + \cdots \bigg). 
\end{align}
where $v_0=(\pi M f_0)^{1/3}$, and $f_0$ is the reference frequency at which $e_0$ is defined\footnote{Note that the eccentricity is a model-dependent parameter that can be converted to the model-independent eccentricity using {\tt PYTHON} package {\tt gw\_eccentricity}~\cite{Shaikh:2023ypz}.}. The coefficients $\varphi_0^e=1$ and $\varphi_1^{e}=\varphi_1^{e,0}=0$ in GR [see Eq.~\eqref{eccentric phase}]. This expansion is known to 3PN order.

A simple way to extend the parametrized test to eccentric binaries is to introduce phenomenological deviation parameters in the eccentric part $\Delta\Psi_{\rm 3PN}^{\rm ecc.}$ via the replacement $\varphi_{i}^{e} \xrightarrow{} \varphi_{i}^{e} (1 + \delta \hat{\varphi}_{i}^{e})$. (To avoid introducing too many new parameters, we assume $\delta \hat{\varphi}_i^{e,0} = \delta\hat{\varphi}_i^{e}$.) In the spirit of null tests, one can choose to only measure the deviation parameters in the eccentric part of the waveform, keeping the deviations in the circular part to zero.

Measuring deviations only in the eccentric part of the waveform would mean that only the eccentric dynamics of the binary is different from GR, and the circular part is described by GR. This assumption is somewhat nonphysical. In principle, if the GW signal is described by some non-GR theory, it is expected to modify both the circular and eccentric dynamics of the binary. However, a GR violation in the signal may be captured by this parametrization, even if suboptimally. In our analysis below, we consider separately the cases where---in addition to the binary parameters (including the eccentricity)---we allow only one eccentric deviation parameter $\delta \hat{\varphi}_i^e$ to vary and where we allow both $\delta \hat{\varphi}_i^e$ and its corresponding circular deviation parameter $\delta \hat{\varphi}_i$ to simultaneously vary; (see Sec.~\ref{sec:parametric deviation} for more details). Though this parametrization is an obvious extension of the one employed in the circular orbits case, this yields weaker constraints on the deviation parameters due to the larger set of parameters that must be estimated (see Sec.~\ref{sec:results parametric deviation}).
  
\subsection{Testing GR using periastron advance in eccentric binaries}
We consider an alternative approach to parametrizing the eccentric waveform in terms of a single non-GR parameter that captures the deviation in the periastron advance rate predicted by GR. While there may be several ways in which the waveform can be parametrized in terms of the periastron advance, we choose to modify the relation between the radial and azimuthal orbital frequencies, which explicitly depends on the periastron advance parameter $k$. Periastron advance leads to a modulation of the GW signal and modifies its phase~\cite{LeTiec:2011bk,Hinderer:2013uwa}. 

Periastron advance has been observed for several planets in the Solar System~\cite{2005AstL...31..340P} and exoplanets~\cite{Pal:2008zh,Blanchet:2019zxv,Gallerati:2022jrl}. In addition to GR dynamics, a periastron shift can also be caused by Newtonian perturbations from other Solar System planets and the solar oblateness. Subtracting these Newtonian effects, the residual periastron shift can be used to constrain deviations from GR. In the weak-field limit, Einstein derived a general relativistic formula for the {\it angular advance per orbit} considering Mercury as a test particle freely falling in the gravitational field of the Sun:
\begin{equation}\label{eq:angular advance}
    \Delta\Phi = \frac{6 \pi G M}{c^2 a (1-e^2)} \,,
 \end{equation}
where $M$ is the central mass (i.e., the Sun), and $a$ and $e$ are the semi-major axis and eccentricity of the orbiting body (i.e., Mercury), respectively. While Eq.~\eqref{eq:angular advance} provides the leading-order periastron advance, PN corrections to this equation must be included when we consider more relativistic regimes like BBHs. In GR, the leading-order corrections due to periastron advance appear at 1PN order in the GW phasing. [Note that the periastron advance angle $\Delta\Phi$ is non-zero in the $e\to0$ limit, and $\Delta\Phi$ vanishes in the Newtonian limit ($c\to \infty$).]

Periastron advance has been observed in different regimes of gravity. The measurement of Mercury's periastron advance of $\sim 43$ arcseconds per century provided the first test of GR~\cite{Richard-William}. Radio observations of binary pulsars test moderately relativistic effects by tracking the binary's orbital evolution over long time periods~\cite{Damour:1988mr,Stairs:2003eg,Weisberg:2004hi,Kramer:2004vio,Kramer:2006nb,Kramer:2021jcw}. For example, the periastron advance rate in the Double Pulsar system agrees with GR with exquisite precision~\cite{Kramer:2021jcw}. Binary pulsars allow periastron advance rate measurements with much larger amplitudes than in the Solar System ($\sim$ a few degrees per year~\cite{Stairs:2003eg}). Periastron advance measurements in relativistic binary pulsar systems~\cite{1999AJ....117..587P,1988NCimB.101..127D} can also be used to constrain the neutron star moment of inertia, which leads to constraints on the neutron star equation of state~\cite{Lattimer:2004nj}. Compact binary mergers, especially BBHs, have characteristic speeds comparable to the speed of light near the merger; this allows for much larger periastron advance rates and probes of the strong-gravity regime. For example, for two BHs with masses $20M_{\odot}$ and $15M_{\odot}$ orbiting with $e=0.2$ and emitting GWs at $1\,\text{Hz}$, the periastron advance rate is $\sim 4$ degrees per second. 

We expand upon the study in \cite{Moore:2016qxz} by parametrizing the waveform phasing in terms of a non-GR parameter related to periastron advance. In particular, we derive the GW phasing formula in the limit of small eccentricity and in terms of a non-GR parameter that explicitly captures deviations in the periastron advance angle per orbit. This is done by modifying the relation between the dimensionless azimuthal frequency $\xi_\phi$ and the dimensionless radial frequency $\xi$, 
 \begin{equation}\label{eq:xirtophi_intro}
\xi_\phi= (1+ k) \xi,
 \end{equation}
  via the following change to the periastron advance parameter $k$:
 \begin{equation}
     k \rightarrow (1+ \Delta \alpha)k\,,
 \end{equation}
where $\Delta \alpha$ is a non-GR parameter that captures the deviation in the periastron advance from the GR value ($\Delta \alpha \rightarrow 0$). Using this modified relation, we recompute the frequency and phase evolution of the binary, resulting in a modified expression for the PN expansion of the SPA phase $\Psi(f)$ in the low-eccentricity limit [see Eq.~\eqref{eq:PsiFTecc}]. This expansion reduces to the GR SPA phase expansion plus $\Delta \alpha$-dependent corrections to both the circular and eccentric pieces:
\begin{equation}
    \Psi(f) \rightarrow \Psi^{\rm GR}(f) + \delta\Psi_{\rm circ.}(f,\Delta \alpha) + \delta\Psi_{\rm ecc.}(f, \Delta\alpha)\,,
\end{equation}
where $\delta\Psi_{\rm ecc.} \sim e_0^2$, and $\delta\Psi_{\rm circ.}$ and $\delta\Psi_{\rm ecc.}$ both depend on $\Delta \alpha$ at each PN order (with $\delta\Psi_{\rm circ.}, \delta\Psi_{\rm ecc.} \rightarrow 0$ as $\Delta\alpha \rightarrow 0$). The phase $\Psi^{\rm GR}$ is given by Eq.~\eqref{eq:spa_intro}. This parametrization depends on only one non-GR parameter ($\Delta \alpha$) instead of many deviation parameters $(\delta \hat{\varphi}_i,\delta \hat{\varphi}_i^e)$ at each PN order (as in the first parametrization). In effect, this approach affects all PN orders in the phasing simultaneously via the single parameter $\Delta \alpha$. Therefore, constraints obtained with the second method are tighter compared to the first parametrization. Note that $\Delta \alpha$ modifies both the circular and eccentric parts of the phasing. This arises as an artifact of the low-eccentricity limit of the periastron advance parameter. Section~\ref{sec:periastron parametrization} explains this method in more detail.

\subsection{Results summary}
We apply the Fisher matrix formalism~\cite{Finn:1992wt,Cutler_Flanagan,Clifford_Will} to both methods to calculate bounds on the non-GR parameters. The results are shown in Sec.~\ref{sec:results}. We consider observations with the LIGO, CE, LISA, and DECIGO detectors for various mass binaries. Figure~\ref{fig:first_parametrization_deviation} shows the $1\sigma$ errors on the 1PN parameters $\delta\hat{\varphi}_{2}$ and $\delta\hat{\varphi}_{2}^e$ as a function of the total mass $M$ of the binary. The deviation parameters $\delta\hat{\varphi}_{i}$ and $\delta\hat{\varphi}_{i}^e$ are not well measured in the LIGO frequency band, either when measured simultaneously or when $\delta\hat{\varphi}_{i}^e$ is considered alone. The best constraints are obtained from multiband sources observed with both LISA and CE: $|\delta\hat{\varphi}_2|\lesssim3\times 10^{-3}$ and $|\delta\hat{\varphi}_2^e|\lesssim 10^{-2}$. For the second parametrization, Fig.~\ref{fig:alpha} shows the $1\sigma$ bound on the periastron deviation parameter $\Delta\alpha$ as a function of the binary mass. LIGO can measure the periastron deviation parameter $\Delta\alpha$ with an error $|\Delta\alpha|\lesssim 5\times 10^{-2}$ for $M=10M_{\odot}$. Hence, the second approach is more effective in placing constraints on non-GR deviations in the LIGO band. Multiband observations of $M=10^2\mbox{--}10^3\, M_{\odot}$ binaries seen with LISA and CE provide a constraint $|\Delta\alpha| \lesssim 5\times 10^{-4}$. DECIGO puts the best constraint on $\Delta\alpha$ with $|\Delta\alpha|\lesssim 8 \times 10^{-6}$.

The rest of the paper is organized as follows: In Section~\ref{params-models} we discuss the post-Newtonian eccentric waveform model {\tt TaylorF2Ecc} that incorporates the leading-order eccentricity corrections to the GW phasing; that model serves as a basis for the parametrized extensions considered here. In Sec.~\ref{sec:parametric deviation} we introduce the first parametrization in which non-GR deviations are added to both the circular and eccentric parts of the {\tt TaylorF2Ecc} phasing. Section~\ref{QKReview} provides a summary of the quasi-Keplerian parametrization; it sets the stage for introducing (in Sec.~\ref{sec:periastron parametrization}) the second parametrization based on deviations in the periastron advance parameter. (Some additional details are relegated to Appendix~\ref{app:full expressions}, and an alternative derivation of the $\Delta\alpha$-dependent SPA phasing is provided in Appendix~\ref{app:alternate method}.) Section~\ref{sec:fisher matrix framework} discusses the Fisher information matrix framework for estimating the projected measurement uncertainties in the waveform parameters. Section \ref{sec:results} applies the Fisher formalism to estimate how well the various non-GR parameters introduced here can be constrained. (Appendices~\ref{app:eccentricity_measurement} and \ref{app_addionalboundresults} examine issues related to those parameter constraints.) Finally, we summarize and conclude our findings in Sec.~\ref{conclude}. Unless explicitly stated, we use units in which $G=c=1$.

\section{\label{params-models}  Waveform Model Preliminaries}
Gravitational waves in GR have two independent polarizations: $h_+$ and $h_\times$. The GW signal $h(t)$ measured in the detector is a linear combination of the polarization amplitudes, 
\begin{equation}\label{strain}
    h(t) = F_{+} (\theta, \phi,\psi) h_{+}(\iota, \beta,t) + F_{\times} (\theta, \phi,\psi) h_{\times}(\iota, \beta, t) \,.
\end{equation}
Here $F_{+,\times} (\theta, \phi,\psi)$ are the antenna pattern functions of the detector, which depend on the sky location angles $(\theta, \phi)$ of the source and a polarization angle $\psi$. The angles $\iota$ and $\beta$ are the inclination angle and the initial orientation of the elliptical orbit, respectively. The angle $\beta$ is an angle from a reference direction that helps define the orientation of the orbital plane or the angular momentum vector.

The {\it Fourier frequency transform} of the strain $h(t)$ in the SPA~\cite{Cutler:1994ys, Droz:1999qx}, can be written as \begin{equation}\label{waveform}
    \Tilde{h}(f) = \mathcal{A}f^{-7/6} e^{i \Psi(f)} \,.
\end{equation}
Averaging over the angles $\theta \in [0,\pi], \phi \in [0,2\pi], \psi \in [0, \pi], \iota \in [0,\pi], \beta \in [0,2\pi]$, the parameter $\mathcal {A}$ in the quadrupole approximation is given by 
\begin{equation}\label{amplitude}
  \mathcal{A} = \frac{1}{\sqrt{30}\pi^{2/3}} \frac{\mathcal{M}^{5/6}}{d_{L}} \,.
\end{equation}
where $\mathcal{M}= M \eta^{3/5}$ is the chirp mass of the system, $M=m_1+m_2$ is the total mass, $\eta = (m_1 m_2)/(m_1+m_2)^2$ is the symmetric mass ratio,   $m_{1,2}$ are the component masses, and $d_{L}$ is the luminosity distance to the source. Here $\mathcal{M}$ and $M$ are the {\it source frame} chirp mass and total mass. These are related to the  {\it detector frame} chirp mass $\mathcal{M}_{\rm det}$ and total mass $M_{\rm det}$ via the redshift $z$ 
\begin{equation}
    \mathcal{M}_{\rm det} = (1+z) \mathcal{M} ,\;\;\; M_{\rm det} = (1+z) M \,.
\end{equation}
For a $\Lambda$CDM universe $d_L$ and $z$ are related by~\cite{Hogg:1999ad}
\begin{equation}
\label{eq:dLz}
d_L(z) = \frac{1+z}{H_0} \int_0^z \frac{dz'}{\sqrt{\Omega_M (1+z')^3 + \Omega_{\Lambda}}}\,.
\end{equation} 
We use flat universe cosmological parameters from the Planck mission~\cite{Planck:2015fie}:
$H_{0}=100h$(km/s)/Mpc, $h=0.6790$, $\Omega_{m}=0.3065$, and $\Omega_{\Lambda}=0.6935$.

\subsection{TaylorF2Ecc waveform model}
Post-Newtonian theory~\cite{Blanchet:1995ez, Blanchet:1995fg, Blanchet:2002av} provides approximate analytic solutions to Einstein's equations in the weak-field, small velocity regime. PN waveform models for binaries in quasicircular orbits with aligned-spins produce analytic frequency-domain expressions up to 3.5PN order~\cite{Kidder:1992fr,Kidder:1995zr, Arun:2004hn,Arun:2008kb,Buonanno:2009zt,Wade:2013hoa,Mishra:2016whh}. These solutions have recently been extended up to 4.5PN order for nonspinning binaries~\cite{Blanchet:2023bwj,Blanchet:2023sbv}. Because of the increased complexity, waveform models for arbitrary eccentricities do not generally admit an analytical expression and must be computed numerically. Eccentric waveforms are computed via a {\it quasi-Keplerian} formalism~\cite{1985AIHPA..43..107D,1988NCimB.101..127D,SCHAFER1993196,Damour:2004bz,Memmesheimer:2004cv,Konigsdorffer:2006zt}, which provides the parametric solution to the conservative part of the PN equations of motion. The radiative evolution is obtained by a set of ordinary differential equations. However, fully analytic expressions for eccentric orbits can be obtained by either ignoring radiation-reaction effects, ignoring PN corrections, or assuming the eccentricity is small.

In Ref.~\cite{Moore:2016qxz} the GW phasing for eccentric binaries was computed up to 3PN order and to leading-order in the eccentricity [i.e., $\mathcal{O}(e_0^2)$ corrections beyond the quasicircular phasing].\footnote{This can be extended to include expressions for aligned spins using inputs from a recent work~\cite{Henry:2023tka}.} A frequency domain (SPA) representation of this waveform is referred to as {\tt TaylorF2Ecc} and is implemented in the LIGO Algorithmic Library (LAL)~\cite{lalsuite}.\footnote{In the literature, several eccentric waveforms are available; see, e.g., \cite{East:2012xq,Huerta:2014eca,Huerta:2016rwp,Huerta:2017kez,Cao:2017ndf,Hinderer:2017jcs,Nagar:2018zoe,Moore:2019xkm,Nagar:2021gss,Ramos-Buades:2021adz}.} The waveform model is expected to be accurate for $e_0\lesssim0.2$. This waveform model consistently accounts for the effects of periastron precession, albeit in the small eccentricity limit. The model incorporates only the secular contribution to the phasing; it ignores oscillatory contributions to the phasing that arise from the radiation-reaction force and the variation in the instantaneous orbital speed along each orbit. These oscillatory contributions are shown to be small for small eccentricities~\cite{Moore:2016qxz}. The waveform amplitude does not contain PN corrections or eccentricity corrections; i.e., it assumes a Newtonian-order (0PN) amplitude for circular orbits. Since a matched-filter based analysis of the GW data is more sensitive to the GW phase than the amplitude, small eccentricity corrections to the amplitude are expected to be less important than the GW phase. The {\tt TaylorF2Ecc} waveform only accounts for the dominant harmonic of the GW frequency; i.e., for small eccentricities this is twice the azimuthal orbital frequency $\omega_\phi$. Terms that oscillate at multiples of the radial orbital frequency $\omega_r$ are negligible for small eccentricities~\cite{Moore:2016qxz}. In Ref.~\cite{Moore:2016qxz}, the phasing of the waveform assumes the spins of compact objects to be zero. However, we include spin corrections to the circular part of the phasing (following the {\tt TaylorF2} waveform)~\cite{Arun:2008kb,Mishra:2016whh}.

Using the PN formalism, the SPA phase $\Psi(f)$ for eccentric binaries with small eccentricity can be decomposed into circular and eccentric parts as~\cite{Moore:2016qxz}
\begin{equation}
\label{phase}
     \Psi(f)  = 2\pi ft_{c} + \phi_{c} -\frac{\pi}{4} +\frac{3}{128 \eta v^{5}} \bigg(\Psi^{\rm circ.}_{\rm 3.5PN} +  \Delta\Psi^{\rm ecc.}_{\rm 3PN}\bigg)  \,,
\end{equation}
where $t_{c}$ and $\phi_{c}$ are the time and phase of coalescence (respectively) and $v=(\pi M f)^{1/3}$ is the orbital velocity parameter. The circular phase $\Psi^{\rm circ.}_{\rm 3.5 PN}$ is expanded as a power series in $v$: 
\begin{equation}\label{eq:circ_phase}
    \Psi^{\rm circ.}_{\rm 3.5PN} = \sum_{i=0}^{7} \big(\varphi_{i}v^{i} 
      +  \varphi_i^{\rm log}  v^{i} \log v\big) \,,
\end{equation}
where $\varphi_i$ and $\varphi_i^{\rm log}$ are the PN coefficients; they are functions of the intrinsic source properties, such as the masses and spins. (We ignore tidal and rotational deformabilities in our analysis.) In GR, 
$\varphi_0=1$, $\varphi_1=0, \varphi_2 = \frac{3715}{756} + \frac{55}{9}\eta$, and $\varphi_3 = -16\pi + \frac{113}{3}(\sqrt{1-4\eta} \chi_a + \chi_s) - \frac{76 \eta \chi_s}{3}$, where $\chi_s = (\chi_1+\chi_2)/2$ and $\chi_a = (\chi_1 - \chi_2)/2$  are the symmetric and antisymmetric combinations of the dimensionless spin parameters $\chi_1$ and $\chi_2$. (Note that we are assuming here and throughout that the spins are aligned or antialigned with the orbital angular momentum. Also, note that $\log$ refers to the natural logarithm.) The 0.5PN $(i=1)$ term is zero in GR. The coefficients $\varphi_i^{\rm log}$ are non-zero only for $i=5$ and $6$, i.e., 2.5PN and 3PN orders. The values of $(\varphi_i, \varphi_i^{\rm log})$ can be found in \cite{Arun:2004hn, Arun:2008kb, Buonanno:2009zt,Mishra:2016whh}. The eccentric part $\Delta\Psi^{\rm ecc.}_{\rm 3PN}$ represents the small eccentricity corrections [$\mathcal{O}(e_0^2)$] computed in Ref.~\cite{Moore:2016qxz}; they can be expressed as 
\begin{align}
    \label{eccentric phase}
    &\Delta\Psi^{\rm ecc.}_{\rm 3PN} = -\frac{2355}{1462} \nonumber e_{0}^{2} \Big(\frac{v_{0}}{v}\Big)^{19/3}\Bigg[1+ \bigg(\frac{ 299076223}{81976608}  
   \\ \nonumber
    &+ \frac{18766963}{2927736} \eta \bigg) v^{2} + \bigg(\frac{2833}{1008} - \frac{197}{36} \eta \bigg)v_{0}^{2} + \cdots + \\ \nonumber
    &  \bigg(\frac{847282939759}{82632420864}  -\frac{718901219  }{368894736}\eta -\frac{3697091711
}{105398496}\eta^2\bigg) \\ 
v^2 v_0^2& + \cdots + \mathcal{O}(v^{7})\Bigg] \,,
\end{align}
where $v_0 = (\pi M f_0)^{1/3}$ and $f_{0}$ is the reference frequency at which the binary's instantaneous time eccentricity $e_t$ equals $e_0$. [See Eq.~(6.26) of \cite{Moore:2016qxz} for the full expression to 3PN order.] Spins are not included in the eccentric piece of the phasing. Since the eccentric piece is a power series expansion in both $v$ and $v_0$, we treat both $v$ and $v_0$ on the same footing when determining the PN order of a particular term. For example, terms proportional to $v^4$, $v^2 v_0^2$ and $v_0^4$ are all considered 2PN terms.

\section{\label{sec:parametric deviation} Extending parametrized tests of GR from circular to eccentric binaries} 
 When applying parametrized tests of GR to compact binaries in circular orbits~\cite{Arun:2004hn, Arun:2006hn,Cornish:2011ys, TIGER:2013upa, Mehta:2022pcn}, the PN coefficients $(\varphi_i, \varphi_i^{\rm log})$ are deformed from their GR values by introducing phenomenological deviation parameters at each PN order in the circular frequency-domain SPA phase:
\begin{subequations}
\label{deformation}
\begin{align}
\varphi_{i} &\xrightarrow{} \varphi_{i} (1 + \delta \hat{\varphi}_{i}) \,, \\
\varphi_i^{\rm log} &\xrightarrow{} \varphi_i^{\rm log} (1 + \delta \hat{\varphi}_i^{\rm log}) \, ,
\end{align}
\end{subequations}
where $\delta\hat{\varphi}_i = \delta\varphi_i/ \varphi_i$ and $\delta\hat{\varphi}_i^{\rm log} = \delta\varphi_i^{\rm log}/ \varphi_i^{\rm log}$ are the fractional deviation parameters which are zero in GR. The parametrized phase now is a function of the following deviation parameters: $\{\delta\hat{\varphi}_0, \delta\hat{\varphi}_1, \delta\hat{\varphi}_2, \delta\hat{\varphi}_3, \delta\hat{\varphi}_4, \delta\hat{\varphi}_5^{\rm log}, \delta\hat{\varphi}_6, \delta\hat{\varphi}_6^{\rm log}, \delta\hat{\varphi}_7\}$. Since the 0.5PN term is zero in GR, $\delta\hat{\varphi}_1$ represents an absolute deviation. Note that there is no GR deviation to the frequency-independent term $\delta\hat{\varphi}_5$, since this is degenerate with the coalescence phase constant $\phi_c$ and can be reabsorbed into the redefinition of $\phi_c$. We also ignore terms corresponding to negative PN orders (e.g., dipole radiation terms). The amplitude of the GW signal is not modified, since detectors are more sensitive to the phase. 

 To extend the parametrized test of GR to eccentric binaries, we introduce phenomenological deviation parameters in the eccentric part of the GW phasing [$\Delta\Psi_{\rm 3PN}^{\rm ecc.}$, Eq.~\eqref{eccentric phase}], analogous to how deviation parameters are introduced in the circular phasing ($\Psi_{\rm 3.5PN}^{\rm circ.}$). From the structure of $\Delta\Psi^{\rm ecc.}_{\rm 3PN}$ given by Eq.~(6.26) of \cite{Moore:2016qxz}, we posit a general 3PN-order expansion of the form:
\begin{multline}
\label{eccentric phase2}
    \Delta\Psi^{\rm ecc.}_{\rm 3PN} = -\frac{2355}{1462} e_{0}^{2} \Big(\frac{v_{0}}{v}\Big)^{19/3}\Bigg[\varphi_0^{\rm e,\, (0,0)} 
    + \left( \varphi_1^{\rm e,\, (1,0)} v \right.
    \\ \left. + \varphi_1^{\rm e,\, (0,1)} v_0 \right)
     +   \varphi_2^{\rm e, \,(2,0)} v^2 + \varphi_2^{\rm e,\, (0,2)} v_0^2 + \left( \varphi_2^{\rm e,\, (1,1)} v v_0 \right)  
    \\+   \varphi_3^{\rm e, \,(3,0)} v^3 +   \varphi_3^{\rm e, \,(0,3)} v_0^3 +   \left(\varphi_3^{\rm e, \,(2,1)} v^2 v_0 +   \varphi_3^{\rm e, \,(1,2)} v v_0^2 \right) 
    \\ + \varphi_4^{\rm e, \,(4,0)} v^4 +   \varphi_4^{\rm e, \,(0,4)} v_0^4 +   \varphi_4^{\rm e, \,(2,2)} v^2 v_0^2
    \\+    \left(\varphi_4^{\rm e, \,(3,1)} v^3 v_0 +   \varphi_4^{\rm e, \,(1,3)} v v_0^3 \right) 
    \\+  \varphi_5^{\rm e, \,(5,0)} v^5 +   \varphi_5^{\rm e, \,(0,5)} v_0^5 + \varphi_5^{\rm e, \,(3,2)} v^3 v_0^2 + \varphi_5^{\rm e, \,(2,3)} v^2 v_0^3
    \\+    \left( \varphi_5^{\rm e, \,(4,1)} v^4 v_0 + \varphi_5^{\rm e, \,(1,4)} v v_0^4 \right) 
    \\ +  \varphi_6^{\rm e, \,(6,0)} v^6 +   \varphi_6^{\rm e, \,(0,6)} v_0^6 + \varphi_6^{\rm e, \,(4,2)} v^4 v_0^2 + \varphi_6^{\rm e, \,(3,3)} v^3 v_0^3 
    \\+ \varphi_6^{\rm e, \,(2,4)} v^2 v_0^4 + \left( \varphi_6^{\rm e, \,(5,1)} v^5 v_0 + \varphi_6^{\rm e, \,(1,5)} v v_0^5 \right) 
    \\ + \varphi_6^{\rm e,\, log,\, (6,0)} v^6 \log v + \varphi_6^{\rm e,\, log,\, (0,6) } v_0^6 \log v_0
    \Bigg]\,.
\end{multline}
Here the PN coefficients $\varphi_i^{\rm e,\, (a,b)}$ multiply the terms proportional to $v^a v_0^b$, with $a+b=i$ for terms of $i/2$ PN order. Unlike the circular phasing at each PN order, there is more than one type of term at a particular PN order in the eccentric phasing. For instance, at 1PN order the phase contains a term proportional to $v^2$ and another proportional to $v_0^2$, as well as a ``mixed'' term proportional to $v v_0$ (which is zero in GR); at 2PN order there are terms proportional to $v^4$, $v_0^2$, and $v^2v_0^2$, as well as $v^3 v_0$ and $v v_0^3$ terms that vanish in GR. (Terms written in parentheses in the above expression correspond to those that vanish in GR.) At 3PN order we also see the presence of logarithmic terms with coefficients $\varphi_i^{\rm e,\, log,\, (a,b)}$, although we ignore the possibility of ``mixed'' log terms where both $a$ and $b$ are non-zero. 

The extension of parametrized GR tests then corresponds to the replacements:
\begin{subequations}
\label{eq:TGReccparams}
\begin{align}
    \varphi_i^{\rm e,\, (a,b)} &\xrightarrow{} \varphi_i^{\rm e,\, (a,b)} (1 + \delta \hat{\varphi}_{i}^{\rm e,\, (a,b)}) \,, \\
 \varphi_i^{\rm e,\, log,\, (a,b)} &\xrightarrow{} \varphi_i^{\rm e,\, log,\, (a,b)} (1 + \delta \hat{\varphi}_{i}^{\rm e,\, log,\,  (a,b)}) \,,
\end{align}
\end{subequations}
in the cases where the GR limit is non-zero. Here, the $\delta \hat{\varphi}_{i}^{\rm e,\, (a,b)}$ and $\delta \hat{\varphi}_{i}^{\rm e,\, log,\, (a,b)}$ are the eccentric fractional deviation parameters. Again, by definition, their values are zero in GR. In the case where the coefficients are zero in GR (terms in parentheses), the appropriate replacement is
\begin{equation}
    \varphi_i^{\rm e,\, (a,b)}  \xrightarrow{} \delta\hat{\varphi}_i^{\rm e,\, (a,b)} \,.
\end{equation}

The above phase expansion introduces 30 free parameters. To prevent this unwieldy proliferation of free parameters, and in the spirit of a null test of GR, we make two simplifying assumptions. First, we ignore terms where the coefficients are zero in GR (terms in parentheses). Second, we introduce only one type of deviation coefficient at each PN order in the eccentric phase. Hence, all the values of the $\delta\hat{\varphi}_i^{\rm e,\, (a,b)}$ for a given $i$ reduce to a single parameter $\delta\hat{\varphi}_i^{\rm e}$ (and similarly for the log terms). For instance, at 2PN order this would mean all three types of terms that are non-zero in GR $(\propto v^4, v_0^2, v^2 v_0^2)$ are corrected by a single fractional deviation parameter $\delta\hat{\varphi}_4^{\rm e}$ that parametrizes a possible GR violation. In this limit, Eq.~\eqref{eq:TGReccparams} simplifies to:
\begin{subequations}
\begin{align}
    \varphi_i^{\rm e,\, (a,b)} &\xrightarrow{} \varphi_i^{\rm e,\, (a,b)} (1 + \delta \hat{\varphi}_{i}^{e}) \,, \\
\varphi_i^{\rm e,\, log,\, (a,b)} &\xrightarrow{} \varphi_i^{\rm e,\, log,\, (a,b)}(1 + \delta \hat{\varphi}_{i}^{\rm e,\,log}) \, .
\end{align}
\end{subequations}
This reduces our eccentric deviation parameters to a set of 7 coefficients in the eccentric part of the waveform: $\{\delta\hat{\varphi}_0^{\rm e}, \delta\hat{\varphi}_2^{\rm e}, \delta\hat{\varphi}_3^{\rm e}, \delta\hat{\varphi}_4^{\rm e}, \delta\hat{\varphi}_5^{\rm e}, \delta\hat{\varphi}_6^{\rm e}, \delta\hat{\varphi}_6^{\rm e, log}\}$.

For a modified theory of gravity, it is expected that both $\delta \hat{\varphi}_{i}$ and $\delta \hat{\varphi}_{i}^{\rm e}$ (for a given PN order) will differ from GR. Therefore, for a particular $(i/2)$PN order, both $\delta \hat{\varphi}_{i}$ and $\delta \hat{\varphi}_{i}^{\rm e}$ should be measured simultaneously. Alternatively, one could consider the (perhaps less likely) case where the modified theory preserves the circular PN phase expansion and modifies only the eccentric dynamics. In that case, GR deviations can be constrained via measurements of a single $\delta \hat{\varphi}_{i}^{\rm e}$. In Sec.~\ref{sec:results parametric deviation} below, we consider both possibilities.

Though this parametrization is an obvious extension of the one employed in the case of circular orbits, the resulting bounds on the deviation parameters are found to be unconstrained for the current generation of GW detectors. This is true in both the case where two deviation parameters are considered $(\delta \hat{\varphi}_{i}, \delta \hat{\varphi}_{i}^{\rm e})$ and just one $(\delta \hat{\varphi}_{i}^{\rm e})$, along with the eccentricity and other source parameters. Therefore, this test is uninformative for detectors comparable to LIGO's sensitivity. This is shown in Sec.~\ref{sec:results parametric deviation} below. 

As an alternative to this multiparameter approach, we consider a waveform parametrization based on a physical effect that is unique to eccentric binaries: the periastron advance due to the gravitational dynamics. As we show in Sec.~\ref{sec:phasing}, such a parametrization contains only a single free parameter in the GW phasing; that parameter characterizes the difference in the periastron advance relative to GR. The next section discusses the details of this parametrization and presents the corresponding phasing.

\section{\label{sec:phasing} Non-GR eccentric GW phasing via parametrization of the periastron advance} 
Here we present a parametrization of the eccentric GW phasing in terms of a modification to the periastron advance. We first recap the quasi-Keplerian parametrization for eccentric orbits; then we discuss how a periastron parametrized phasing formula can be derived.

\subsection{Review of quasi-Keplerian representation}\label{QKReview}
\begin{figure}[t]
   \centering 
  {\includegraphics[width=0.48\textwidth]{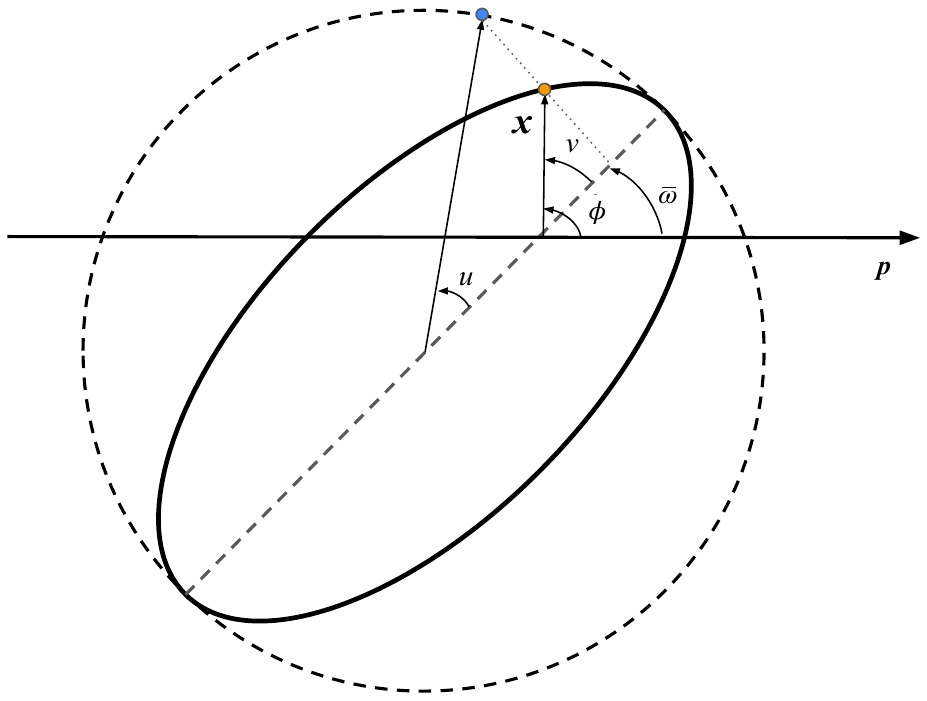}}
  \caption{(Color online) A geometric representation of the various orbital elements appearing in the quasi-Keplerian parametrization for eccentric orbits. The auxiliary circle circumscribes the orbital ellipse with semi-major axis $a$. The vector $\bm{x}$ represents the relative separation vector pointing from the binary's center of mass to the reduced mass particle. The angle $\phi$ from the unit vector $\bm{p}$ to $\bm{x}$ represents the orbital phase angle; $\bm p$ points in an arbitrary reference direction. The angle $\nu$ from the pericenter of the ellipse is the true anomaly. The angle $u$ from the pericenter of the ellipse to the projection of the reduced mass position to the circumscribing circle (and perpendicular to the major axis) denotes the eccentric anomaly. The angle $\Bar{\omega}$ is the argument of the pericenter. The mean anomaly $l$ (not shown) is the fraction of the radial orbital period (measured in angle) that has elapsed since the last pericenter passage; it does not have a geometric interpretation.
  }
  \label{fig:eccentric orbit}
\end{figure}
In this section we briefly summarize the {\it quasi-Keplerian} parametrization~\cite{1985AIHPA..43..107D,1988NCimB.101..127D,SCHAFER1993196,Damour:2004bz,Memmesheimer:2004cv,Konigsdorffer:2006zt,Trestini:2024zpi}; that formalism provides a parametric solution to the conservative part of the PN equations of motion for bound elliptical orbits. We closely follow the summary provided in Ref.~\cite{Moore:2016qxz}. Figure~\ref{fig:eccentric orbit} shows the orbital quantities that describe the geometry of an eccentric orbit. In the quasi-Keplerian framework the primary orbital variables are the binary separation $r$ and the orbital phase angle $\phi$. Along with their time derivatives $\Dot{r}$ and $\Dot{\phi}$, they are expressed in terms of several parametric angles: the eccentric anomaly $u$, the true anomaly $\nu$, and the mean anomaly $l$. When supplied with a numerical solution to the PN extension of Kepler's equation, this formalism determines the time-evolution of the orbital variables $(r, \phi)$ without actually solving ordinary differential equations (see Sec.~III of Ref.~\cite{Moore:2016qxz} for further details). The parametric equations for $r$, $\phi$, $\dot{r}$, and $\dot{\phi}$ for the conservative contributions to the equations of motion up to 3PN order (and ignoring radiation reaction) can be expressed as~\cite{Konigsdorffer:2006zt,Moore:2016qxz}: 
\begin{subequations}\label{quasiKeplEqns}
    \begin{align}
        r &=  S(l;n,e_t) = a_r(1-e_r \cos u) \,, \\[1ex]
        \dot{r} &= n \frac{\partial S}{\partial l}(l;n,e_t) \,, \\[1ex]
        \phi &= \lambda + W(l;n,e_t) \,, \\[1ex]
        \lambda &= (1+k)n(t-t_0) + c_\lambda \,,\\[1ex]
        W&= (1+k)(\nu-l)+ \bigg(\frac{f_{4\phi}}{c^4}+ \frac{f_{6 \phi}}{c^6} \bigg)\sin 2\nu \\[1ex] \nonumber 
        &+ \bigg(\frac{g_{4\phi}}{c^4} + \frac{g_{6 \phi}}{c^6} \bigg)\sin 3\nu + \frac{i_{6 \phi}}{c^6} \sin 4\nu \\[1ex] \nonumber
        &+ \frac{h_{6\phi}}{c^6} \sin 5\nu \,,\\[1ex]
        \dot{\phi} &= (1+k)n + n \frac{\partial W}{\partial l}(l;n,e_t) \label{eq:phidot}\,, \\[1ex]
        l &= n(t-t_0) + c_l = u - e_t \sin u \\[1ex] \nonumber
        &+ \bigg(\frac{g_{4t}}{c^4}+ \frac{g_{6t}}{c^6} \bigg) (\nu-u) + \bigg(\frac{f_{4t}}{c^4}+ \frac{f_{6t}}{c^6} \bigg) \sin \nu \\[1ex] \nonumber 
        & + \frac{i_{6t}}{c^6} \sin 2\nu + \frac{h_{6t}}{c^6} \sin 3\nu \,,\\[1ex]
            \nu &= V (u) \equiv 2 \arctan \bigg[\bigg(\frac{1+e_{\phi}}{1-e_{\phi}} \bigg)^2 \tan \bigg(\frac{u}{2}\bigg)\bigg] \,,
            \end{align}
\end{subequations}
 where $a_r = (M/n^2)^{1/3}$ is the semimajor axis of the orbit in geometric units, $c$ is the speed of light, $n \equiv 2\pi/P$ is the {\it mean motion}, and $P$ is the radial orbit period (periastron-to-periastron time). The phase angle $\lambda$ is the linearly accumulating piece of the orbital phase; in the Newtonian limit it reduces to the mean anomaly $l$. The functions $g_{4t}$, $g_{6t}$, $f_{4t}$, $f_{6t}$, $i_{6t}$, $h_{6t}$, $f_{4 \phi}$, $g_{4 \phi}$, $g_{6 \phi}$, $f_{6 \phi}$, $g_{6 \phi}$, $i_{6 \phi}$, and $h_{6 \phi}$ can be found in Refs.~\cite{Damour:2004bz,Memmesheimer:2004cv,Konigsdorffer:2006zt}. The functions $S(l)$ and $W(l)$ are periodic in $l$ with a period of $2\pi$. The constants $c_l$ and $c_\lambda$ are the two positional constants that provide the value of $l$ and $\lambda$ at time $t_0$. These two angles define the argument of pericenter, $\Bar{\omega} = c_l - c_\lambda$. The formalism also introduces three eccentricity parameters $(e_t, e_r, e_{\phi})$; we express our results in terms of $e_t$ using relations found in the above references.

A conservative orbit requires four initial conditions $[r(t_0), \phi(t_0), \dot{r}(t_0), \dot{\phi}(t_0)]$ or equivalently $[n, e_t, c_l, c_\lambda]$; here $n$ and $e_t$ are the intrinsic constants of motion that determine the orbit's shape, and $c_l$ and $c_\lambda$ are the two extrinsic constants of motion that define the orientation of the orbit and the orbital phase at $t_0$. When radiation reaction is taken into account, these constants are no longer fixed and evolve with time. In that case, the complete 3.5PN solution is obtained using the method of \emph{variation of constants}: the conservative 3PN solution is considered as the leading-order solution, while the 2.5PN and 3.5PN dissipative pieces of the equations of motion are treated as perturbations that cause the constants of motion to vary with time. These time-varying constants have both fast periodic variations and slowly varying (secular) contributions due to radiation reaction.\footnote{Note that the secular contributions to the varying constants of motion are traditionally denoted by overbars as in $[\bar{n}, \bar{e}_t, \bar{c}_l, \bar{c}_\lambda]$; however, since the periodic variations are negligible for our purposes, we drop the overbars from the secularly varying quantities in the sections that follow.} Fast periodic variations are negligible in the limit of small eccentricity, and the secular variations of the positional constants of motion vanish ($\dot{\bar{c_l}}= \dot{\bar{c_\lambda}}=0$). We are left with only the secular variations of the intrinsic constants ($\dot{\bar{n}}$ and $\dot{\bar{e_t}}$) which are required for calculating the SPA phase as discussed in Sec.~\ref{sec:periastron parametrization}. 

\subsection{Computation of the parametrized phasing}\label{sec:periastron parametrization}
The secular effect of periastron advance is captured by the periastron advance parameter $k= \Delta\Phi/2\pi$, where $\Delta\Phi$ is the advance of the periastron angle in one radial orbit period $P$ (the time between adjacent periastron passages). In the quasi-Keplerian formalism, elliptical orbit variables are expanded in a PN series and are most naturally expressed in terms of the radial orbit angular frequency $\omega_r \equiv n \equiv \xi / M = 2\pi/P$. However, the orbital variables for circular orbits are more naturally expressed in terms of the azimuthal frequency $\omega_\phi$ (frequency corresponding to the period to return to the same angular coordinate $\phi$). Equation~\eqref{eq:phidot} relates the azimuthal frequency $\xi_\phi$ and radial frequency $\xi_r\equiv \xi$ to the periastron advance parameter $k$. 
Taking the orbit-average of Eq.~\eqref{eq:phidot} leads to 
\begin{equation} \label{eq:azimuthal frequency}
\omega_\phi \equiv  \xi_\phi /M \equiv \langle \dot{\phi} \rangle = \frac{d\lambda}{dt} = (1+k) \xi/M \,.
\end{equation}
Note that the oscillatory term $\frac{\partial W}{\partial l}$ in Eq.~\eqref{eq:phidot} vanishes after orbit averaging and does not appear in the above equation. 

In a modified theory of gravity, the periastron is expected to advance at a different rate than predicted by GR. As an alternative to the parametrized test of GR formalism discussed in Sec.~\ref{sec:parametric deviation} (which modifies each coefficient of the PN phase expansion), we consider the modification of the periastron advance as the basis for a new parametrization of the GW phasing. We introduce this parametrization via a modification of Eq.~\eqref{eq:azimuthal frequency} above: 
  \begin{equation}\label{eq:xirtophi}
     \xi_\phi= [1+ k(1+\Delta\alpha)] \xi,
 \end{equation}
where $\Delta \alpha$ is introduced as a non-GR parameter that captures the fractional deviation in the periastron advance parameter. In the GR limit $\Delta \alpha \to 0$. We will refer to $\Delta \alpha$ as the {\it periastron deviation parameter}. For mathematical convenience in the expansions below, we also define an intermediate parameter $\alpha \equiv 1+\Delta \alpha$. Most of our calculations and intermediate results are performed in terms of $\alpha$ (with $\alpha \rightarrow 1$ in the GR limit). However, our final result for the SPA phasing [Eq.~\eqref{eq:PsiFTecc} below] is presented in terms of the fractional deviation parameter $\Delta \alpha$.

Our task is now to obtain the SPA phasing in terms of this new parameter $\Delta \alpha$. However, it is more natural to express the orbital variables in terms of the azimuthal frequency $\xi_{\phi}$ rather the radial frequency $\xi$; this allows a straightforward comparison with expressions written in the circular limit (where $\xi_{\phi}$ is the natural frequency variable). The azimuthal frequency is also related to the standard PN expansion parameter $v$ (the relative orbital speed) via $\xi_{\phi} = M \omega_{\phi} = v^3$. It is also convenient to define another commonly used PN expansion parameter, $x\equiv \xi_{\phi}^{2/3}$. Hence, our goal is to determine the SPA phasing as a function $\Psi = \Psi(\xi_\phi, e_t, \Delta\alpha)$ (along with its dependence on the source parameters that enter the circular case). Some of the expressions for various quantities that are needed to compute $\Psi$ are available in the literature in the Arnowitt-Deser-Misner (ADM) gauge, whereas some are available in modified harmonic (MH) gauge. Since the {\tt TaylorF2Ecc} waveform model (the basis for our periastron parametrization) is obtained using the MH gauge, we convert the expressions from ADM to MH gauge (and vice versa) whenever required.

The periastron parameter $k$ in Eq.~\eqref{eq:xirtophi} is a function of $\xi$ and $e_t$. The expression for $k$ in terms of the dimensionless energy $\epsilon$ and dimensionless angular momentum $j$ is given in Eq.~(6.2b) of Ref.~\cite{ABIS09}. Further, Eqs.~(6.5a) and (6.5b) of Ref.~\cite{ABIS09} express $\epsilon$ and $j$ in terms of the parameters $x$ and $e_t$ in ADM coordinates. Further, $x$ is expressed in terms of the radial frequency $x=x(\xi, e_t^{\rm ADM})$ via Eq.~(4.17) of Ref.~\cite{ABIS09}. Hence, substituting Eqs.~(6.5a), (6.5b), and (4.17) in Eq.~(6.2b) of Ref.~\cite{ABIS09} gives the expression for $k = k(\xi, e_t^{\rm ADM})$. Taylor expanding in $\xi$ and retaining the terms up to 3PN order yields:
\begin{widetext}
 \begin{multline}
\label{eq:k3pnADM}
k(\xi, e_t^{\rm ADM}) = \frac{ 3 \xi ^{2/3} }{ 1 - e_t^2 }
+ \left[ 78  - 28 \eta + ( 51 - 26 \eta ) e_t^2 \right]\frac{ \xi ^{4/3} }{ 4 ( 1 - e_t^2 )^2 }
 + \Big \{
18240 - 25376 \eta + 492 \pi^2 \eta + 896 \eta ^2
\\
+ ( 27936 - 31104 \eta 
+ 123 \pi^2 \eta + 5120 \eta^2) e_t^2
+ ( 2496 - 1760 \eta + 1040 \eta^2  ) e_t^4
\\
+ \left[ 1920 - 768 \eta + ( 3840 - 1536 \eta ) e_t^2 \right] \sqrt{1 - e_t^2}
\Big \}\frac{ \xi^2 }{ 128 ( 1 - e_t^2 )^3 } \,.
\end{multline}
The corresponding expression for $k$ in MH gauge is (see Eq.(3.3) of~\cite{Moore:2016qxz}):
\begin{multline}
\label{eq:k3pnMH}
k(\xi, e_t^{\rm MH}) = \frac{ 3 \xi ^{2/3} }{ 1 - e_t^2 }
+ \left[ 78  - 28 \eta + ( 51 - 26 \eta ) e_t^2 \right]\frac{ \xi ^{4/3} }{ 4 ( 1 - e_t^2 )^2 }
+ \Big \{
18240 - 25376 \eta + 492 \pi^2 \eta + 896 \eta ^2
\\
+ ( 28128 - 27840 \eta 
+ 123 \pi^2 \eta + 5120 \eta^2 ) e_t^2
+ ( 2496 - 1760 \eta + 1040 \eta^2  ) e_t^4
\\
+ \left[ 1920 - 768 \eta + ( 3840 - 1536 \eta ) e_t^2 \right] \sqrt{1 - e_t^2}
\Big \}\frac{ \xi^2 }{ 128 ( 1 - e_t^2 )^3 } \,.
\end{multline}
Note that the ADM and MH expressions are identical up to 1PN order and only differ at 2PN or higher orders. Substituting Eq.~\eqref{eq:k3pnADM} for $k(\xi,e_t^{\rm ADM})$ in Eq.~\eqref{eq:xirtophi} [using the form $\xi_{\phi} = (1+\alpha k)\xi$] and inverting the resulting equation yields a series for $\xi(\xi_\phi, e_t^{\rm ADM}, \alpha)$:
\begin{multline}
\label{eq:zeeADM}
\xi(\xi_\phi, e_t^{\rm ADM}, \alpha) = \xi_{\phi} \Big \{1 - \frac{ 3 \alpha \xi_\phi ^{2/3} }{ 1 - e_t^2 }
- \left[  78 \alpha -60 \alpha ^2 -28 \alpha  \eta +e_t ^2 \left(51 \alpha -26 \alpha  \eta \right) \right]\frac{ \xi_\phi ^{4/3} }{ 4 ( 1 - e_t^2 )^2 }
\\
- \Big \{
18240 \alpha -29952 \alpha ^2 +11520 \alpha ^3 -25376 \alpha \eta +492 \pi ^2 \alpha  \eta +10752 \alpha ^2 \eta +896 \alpha  \eta ^2
\\
+ ( 27936 \alpha -19584 \alpha ^2 -31104 \alpha  \eta +123 \pi ^2 \alpha  \eta + 9984 \alpha ^2 \eta  +5120 \alpha  \eta ^2 ) e_t^2
+ ( 2496 \alpha -1760 \alpha  \eta +1040 \alpha  \eta ^2 ) e_t^4
\\
+ \left[ 1920 \alpha -768 \alpha  \eta + ( 3840 \alpha -1536 \alpha  \eta) e_t^2 \right] \sqrt{1 - e_t^2}
\Big \}\frac{ \xi_{\phi}^2 }{ 128 ( 1 - e_t^2 )^3 } \Big \} \,.
\end{multline}
Similarly one can obtain a power series for $\xi(\xi_\phi, e_t^{\rm MH}, \alpha)$ by substituting Eq.~\eqref{eq:k3pnMH} $[k(\xi, e_t^{\rm MH})]$ in \eqref{eq:xirtophi} and inverting the resulting equation:
\begin{multline}
\label{eq:zeeMH}
\xi(\xi_\phi, e_t^{\rm MH}, \alpha) =\xi_{\phi} \Big \{ 1 - \frac{ 3 \alpha \xi_\phi ^{2/3} }{ 1 - e_t^2 }
- \left[  78 \alpha -60 \alpha ^2 -28 \alpha  \eta +e_t ^2 \left(51 \alpha -26 \alpha  \eta \right) \right]\frac{ \xi_\phi ^{4/3} }{ 4 ( 1 - e_t^2 )^2 }
\\
- \Big \{
18240 \alpha -29952 \alpha ^2 +11520 \alpha ^3 -25376 \alpha \eta +492 \pi ^2 \alpha  \eta +10752 \alpha ^2 \eta +896 \alpha  \eta ^2
\\
+ ( 28128 \alpha -19584 \alpha ^2-27840 \alpha  \eta +123 \pi ^2 \alpha  \eta +9984 \alpha ^2 \eta +5120 \alpha 
   \eta ^2) e_t^2
+ ( 2496 \alpha -1760 \alpha  \eta +1040 \alpha  \eta ^2 ) e_t^4
\\
+ \left[ 1920 \alpha -768 \alpha  \eta + ( 3840 \alpha -1536 \alpha  \eta) e_t^2 \right] \sqrt{1 - e_t^2}
\Big \}\frac{ \xi_{\phi}^2 }{ 128 ( 1 - e_t^2 )^3 } \Big \} \,.
\end{multline}
Note that in the limit $e_t\to 0$ the frequencies $\xi$ and $\xi_\phi$ are not identical. This is because in the $e_t\to 0$ limit, the periastron advance angle is independent of eccentricity. In the $e_t\to 0$ limit, PN expressions for eccentric orbits reduce to the standard circular results only if they are a function of $\xi_\phi$~\cite{Moore:2016qxz}. 

To incorporate the effects of radiation reaction, we next compute the secular evolution of the constants of motion $(\xi_{\phi}, e_t)$, which will eventually be used to obtain the SPA phase parametrized in terms of $\alpha$. The expressions for $\dot{\xi}_{\phi}$ and $\dot{e}_t$ have been calculated to 3PN order in ADM coordinates in Ref.~\cite{ABIS09}. Since we work in MH coordinates, we need to convert those results from ADM to MH coordinates. Equation~(4.15) of \cite{ABIS09} relates $e_t^{\rm ADM}$ to $e_t^{\rm MH}$ via $e_t^{\rm ADM} = e_t^{\rm ADM}(e_t^{\rm MH}, x)$. Substituting Eq.~(C14) of Ref.~\cite{ABIS09} [$x=x(e_t^{\rm MH}, \xi)$] in Eq.~(4.15) of~\cite{ABIS09}, we get:
\begin{multline}
\label{eq:eADM}
   e_t^{\rm ADM}(\xi, e_t^{\rm MH}) = e_t^{\text{MH}}\left\{1+\left(\frac{1}{4}+\frac{17}{4}\eta\right)\frac{\xi^{4/3}}{1-e_t^2}+\left[\frac{3}{2} +\left(\frac{45299}{1680}-\frac{21}{16} \pi^2 \right)\eta -\frac{83}{24} \eta^2
 \right. \right.  \\ \left. \left.
 +\left(\frac{1}{2}+\frac{249}{16}\eta-\frac{241}{24}\eta^2\right)e_t^2\right]\frac{\xi^2}{\left(1-e_t^2\right)^2}\right\},
\end{multline}
where $e_t=e_t^{\rm MH}$ on the right-hand side. Similarly, substituting Eq.~(4.17) of \cite{ABIS09} [$x=x(e^{\rm ADM}, \xi)$] into Eq.~(8.21) of~\cite{ABIS08} [which relates $e_t^{\rm MH}$ to $e_t^{\rm ADM}$ via $e_t^{\rm MH}=e_t^{\rm MH}(e_t^{\rm ADM}, x)$], we get:
\begin{multline}
\label{eq:eMH}
e_t^{\rm MH}(\xi, e_t^{\rm ADM}) = e_t^{\rm ADM}\left\{1-\left(\frac{1}{4}+\frac{17}{4}\eta\right)\frac{\xi^{4/3}}{(1-e_t^2)} -\left[\frac{3}{2} +\left(\frac{45299}{1680}-\frac{21}{16} \pi ^2\right)
\eta -\frac{83 }{24}\eta ^2
\right. \right.  \\ \left. \left.
+ \left(\frac{1}{2}+\frac{249 }{16} \eta-\frac{241}{24} \eta ^2\right)e_t^2\right]\frac{\xi^2}{\left(1-e_t^2\right)^2} \right\}\,,
\end{multline}
where $e_t=e_t^{\rm ADM}$ on the right-hand side. For the two time eccentricity relations above, $\xi$ can be expressed as a power series in $\xi_\phi$ i.e. $\xi=\xi(\xi_\phi, e_t, \alpha)$ via substituting Eqs.~\eqref{eq:zeeMH} and \eqref{eq:zeeADM} in Eqs.~\eqref{eq:eADM} and \eqref{eq:eMH} (respectively), obtaining:
\begin{multline}
\label{eq:ADMtoMH}
e_t^{\rm ADM}(\xi_\phi, e_t^{\rm MH}, \alpha) = e_t^{\text{MH}}\left\{1+\left(\frac{1}{4}+\frac{17}{4}\eta\right)\frac{\xi_{\phi}^{4/3}}{1-e_t^2}+ \left[\left(\frac{3}{2} - \alpha\right)+\left(\frac{45299}{1680}-17\alpha-\frac{21}{16} \pi^2 \right)\eta
\right. \right.  \\ \left. \left.
-\frac{83}{24} \eta^2 +
 \left(\frac{1}{2}+\frac{249}{16}\eta-\frac{241}{24}\eta^2\right)e_t^2\right]\frac{\xi_{\phi}^2}{\left(1-e_t^2\right)^2}\right\},
\end{multline}
where $e_t=e_t^{\rm MH}$ on the right-hand side. The corresponding inverse transformation is:
\begin{multline}
\label{eq:MHtoADM}
e_t^{\rm MH}(\xi_\phi, e_t^{\rm ADM}, \alpha) = e_t^{\rm ADM}\left\{1-\left(\frac{1}{4}+\frac{17}{4}\eta\right)\frac{\xi_{\phi}^{4/3}}{(1-e_t^2)} -\left[\left(\frac{3}{2}-\alpha \right)+\left(\frac{45299}{1680}-17 \alpha-\frac{21}{16} \pi ^2\right)
\eta 
\right. \right.  \\ \left. \left.
-\frac{83 }{24}\eta ^2+ \left(\frac{1}{2}+\frac{249 }{16} \eta-\frac{241}{24} \eta ^2\right)e_t^2\right]\frac{\xi_{\phi}^2}{\left(1-e_t^2\right)^2} \right\},
\end{multline} 
where $e_t = e_t^{\rm ADM}$ on the right-hand side. 
 
To compute $\Psi$, we need the equations describing the time derivatives of the azimuthal frequency $d\xi_{\phi}/dt$ and the time eccentricity $de_t/dt$. To obtain the time evolution of $\xi_{\phi}$, we take the time derivative of Eq.~\eqref{eq:xirtophi}
\begin{equation}\label{eq:xiphievolution}
      \frac{d\xi_\phi}{dt}= (1+\alpha k) \bigg(\frac{d\xi}{dt}\bigg)_{\rm ADM}+\alpha \xi \bigg(\frac{dk}{dt}\bigg)_{\rm ADM} \, ,
\end{equation} 
where $k$ and $\xi$ on the right-hand side of the above equation are in ADM coordinates. Since the expressions for $d\xi/dt$ and $dk/dt$ are available in ADM gauge, we first express the above equation in terms of ADM coordinates and then use transformations (ADM $\to$ MH) to convert $d\xi_{\phi}/dt$ to MH coordinates. 

For $(d\xi/dt)_{\rm ADM}$ and $(dk/dt)_{\rm ADM}$ in Eq.~\eqref{eq:xiphievolution}, we use Eqs.~(7.6a) and (7.6b) of Arun \emph{et al.}~\cite{ABIS09}, specializing their results to leading order in $e_t$ $[\mathcal{O}(e_t^2)]$. These equations are given in terms of the parameters $x$ and $e_t$ in ADM gauge:
\begin{subequations}
    \begin{align}
           \bigg(\frac{d\xi}{dt}\bigg)_{\rm ADM} &=\frac{96 \eta }{5 M } x^{11/2}\,\left\{1+\left(-\frac{2423}{336} -\frac{11}{4} \eta
\right)x+ \cdots +\mathcal{O}(x^{3})
+e_t^2\left[\frac{157}{24}+\left(-\frac{2801}{112} -\frac{673}{16} \eta
\right)x+ \cdots +\mathcal{O}(x^{3}) \right] \right\}\,, \\  
        \bigg(\frac{dk}{dt}\bigg)_{\rm ADM} &= 
\frac{192 \eta}{5 M} x^5\left\{ 1+
\left(\frac{2281}{336}-\frac{89 }{12}\eta \right) x+ \cdots + +\mathcal{O}(x^{2})  
+e_t^2\left[\frac{35}{8}+ \left(\frac{11\,595}{224}-\frac{851}{16} \eta
\right) x+ \cdots  +\mathcal{O}(x^{2}) \right]\right\} \,.
    \end{align}
\end{subequations}
Substituting Eq.~(4.17) of Ref.~\cite{ABIS09} for $x=x(\xi, e_t^{\rm ADM})$, Eq.~\eqref{eq:k3pnADM} for $k=k(\xi,e_t^{\rm ADM})$, and Eq.~\eqref{eq:zeeADM} for $\xi=\xi(\xi_\phi, e_t^{\rm ADM},\alpha)$ in Eq.~\eqref{eq:xiphievolution}, we get an expression for $d\xi_{\phi}/dt=d\xi_{\phi}/dt(\xi_\phi, e_t^{\rm ADM},\alpha)$. Finally, we use the transformation $e_t^{\rm ADM} = e_t^{\rm ADM}(\xi_\phi,e_t^{\rm MH}, \alpha)$  [Eq.~\eqref{eq:ADMtoMH}] to transform that expression to MH gauge: $d\xi_{\phi}/dt=d\xi_{\phi}/dt(\xi_\phi, e_t^{\rm MH}, \alpha)$. Simultaneously expanding the resulting equation in $e_t^{\rm MH}$ and $\xi_\phi$ and keeping only terms up to $\mathcal{O}(e_t^2)$ and $\mathcal{O}(\xi_\phi^2)$, we get an appropriately expanded expression for $d\xi_\phi/dt$ in MH gauge to 3PN order in the small eccentricity limit. This expression can be written as:
\begin{multline} 
\label{eq:dxidt-MH-lowe}
\frac{d\xi_{\phi}}{dt} = M \frac{d\omega_{\phi}}{dt} = \frac{96 \eta  \xi_{\phi} ^{11/3}}{5 M} \bigg\{1+\left(\frac{1273}{336}-6 \alpha -\frac{11 \eta }{4}\right) \xi_{\phi} ^{2/3} + \cdots +\mathcal{O}(\xi_\phi^{7/3}) 
+ e_t^2 \bigg[\frac{157}{24}+\left(\frac{19471}{336}-\frac{619 \alpha }{12}-\frac{673 \eta }{16}\right) \xi_{\phi}^{2/3}  \\ +  \cdots+ \mathcal{O}(\xi_\phi^{7/3}) \bigg] \bigg\} \, .
\end{multline} 
The full expression for $d\xi_\phi/dt$ can be found in Appendix~\ref{app:full expressions}. In the limit $\alpha\to1$, all the expressions reduce to their GR limits and match with the expressions in Ref.~\cite{Moore:2016qxz}. 

To compute $de_t/dt$ in MH gauge, we take the time derivative of Eq.~\eqref{eq:eMH}: 
\begin{equation}\label{edotMH}
    \frac{de_t^{\rm MH}}{dt} = \frac{\partial{e_t^{\rm MH}} }{\partial{e_t^{\rm ADM}} } \bigg(\frac{de_t^{\rm ADM}}{dt}\bigg)+ \frac{\partial{e_t^{\rm MH}}}{\partial{\xi}} \bigg(\frac{d \xi}{dt}\bigg)_{\rm ADM}\,.
\end{equation}
The expressions for $de_t^{\rm ADM}/dt$ and $(d\xi/dt)_{\rm ADM}$ in the above equation are given in Eqs.~(7.6e) and (7.6a) of Ref.~\cite{ABIS09} in terms of $x$ and $e_t^{\rm ADM}$. We use Eq.~(4.17) for $x=x(\xi,e_t^{\rm ADM})$ in Ref.~\cite{ABIS09} and $\xi = \xi(\xi_\phi, e_t^{\rm ADM}, \alpha)$ in Eq.~\eqref{eq:zeeADM} above to get $de^{\rm ADM}_t/dt = de_t/dt(\xi_\phi,\alpha, e_t^{\rm ADM})$. Finally, we use Eq.~\eqref{eq:ADMtoMH} $[e_t^{\rm ADM} = e_t^{\rm ADM}(\xi_\phi, e_t^{\rm MH}, \alpha)]$ to get $de^{\rm ADM}_t/dt$ in terms of $\xi_\phi$, $e_t^{\rm MH}$, and $\alpha$. The resulting equation is simultaneously expanded to $\mathcal{O}(e_t^2)$ and to 3PN order $\mathcal{O}(\xi_{\phi}^2)$: 
\begin{multline}
\label{eq:dedt-MH-lowe}
\frac{de_t}{dt} =  -\frac{304 e_t \eta  \xi_{\phi}^{8/3}}{15 M} \left\{1+\left( \frac{14207}{2128}-8 \alpha -\frac{1021 \eta }{228} \right) \xi_{\phi}^{2/3} + \cdots + \mathcal{O}(\xi_\phi^{7/3}) +e_t^2 \left[\frac{881}{304}+\left(\frac{172835}{4256}-\frac{1185 \alpha}{38}
 \right. \right. \right.  \\  \left. \left. \left.
-\frac{51847 \eta }{1824} \right) \xi_{\phi}^{2/3} + \cdots +\mathcal{O}(\xi_\phi^{7/3}) \right]\right\}. 
\end{multline}
Note that from here onwards $e_t$ refers to $e^{\rm MH}_t$.

Using Eqs.~\eqref{eq:dxidt-MH-lowe} and \eqref{eq:dedt-MH-lowe} we can obtain the differential equation for $d\xi_{\phi}/de_t$ in MH coordinates. The resulting 3PN equation is not separable, so an exact solution is not easily found for arbitrary eccentricity. However, an analytic result can be obtained if we only include the leading-order eccentricity terms $\mathcal{O}(e_t)$. Expanding the low-eccentricity limit of $de_t/d\xi_{\phi}$ in $\xi_{\phi}$ to order $\mathcal{O}(\xi_{\phi}^2)$ gives:
\begin{equation}
\label{eq:dedxi-lowe}
\frac{de_t}{d\xi_{\phi}} = -\frac{19}{18}\frac{e_t}{\xi_{\phi}}\left[1+\left(\frac{9217}{3192}-2 \alpha -\frac{197 \eta }{114}\right) \xi_{\phi} ^{2/3}+ \frac{377}{152}\pi  \xi_{\phi} +\cdots +\mathcal{O}(\xi_\phi^{7/3})\right].
\end{equation}
Integrating the above equation after the separation of variables, and expanding the resulting equation in $\xi_{\phi}$ to 3PN order $\mathcal{O}(\xi_{\phi}^2)$ gives 
\begin{equation}
\label{eq:etofxi}
e_t = e_0 \left( \frac{\xi_{\phi,0}}{\xi_{\phi}} \right)^{19/18} \frac{{\mathcal E}(\xi_{\phi})}{{\mathcal E}(\xi_{\phi,0})},
\end{equation}
where
\begin{equation}
\label{eq:E}
{\mathcal E}(\xi_{\phi}) = \bigg[1+\left(-\frac{9217}{2016}+\frac{19 \alpha }{6}+\frac{197 \eta }{72}\right) \xi_{\phi} ^{2/3}  + \cdots + \mathcal{O}(\xi_\phi^{7/3}) \bigg] \,,
\end{equation}
and $e_0=e_t(\xi_{\phi,0})$ is the initial condition that determines the constant of integration.

Using Eqs.~\eqref{eq:dxidt-MH-lowe} and \eqref{eq:etofxi}, we can obtain the evolution of frequency with time in the small eccentricity limit: 
\begin{multline}
\label{eq:dxidtexpand2}
\frac{d\xi_{\phi}}{dt} = \frac{96 \eta  \xi_{\phi}^{11/3}}{5M} \left\{1-\left(-\frac{1273}{336}+6 \alpha +\frac{11 \eta }{4}\right) \xi_{\phi} ^{2/3}+ \cdots +\mathcal{O}(\xi_\phi^{7/3})+\frac{157}{24} e_0^2 \left(\frac{\xi_{\phi,0}}{\xi_{\phi} }\right)^{19/9}
\left[1- \left(\frac{6451}{22608}+\frac{731 \alpha }{471} \right. \right. \right. \\ \left. \left. \left.
+\frac{5413 \eta }{5652}\right) \xi_{\phi}^{2/3}+\left(\frac{9217}{1008}-\frac{19 \alpha }{3}-\frac{197 \eta }{36}\right)
\xi_{\phi,0}^{2/3}+ \cdots +\mathcal{O}(\xi_{\phi,0}^{7/3})+\cdots+ \mathcal{O}(\xi_\phi^{7/3}) \right] \right\} ,
\end{multline}
The full expressions for Eq.~\eqref{eq:dxidtexpand2} and other intermediate results in this section can be found in Appendix \ref{app:full expressions}. The above equation determines the frequency evolution and allows us to compute the phase evolution. The frequency and phase evolution can be used to compute the SPA phasing in the small eccentricity limit. The time to coalescence $t(v)$ and phase $\langle\phi\rangle(v)$ are obtained by integrating:
\begin{subequations}
\label{eq:approx_freqdom}
\begin{align}
\label{eq:appox_freqdom_t_b}
dt &= \frac{dv}{(dv/dt)} \, , \\
\label{eq:approx_freqdom_phi}
\frac{d\langle\phi\rangle}{dv} &= \frac{d\langle\phi\rangle}{dt}\frac{dt}{dv}=-\frac{v^3}{M} \frac{dt}{dv} \,.
\end{align}
\end{subequations}
To calculate $dv/dt$, we substitute $v=\xi_\phi^{1/3}, v_0 = \xi_{\phi,0}^{1/3}$ in Eq.~\eqref{eq:dxidtexpand2}. To compute $t(v)$ and $\langle\phi\rangle (v)$, we invert Eq.~\eqref{eq:dxidtexpand2} and simultaneously expand the resulting series to $\mathcal{O}(e_0^2)$ and to 3PN order in $v$ and $v_0$. Integrating the resulting equation and keeping the $\mathcal{O}(e_0^2)$ and 3PN terms yields:
\begin{subequations}\label{eq:t(v)}
\begin{align}
     t_{c}-t=\frac{5}{256 } \frac{M}{\eta} \frac{1}{v^8} {\mathcal T}(v,v_0,e_0) \,, \;\;\; \text{where}
 \end{align}    
\begin{multline}
\label{eq:T_PNapprox_b}
{\mathcal T}(v,v_0,e_0) =  \left\{1+ \left(-\frac{1273}{252}+8 \alpha +\frac{11 \eta }{3}\right)v^2+ \cdots + \mathcal{O}(v^7) -\frac{157}{43} e_0^2 \left(\frac{v_0}{v}\right)^{19/3} \left[1+\left(-\frac{53505889}{5855472}+\frac{5719 \alpha }{471} \right.  \right. \right. \\ \left. \left. \left.  
+\frac{1103939 \eta }{209124}\right)v^2 
+\left(\frac{9217}{1008}-\frac{19 \alpha }{3} 
-\frac{197 \eta }{36} \right) v_0^2 + \cdots + \mathcal{O}(v^7)+ \cdots+ \mathcal{O}(v_0^7) \right] \right\} \,.
\end{multline}
\end{subequations}
Similarly, the equation for $\langle\phi\rangle(v)$ can be written as:
\begin{subequations}
\label{eq:phi(v)}
\begin{equation}
\label{eq:taylort2_phi_a}
\langle\phi\rangle-\phi_{c}=-\frac{1}{32 v^5 \eta } \Lambda_f(v,v_0,e_0) \,,  \;\;\; \text{where}
\end{equation}
\begin{multline}
\label{eq:Lambdaf_PNapprox}
\Lambda_f(v,v_0,e_0) = \left\{1+ \left(-\frac{6365}{1008}+10 \alpha +\frac{55 \eta }{12}\right)v^2 + \cdots +\mathcal{O}(v^7)
-\frac{785}{272} e_0^2 \left(\frac{v_0}{v}\right)^{19/3}
\left[1
+\left(-\frac{21153491}{2215584}+\frac{11951 \alpha }{942} \right. \right. \right. \\ \left. \left. \left. +\frac{436441 \eta }{79128}\right)v^2 +\cdots +\mathcal{O}(v^7) + \cdots +\mathcal{O}(v_0^7)\right]\right\} \,.
\end{multline}
\end{subequations}
Here $t_c$ and $\phi_c$ are the time and phase of coalescence. 
\end{widetext}

The frequency and phase evolution equations can be used to compute the SPA phasing in the small eccentricity limit (next subsection).
The parametric solution for $[t(v), \langle\phi\rangle(v)]$ can also be obtained using the adiabatic energy $E(v)$ and flux $\mathcal{F}(v)$ equations:
\begin{subequations}
\label{eq:approx_freqdom_b}
\begin{align}
\label{eq:approx_freqdom_phi_b}
\frac{d\langle\phi\rangle}{dv} &= \frac{d\langle\phi\rangle}{dt}\frac{dt}{dv}=-\frac{v^3}{M} \frac{dE(v)/dv}{{\mathcal F}(v)} \,, \\
\label{eq:appox_freqdom_t_c}
\frac{dt}{dv} &= -\frac{dE(v)/dv}{{\mathcal F}(v)} \, .
\end{align}
\end{subequations}
The expressions obtained with both methods are consistent with each other. Appendix~\ref{app:alternate method} provides the intermediate steps to obtain the SPA phasing using energy and flux expressions. 

\subsection{\label{subsec:SPAphase}SPA Phasing}
Now we can obtain the {\tt TaylorF2} approximant which is a frequency-domain waveform evaluated using the SPA phase. Equation~\eqref{strain} can be written as
\begin{subequations}
\begin{align}
h(t) &=  A(t) \cos[\Phi(t)]=A(t) \cos[2\phi(t) - 2\Phi_0]\,,\\
    \text {where} \;
        A(t) &= -\frac{4 \eta M}{d_L} [v(t)]^2  \nonumber \bigg[\Big(\frac{1+\cos^2\iota}{2}\Big)^2 F_+^2 \\ 
        &+ \cos^2\iota F_\times^2\bigg]^{1/2} \,, \\
        \text{and} \;\Phi_0 &= \frac{1}{2} \arctan\bigg[\frac{2 F_\times \cos\iota}{F_+ (1+\cos^2\iota)}\bigg] \,.
    \end{align}
\end{subequations}
Defining the Fourier transform of $h(t)$ via
\begin{align}
    \tilde{h}(f) \equiv \int_{-\infty}^{\infty} h(t)e^{2\pi i f t} dt\,,
\end{align}
and applying the SPA approximation, $\tilde{h}(f)$ can be written as
\begin{align}
\tilde{h}(f) &= {\mathcal A} f^{-7/6} e^{i \Psi}\,,
\end{align}
where the pattern-inclination-averaged amplitude $\mathcal{A}$ is given by Eq.~\eqref{amplitude} and the SPA phase $\Psi$ reads as 
\begin{align}\label{eq:PsiFT}
\Psi = 2 \pi f t_0(f) - 2 \phi[t(f)] + 2\Phi_0 - \frac{\pi}{4}.
\end{align}
Here $t=t_0$ is the location of the stationary point which is given by $2\pi f = d\Phi/dt(t_0)$, and $\phi$ in the above equation is only the secular contribution to the orbital phase (i.e., we are ignoring all oscillatory terms). Using Eqs.~\eqref{eq:t(v)} and \eqref{eq:phi(v)} for $t_0$ and $\phi$, converting to the $\Delta\alpha$ parameter, and simplifying, we obtain the final result for the SPA phase parametrized in terms of $\Delta \alpha$:
\begin{equation}
\label{eq:PsiFTecc}
\Psi = \Psi_{\rm GR}(f) + \delta\Psi_{\rm circ.}(f, \Delta\alpha) + \delta\Psi_{\rm ecc.}(f, \Delta\alpha).
\end{equation}
Here $\Psi_{\rm GR}(f)$ is the complete 3PN SPA phasing in GR given by the {\tt TaylorF2Ecc} waveform [Eqs.~\eqref{phase}, \eqref{eq:circ_phase} and \eqref{eccentric phase} here, or Eq.~(6.26) in \cite{Moore:2016qxz}]. The $\Delta\alpha$-dependent corrections modify both the circular and eccentric pieces of the {\tt TaylorF2Ecc} waveform. They are given by:
\begin{widetext}
\begin{subequations}\label{eq:deltaPsi}
    \begin{multline}
    \label{eq:deltaPsicirc}
    \delta\Psi_{\rm circ.}(f, \Delta\alpha) = \frac{3}{128 v^5 \eta } \left\{ \Delta \alpha \left[\frac{40}{3} v^2+ \left(\frac{6235}{42}+\frac{50 \eta }{3}\right)v^4+\frac{160}{3}  \left(1+\log v^3 \right) \pi  v^5+ \left(-\frac{55006045}{127008}+\frac{10005 \eta }{7}
   \right. \right. \right. 
   \\ \left. \left. \left.
   -\frac{205 \pi ^2 \eta }{4}-\frac{145 \eta ^2}{18}\right)v^6
   \right]  + \Delta \alpha ^2 \left[20 v^4+ \left(\frac{8755}{84}-\frac{115 \eta }{3}\right)v^6 \right] -\frac{80}{3} \Delta \alpha^3  v^6  \right\} \,,
    \end{multline}
    \begin{multline}
    \label{eq:deltaPsiecc}
    \delta\Psi_{\rm ecc.}(f, \Delta\alpha) = \frac{3}{128 v^5 \eta} \left[ -\frac{2355 }{1462} \left(\frac{v_0}{v}\right)^{19/3} e_{0}^2 \right] \left\{ \Delta \alpha \left[ \frac{13889 }{942} v^2-\frac{19 }{3}v_0^2+\left(\frac{354988951}{5222448}+\frac{8249335 \eta }{186516}\right) v^4
     \right. \right.  
     \\  
    +\left(-\frac{99553}{3024}+\frac{6521 \eta
   }{108}\right) v_0^4+\left(\frac{2254281923}{122964912} 
   -\frac{133153843 \eta }{1097901}\right) v^2 v_0^2 -\frac{36258331}{322164} \pi v^5 -\frac{2639}{54} \pi v_0^5 +\frac{53563337}{847800} \pi v^3 v_0^2
     \\  
   +\frac{5236153}{67824} \pi v^2 v_0^3 +\left(\frac{1373939088647}{11485587456}-\frac{2798996807 \eta }{9766656}+\frac{2367709 \pi ^2 \eta }{120576}+\frac{346466953
   \eta ^2}{4883328}\right) v^6 
   +\left(-\frac{544891835}{9144576} 
   \right.
   \\ \left.
  +\frac{6156943 \eta }{13608} -\frac{779 \pi ^2 \eta }{96}-\frac{1033961 \eta ^2}{3888}\right) v_0^6 + \left(-\frac{46807114518979}{371845893888} -\frac{36344458151 \eta}{368894736}
   +\frac{15640621469 \eta ^2}{26349624}\right) v^2 v_0^4
   \\ \left.
   +\left(\frac{39205650406583}{244786582656}-\frac{393415414111 \eta}{971375328}-\frac{2224117939 \eta^2}{5781996}\right) v^4 v_0^2 \right] 
   +\Delta \alpha ^2 \left[ \frac{79679}{1413} v^4 +\frac{380 }{9} v_0^4 -\frac{263891}{2826} v^2 v_0^2
  \right.   \\  
   +\left(\frac{357848623}{1424304}+\frac{679099 \eta }{50868}\right) v^6  +\left(\frac{2950463}{9072}-\frac{186691 \eta }{324}\right) v_0^6
   + \left(-\frac{88786529}{326403} -\frac{54900293 \eta }{93258}\right) v^4 v_0^2
\\ \left. \left.
   +\left(-\frac{81489388021}{245929824}+\frac{3398791079 \eta }{2927736}\right) v^2 v_0^4 
   \right] + \Delta \alpha^3 \left(\frac{434945}{12717}v^6 -\frac{24035}{81} v_0^6 -\frac{1513901} {4239} v^4 v_0^2 +\frac{2638910 }{4239} v^2 v_0^4 \right)  \right\} \,.
    \end{multline}
\end{subequations}
\end{widetext}

 The above phasing is accurate up to 3PN order $[\mathcal{O}(v^6)]$ in both the circular and eccentric parts. In the GR limit where $\Delta\alpha 
\rightarrow 0$, the phasing reduces to the standard {\tt TaylorF2Ecc} waveform \cite{Moore:2016qxz} and to the standard {\tt TaylorF2} circular waveform when $\Delta\alpha \to 0$ and $e_0 \to 0$ simultaneously. However, it does not reduce to the circular {\tt TaylorF2} waveform when $e_0 \to 0$ for $\Delta\alpha \neq 0$. This ultimately arises from Eq.~\eqref{eq:xirtophi} and the fact that $\xi_{\phi} \neq \xi$ when $e_0\rightarrow 0$. For binaries with very small (but non-vanishing) eccentricity, the periastron advance parameter $k$ becomes independent of eccentricity. [This behavior is also seen in the periastron advance angle per orbit, Eq.~\eqref{eq:angular advance}.] Because $\alpha\equiv 1+\Delta\alpha$ multiplies $k$ and relation \eqref{eq:xirtophi} forms the basis of our parametrization, the resulting waveform phasing inherits this property.  In GR, we interpret the $e_0\rightarrow 0$ limit of the $\xi_{\phi} = (1+k) \xi$ relation as having meaning only for binaries that have a finite but very small eccentricity. (Only in that sense are $\xi_{\phi}$ and $\xi$ distinct quantities.)\footnote{ To see this, consider the circular limit of Eq.~\eqref{quasiKeplEqns}. In that limit $W\rightarrow 0$, $\nu = u=l$, and $\phi(t) = (1+k) n (t-t_0) + c_{\lambda}$. Here, we see that $\phi$ depends only on a single frequency equal to $(1+k)n = \xi_{\phi}/M$. Intuitively, a perfectly circular orbit can only be characterized by a single frequency.} In a similar way, the phase expansion above should be interpreted as only applying to binaries that are eccentric. The term $\delta\Psi_{\rm circ.}$ is not a non-GR extension that applies to circular orbits. Rather, in the case of elliptical binaries with very small eccentricity, the $\Delta\alpha$-dependent phase corrections are dominated by a term $\delta\Psi_{\rm circ.}$ that has no explicit eccentricity dependence [analogous to the periastron advance angle in Eq.~\eqref{eq:angular advance}].

The corrections due to $\Delta\alpha$ first appear at relative 1PN order in $\delta\Psi_{\rm circ.}$ and $\delta\Psi_{\rm ecc.}$; no corrections appear at 1.5PN order. This is consistent with the PN behavior of the periastron advance constant $k$. Both $\delta\Psi_{\rm circ.}$ and $\delta\Psi_{\rm ecc.}$ are cubic polynomials in $\Delta\alpha$ and vanish in the $\Delta\alpha \rightarrow 0$ limit. 

The SPA phase expansion in Eq.~\eqref{eq:PsiFTecc} represents an alternative approach to testing GR for eccentric binaries. The phasing is modified at 1PN and higher-orders. However, unlike the parametrized testing GR formalism that introduces free parameters at each PN order, a single deviation parameter $\Delta\alpha$ controls the GR deviations at all PN orders in the phasing (starting at 1PN order). The fact that only a single parameter is needed and that it affects the phasing at multiple PN orders allows this parametrization to serve as a more sensitive null test of GR in comparison to the approach in Sec.~\ref{sec:parametric deviation}.

\section{Parameter measurement uncertainty}\label{sec:fisher matrix framework}
\subsection{Fisher Matrix Formalism}
We use the Fisher information matrix framework~\cite{Finn:1992wt,Cutler_Flanagan,Clifford_Will} for estimating the projected statistical errors on binary black hole parameters. The Fisher matrix framework is valid under the assumption of Gaussian and stationary detector noise and the high signal-to-noise ratio (SNR) limit~\cite{vallisneri2008use,Rodriguez:2013mla}. We only consider high SNR sources in our study so that our Fisher estimates are reliable. 

The Fisher framework yields the $1\sigma$ widths of the projected marginalized posterior probability densities for the binary parameters. Given the detector output $s(t)$, the projected posterior probability density for the binary parameters ${\bm \theta} \equiv \theta^a$ can be approximately expressed as
\begin{equation}\label{posterior}
 p({\bm \theta}|s) \propto p^{0}({\bm \theta}) \exp\left[ -\frac{1}{2} \Gamma_{ab} (\theta^{a} - \hat{\theta}^{a}) (\theta^{b} - \hat{\theta}^{b}) \right]\,, \end{equation}
where $p^{0}({\bm \theta})$ denotes the prior probability distribution on the signal parameters characterized by $\bm\theta$. The values $\hat{\theta}^{a}$ denote the ``best-fit'' parameter values obtained by maximizing the likelihood.  The Fisher information matrix $\Gamma_{ab}$ can be expressed as 
\begin{equation}\label{fisher}
     \Gamma_{ab} = \bigg(\frac{\partial{h}}{\partial\theta^a}\bigg|\frac{\partial{h}}{\partial\theta^b}\bigg)\, ,
\end{equation}
where $(\cdots|\cdots)$ represents the noise-weighted inner product. Given two time-domain signals $F(t)$ and $G(t)$, the inner product is defined as
\begin{equation}\label{inner product}
(F|G)= 2 \int^{f_{\rm high}}_{f_{\rm low}}df\,\frac{\tilde F^{*}(f)\,\tilde G(f)+\tilde F(f)\, 
\tilde G^{*}(f)}{S_n(f)}\, ,
\end{equation}
where $\tilde{F}(f)$ is the Fourier transform of $F(t)$, $\ast$ represents complex conjugation, and $S_n(f)$ is the one-sided noise power spectral density (PSD) of the detector. The Fisher matrix $\Gamma_{ab}$ is evaluated at the best-fit value $\hat{\theta}^{a}$ of the binary parameters $\bm \theta$.  The prior probability distribution of the binary parameters $\bm \theta$ is assumed to be Gaussian and centered around $\bar{\theta}^a$,
\begin{equation}
 p^0({\bm \theta}) \propto \exp\bigg[ -\frac{1}{2} \Gamma^0_{ab} (\theta^a-\bar{\theta}^a) (\theta^b-\bar{\theta}^b) \bigg],
\end{equation}
where $\Gamma^0_{ab}$ is the prior matrix. The covariance matrix corresponding to the covariance of the Gaussian-approximated posterior probability distribution, under the assumption $\bar{\theta}^a \approx \hat{\theta}^a$,
is given by
\begin{equation}\label{covariance}
\Sigma_{ab} = (\Gamma_{ab}+\Gamma_{ab}^0)^{-1} \,. 
\end{equation}
The $1\sigma$ width of the projected posterior probability distribution for the binary parameter $\theta^a$ is given by the square root of the diagonal element of the covariance matrix, $\sigma_a = \sqrt{\Sigma_{aa}}$. 

To combine information from different detectors, we add the Fisher matrices from each detector as 
\begin{equation}\label{multiband}
    \Gamma_{ab}^{\rm MB} =\sum_{j=1}^{J}\Gamma_{ab}^{(j)}\;,
\end{equation}
where $\Gamma_{ab}^{\rm MB}$ is the multiband Fisher matrix. The index $j=1, \cdots, J$ sums over the number of detectors $J$, with $\Gamma_{ab}^{(j)}$ denoting the Fisher matrix for the $j^{\rm th}$ detector [as defined in Eq.~\eqref{fisher}]. The multiband SNR $\rho_{\rm MB}$ is given by the quadrature sum of the individual detector SNRs:
\begin{equation}\label{multibanding SNR}
\rho_{\rm MB}^2 = \sum_{j=1}^{J}\rho_{j}^2 \,,
\end{equation}
where the single-detector SNR ($\rho_j$) is defined via 
\begin{equation}
  \rho_j^2 = (h|h) = 4 \int_{0}^{\infty} \frac{|\Tilde{h}(f)|^2}{S_n(f)} df\,,
\end{equation}
with quantities on the right-hand-side evaluated for detector $j$. For multibanding we assume that BBHs are simultaneously observed in all the detectors. 

For the parametric deviations in Sec.~\ref{sec:parametric deviation}, the parameters of the phasing are
\begin{equation}\label{first parameter set}
 \theta^{a} = \{t_{c}, \, \phi_{c}, \,  \log \mathcal{M} ,\log \eta, \chi_{1}, \, \chi_{2}, \,  \, \log e_0, \, \delta\hat{\varphi}_i, \, \delta\hat{\varphi}_i^e  \} \,.
 \end{equation}
Here the index $i$ refers to a single PN order. We consider two cases: (i) when both $\delta\hat{\varphi}_{i}$ and $\delta\hat{\varphi}_{i}^e$ are measured simultaneously [e.g., $(\delta\hat{\varphi}_2,\delta\hat{\varphi}^e_2)$]; and (ii) when only the $\delta\hat{\varphi}_{i}^e$ [e.g., $\delta\hat{\varphi}^e_2$] is measured fixing the $\delta\hat{\varphi}_{i}$ to zero. However, if a GW signal is described by a specific modified theory of gravity, a deviation from GR would naturally be expected to manifest in deviations of the coefficients $(\delta\hat{\varphi}_i, \, \delta\hat{\varphi}_i^e)$ across multiple PN orders at once. Ideally, all $(\delta\hat{\varphi}_i, \, \delta\hat{\varphi}_i^e)$ should be measured simultaneously---along with other system parameters---to test the true nature of the deviations~\cite{Arun:2006yw}. Tests in which multiple deviation parameters (across multiple PN orders) are measured simultaneously are called {\it multiparameter tests}. With current detector sensitivities, multiparameter tests give uninformative posteriors due to correlations between the deviation and the system parameters~\cite{Datta:2020vcj}. However, multiparameter tests will be possible via multiband observations involving 3G ground-based detectors. Moreover, principal component analysis can solve the degeneracy problem by constructing certain linear combinations of the PN deviation parameters that are better measured~\cite{Saleem:2021nsb, Datta:2022izc}. Here, we only focus on the case where a single PN order pair $(\delta\hat{\varphi}_i, \, \delta\hat{\varphi}_i^e)$ is considered. 

For the periastron advance parametrization derived in Sec.~\ref{sec:periastron parametrization},  the parameters of the phasing are
\begin{equation}\label{second parameter set }
 \theta^{a} = \{t_{c}, \, \phi_{c}, \, {\log \mathcal{M}},{\log \eta}, \,  \, \log e_0, \, \Delta\alpha \} \,.
\end{equation}
For both cases, the physically allowed values of the coalescence phase $\phi_{c}$ and spins $\chi_{1,2}$ are restricted to the ranges $\phi_{c} \in [-\pi, \pi]$, $\chi_{1,2} \in [-1, 1]$. This is taken into account by adopting Gaussian priors on $\phi_c$ and $\chi_{1,2}$ with zero means and 1$\sigma$ widths $\delta \phi_c =\pi$ and $\delta \chi_{1,2} =1$. To incorporate these priors, a prior matrix with the non-zero components, $\Gamma^0_{\rm \phi_c,\phi_c}=1/\pi^2$ and $\Gamma^0_{\chi_i,\chi_i}=1$, is added to the Fisher matrix as shown in Eq.~\eqref{covariance}.

\subsection{GW detector configurations}
We obtain bounds on the various deviation parameters using sensitivity curves for different GW detectors. We consider two ground-based and two space-based detector configurations as detailed below: 
\begin{enumerate}
    \item A LIGO detector operating at design sensitivity~\cite{LIGOScientific:2014pky} with lower cutoff frequency $f_{\rm low} = 10\,{\rm Hz}$.
    \item Cosmic Explorer (CE) with $40 {\rm km}$ arm length at its design sensitivity~\cite{LIGOScientific:2016wof, Reitze:2019iox} and with $f_{\rm low} = 5\,{\rm Hz}$.
    \item The Laser Interferometer Space Antenna (LISA), a planned space-based mission that will observe in the frequency range $[10^{-4} \mbox{--}0.1\,{\rm Hz}]$~\cite{Babak:2017tow}. 
    \item The Deci-hertz Interferometer Gravitational Wave Observatory (DECIGO)~\cite{Yagi:2013du, Kawamura:2020pcg}, a future Japanese space mission with a frequency-band of $[0.1 \mbox{--} 10\,{\rm Hz}]$.
\end{enumerate} 
We have not considered Virgo and KAGRA separately, as the LIGO sensitivity is representative of all 2nd generation detectors for our purposes. It is possible that LISA and CE will be operational at the same time. This opens up the opportunity for multiband observations of GW sources. The mergers of supermassive black holes (SMBHs) will be the main targets for LISA. However, LISA is capable of observing the inspiral of a few heavy stellar/intermediate-mass BBHs before merging in the frequency band of ground-based detectors~\cite{Sesana:2016ljz,Sesana:2017vsj,Gerosa:2019dbe,Carson:2019kkh,Gupta:2020lxa,Datta:2020vcj,Nakano:2021bbw,Sedda:2021yhn,Klein:2022rbf,Baker:2022eiz}, thus allowing for multiband observations of that class of GW sources. 

Analytical fitting formulas of the noise PSDs for LIGO, CE wideband (CE-wb), and DECIGO are taken from Eq.~(4.7) of Ref.~\cite{Ajith:2011ec}, Eq.~(3.7) of Ref.~\cite{Kastha:2018bcr}, and Eq.~(5) of Ref.~\cite{Yagi:2011wg}, respectively. Since we do not take into account the orbital motions of LISA and DECIGO, we use the pattern-averaged noise PSDs in our calculation. The noise PSD for LISA consists of instrumental noise and confusion noise due to galactic white dwarf binaries. The sky-averaged instrumental noise is taken from Eq.~(1) of Ref.~\cite{babak2017science}, and the galactic confusion noise can be found in Eq.~(4) of Ref.~\cite{babak2017science}. Note that the noise PSD in Ref.~\cite{babak2017science} is averaged over sky and polarization angles and accounts for the $60^{\circ}$ angle due to the triangular shape of LISA. To account for averaging over the inclination angle, we multiply by a factor of $\sqrt{4/5}$ in the angle non-averaged GW amplitude \footnote{In Eq.~\eqref{waveform}, the non-averaged $\mathcal{A}$ is given by $\mathcal{A}= \mathcal{C}(\theta, \phi, \psi, \iota)\sqrt{\frac{5}{24}} \frac{\mathcal{M}^{5/6}}{\pi^{2/3} d_{L}}$.}. We also divide the LISA noise PSD by a factor of $2$ to account for the effective number of L-shaped detectors in LISA. Note that the DECIGO noise PSD in Eq.~(5) of Ref.~\cite{Yagi:2011wg} is non-pattern-averaged, but includes the geometrical factor of $(\sqrt{3}/2)^{-2}$ accounting for DECIGO's triangular configuration. We divide this noise PSD by a factor of $8$ to account for the effective number of L-shaped detectors in DECIGO. Since the noise PSD for DECIGO is non-pattern-averaged, we use pattern-averaged amplitude for DECIGO [Eq.~\eqref{amplitude}].

To calculate the inner product in Eq.~\eqref{inner product}, the lower cutoff frequency limits $f_{\rm low}$ for LIGO, CE, LISA, and DECIGO are taken as $10\,{\rm Hz}$, $5\,{\rm Hz}$, ${\rm max}(10^{-4}\,{\rm Hz}, f_{\rm year})$ and ${\rm max}(10^{-2}\,{\rm Hz}, f_{\rm year})$, respectively. Here $f_{\rm year}$ is given as~\cite{berti2005estimating}
\begin{equation}\label{eq:fyear}
f_{\rm year} = 4.149\times 10^{-5}\left(\frac{{\cal M} (1+z)}{10^6 M_\odot}\right)^{-5/8}
\left(\frac{T_{\rm obs}}{1 \rm \text{year}}\right)^{-3/8} \, .   
\end{equation}
The time $T_{\rm obs}$ denotes the observation time before a binary reaches the ISCO (innermost stable circular orbit) and is assumed to be $T_{\rm obs}=4$ years for both LISA and DECIGO. The upper frequency limit $f_{\rm high}$ for spinning and nonspinning BBHs in space-based detectors LISA and DECIGO and nonspinning BBHs in LIGO and CE is chosen to be ${\rm min}(0.1\,{\rm Hz},f_{\rm ISCO})$ or ${\rm min}(10\,{\rm Hz},f_{\rm ISCO})$, respectively, where $f_{\rm ISCO}$ is the GW frequency corresponding to the Schwarzschild ISCO and can be expressed as
\begin{equation}\label{eq:fisco_schw}
    f_{\rm ISCO} = \frac{1}{6^{3/2}\pi (1+z) M}\,.
\end{equation}  
The upper cut-off frequency $f_{\rm high}$ for spinning BBHs in LIGO and CE corresponds to the ISCO frequency of the remnant Kerr BH: \footnote{We use Eq.~\eqref{eq:fisco_schw} for space-based detectors because it simplifies the calculation of the frequency corresponding to a given time before ISCO using Eq.~\eqref{eq:fyear}. For ground-based detectors, the GW signal sweeps the entire frequency band.} 
\begin{equation}\label{eq:fisco_kerr}
    f_{\rm ISCO} = \frac{1}{1+z} \frac{\hat{\Omega}_{\rm ISCO}(\chi_f)}{\pi M f} \,.
\end{equation}
The detailed expression for $\hat{\Omega}_{\rm ISCO}(\chi_f)$ is given in Appendix C of Ref.~\cite{Favata:2021vhw}. For binary neutron stars (BNSs) in LIGO and CE, we restrict $f_{\rm ISCO} = 1500$ Hz.


\section{Bounds on non-GR parameters}\label{sec:results}
After extending the parametrized PN phasing to low-eccentricity binaries in Sec.~\ref{sec:parametric deviation} and developing the periastron advance parametrization in Sec.~\ref{sec:periastron parametrization}, we next calculate the projected bounds on these deviation parameters using the Fisher information matrix framework as discussed in Sec.~\ref{sec:fisher matrix framework}. This is done for LIGO, CE, LISA, and DECIGO considering various mass binaries and covering a frequency range from $10^{-4}\,{\rm Hz}$ to $1500\,{\rm Hz}$. For all BBHs considered, the mass-ratio is fixed to be $q=2$. The spin vectors for spinning systems are aligned with the orbital angular momentum, and the dimensionless spin magnitudes are chosen to be $\chi_1=0.5$ and $\chi_2=0.4$. We choose moderate values of the mass-ratio and spins as a representative example of the bounds that can be placed on the non-GR parameters. For LIGO and CE, we consider systems with total masses in the range $10 M_{\odot}$ to $100 M_{\odot}$ and located at $d_L=500$ Mpc. These sources have SNR $\mathcal{O}(8\mbox{--}50)$ in LIGO and $\mathcal{O}(294\mbox{--}1945)$ in CE. For LISA, we consider intermediate and supermassive BBHs between $10^3 M_{\odot}$ to $10^6 M_{\odot}$ located at $d_L=3$ Gpc. For $4$ years of LISA observation, these sources have SNR $\mathcal{O}(38\mbox{--}3867)$. For DECIGO, we consider $10 M_{\odot}$ to $10^3 M_{\odot}$ binaries at $3$ Gpc. The SNR of these sources is $\mathcal{O}(300\mbox{--}13775)$. We also consider the possibility of multibanding between LISA and CE for BBHs with total masses between $10^2$--$10^3\,M_{\odot}$. The inspiral phase will be visible in LISA only for those sources that are nearby. We fix the distance $d_L=500$ Mpc for these potential multiband sources, which implies SNR $~\mathcal{O}(10\mbox{--}172)$ in LISA and $\mathcal{O}(37\mbox{--}5152)$ in CE. We also consider a binary neutron star (BNS) system in LIGO and CE at $d_L=100$ Mpc with $m_1=1.4\, M_{\odot}$, $m_2=1.2\, M_{\odot}$, and spins $\chi_{1,2}=0.05$. The BNS system has SNR $\sim 14$ in LIGO and $\sim 476$ in CE.

For LIGO and CE, the initial eccentricity parameter $e_0$ is defined at a reference frequency $f_0=10\,{\rm Hz}$. For LISA, $e_0$ is defined at a reference frequency $f_0$ corresponding to the GW frequency of the binary $4$ years prior to its coalescence. If the binary is outside the LISA lower cut-off frequency at that time, we set $f_0=10^{-4}\,{\rm Hz}$ for such binaries (LISA's low-frequency limit). For DECIGO, $f_0$ is set to that detector's low-frequency limit $(f_0=0.1\,{\rm Hz})$. For multiband sources, $f_0$ is defined as for LISA-band binaries (i.e., at a frequency corresponding to $4$ years before the binary merges in the CE band).   

 In this section we focus on the projected bounds that can be placed on the non-GR parameters $(\delta\hat{\varphi}_2,\delta\hat{\varphi}^e_2)$ and $\Delta\alpha$. Appendix~\ref{app:eccentricity_measurement} considers the constraints on the eccentricity parameter $e_0$ when these parametrizations are used.

\subsection{Bounds on parametric deviations}\label{sec:results parametric deviation}
In this section, we calculate bounds using the parametrized PN phasing for eccentric binaries derived in Sec.~\ref{sec:parametric deviation}. Note that we consider aligned-spin binaries for this parametrization. A parameter pair consisting of $(\delta\hat{\varphi}_i, \delta\hat{\varphi}_{i}^e)$ for a particular PN-order is measured. We separately calculate the bound on a single eccentric deformation parameter $\delta\hat{\varphi}_{i}^e$, fixing the corresponding circular part to its GR value $(\delta\hat{\varphi}_i=0)$. \footnote{We do not consider the case when only circular deviations $\delta\hat{\varphi}_i$ are measured. That case has been extensively analyzed in the literature~\cite{GWTC3:2021sio}. In our recent paper~\cite{Saini:2022igm}, we also show the errors on the circular deviations for binaries with similar properties measured by both LIGO and CE.}
\begin{figure*}
    \centering
    \begin{subfigure}{\includegraphics[width=0.235\textwidth]{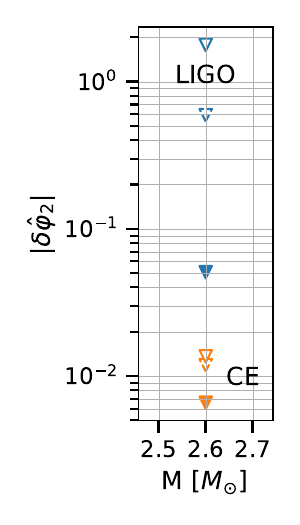}}
    \end{subfigure}
   \begin{subfigure}{\includegraphics[width=0.71\textwidth]{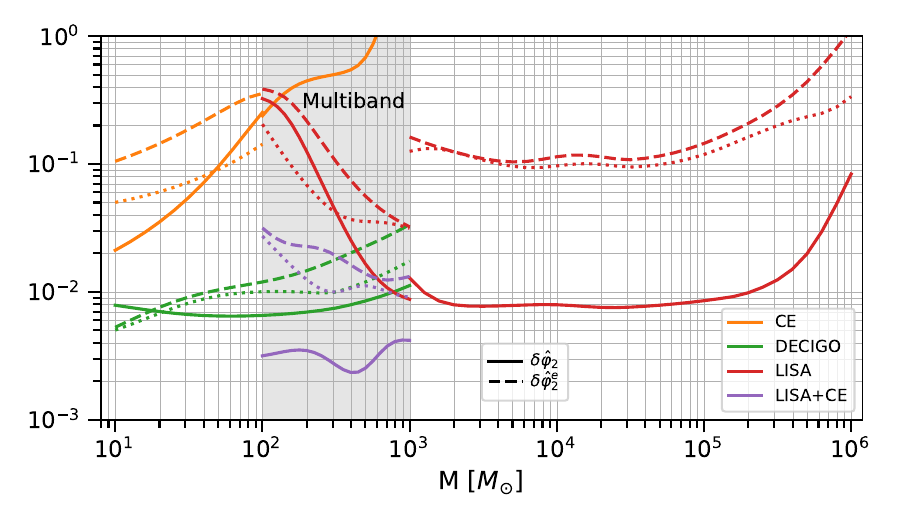}}
   \end{subfigure}
\caption{(Color online) $1\sigma$ errors on the 1PN circular deviation $\delta\hat{\varphi}_2$ and eccentric deviation $\delta\hat{\varphi}_{2}^e$ parameters. Different colors represent different detectors (see legend). Solid and dashed lines indicate the errors on $\delta\hat{\varphi}_2$ and $\delta\hat{\varphi}_2^e$, respectively, when both are measured simultaneously. Dotted lines denote the errors on $\delta\hat{\varphi}_2^e$ when it is measured individually. The shaded region represents the mass range where multiband observations with LISA and CE are possible (purple curves only). The mass ratio for BH binaries is fixed at $q=2$. The BBHs are located at luminosity distances of $d_L=500\,{\rm Mpc}$ for CE sources and $d_L=3\,{\rm Gpc}$ for DECIGO and LISA sources. Multiband sources $(M=10^2\mbox{--}10^3\,M_{\odot})$ are located at $d_L=500\,{\rm Mpc}$. For BBHs, the dimensionless spin magnitudes are $\chi_1=0.5$ and $\chi_2=0.4$. The small left-hand panel displays bounds from a typical BNS system in LIGO and CE. The BNS system has component masses $m_1=1.4 \, M_{\odot}$, $m_2=1.2 \, M_{\odot}$ and spins $\chi_{1,2}=0.05$. Filled triangles correspond to the circular 1PN deviation parameter $\delta\hat{\varphi}_2$; unfilled is the 1PN eccentric parameter $\delta\hat{\varphi}_2^e$. The unfilled-solid triangle is the value of $\delta\hat{\varphi}_2^e$ when it is simultaneously measured with $\delta\hat{\varphi}_2$. The unfilled-dashed triangle represents the bound when $\delta\hat{\varphi}_2^e$ is measured individually. (Blue is LIGO and orange is CE.) The BNS luminosity distance is fixed at $100$ Mpc. The initial eccentricity for all binaries is chosen to be $e_0=0.2$ (with the reference frequency $f_0$ chosen differently for each detector; see the main text for details).}
\label{fig:first_parametrization_deviation}
\end{figure*}

Figure~\ref{fig:first_parametrization_deviation} shows the $1\sigma$ bounds on $\delta\hat{\varphi}_{2}$ and $\delta\hat{\varphi}_{2}^e$ as a function of the total mass $M$ of the binary in different detectors. The right panel is for BBHs while the left panel represents the bound on a typical BNS system. While we considered multiple PN orders, we only show errors for the 1PN terms for better visibility in the plot. The 1PN term is best measured in most of the cases. The errors on the 0PN and 1.5PN parameters show similar trends with the total mass. (To see the detailed behavior of circular deviation parameters for various binary systems, see \cite{Saini:2022igm}.) Different colors represent the bounds in different detectors. The solid lines denote the errors on $\delta\hat{\varphi}_{2}$ while the dashed lines denote the errors on $\delta\hat{\varphi}_{2}^e$ when both $\delta\hat{\varphi}_{2}$ and $\delta\hat{\varphi}_{2}^e$ are measured simultaneously. Dotted lines denote the error on $\delta\hat{\varphi}_{2}^e$ for the case when it is measured individually (i.e, $\delta\hat{\varphi}_{2}=0$). The shaded region shows errors on multiband sources observed by both LISA and CE. For all the detectors, the initial eccentricity is chosen to be $e_0=0.2$. 

 The errors on the deviation parameters $(\delta\hat{\varphi}_2, \delta\hat{\varphi}_2^e)$ for BBHs in LIGO lie outside the plot as they are greater than $1$. For $10 \,M_{\odot}$ in LIGO, the 1PN deviation parameters can be measured simultaneously with errors $\delta\hat{\varphi}_2 \lesssim 1.2$,  $\delta\hat{\varphi}_2^e \lesssim 18.5$. When only $\delta\hat{\varphi}_2^e$ is measured, the errors reduce to $\delta\hat{\varphi}_2^e \lesssim 5$. The errors on $\delta\hat{\varphi}_{2}^e$ are larger compared to the errors on $\delta\hat{\varphi}_{2}$ since we work in the small eccentricity regime. When only $\delta\hat{\varphi}_{2}^e$ is measured along with other binary parameters (keeping $\delta\hat{\varphi}_{2}=0$), the errors on $\delta\hat{\varphi}_{2}^e$ decrease relative to the case when $\delta\hat{\varphi}_{2}$ is also a free parameter. For $10\mbox{--}100\,M_{\odot}$ binaries in CE, the errors on the deviation parameters increase with increasing binary mass; this is predominantly due to the decrease in the number of GW cycles in the detector frequency band as the mass increases. For binaries with $M=10 M_{\odot}$ in CE, the circular 1PN deviation parameter can be constrained with $|\delta\hat{\varphi}_{2}|\lesssim2\times 10^{-2}$ and the eccentric deviation parameter can be measured with $|\delta\hat{\varphi}_{2}^e|\lesssim 10^{-1}$, when both deviations are measured simultaneously; this reduces to $|\delta\hat{\varphi}_{2}^e|\lesssim 5\times10^{-2}$ when it is measured individually. For the BNS system in the LIGO band (left panel), the circular and eccentric 1PN deviation parameters (together) can be measured with $|\delta\hat{\varphi}_{2}|\lesssim 5\times 10^{-2}$ and $|\delta\hat{\varphi}_{2}^e|\lesssim 2$. The bound on the eccentric deviation improves to $|\delta\hat{\varphi}_{2}^e|\lesssim 0.5$ when measured alone. In CE, the circular and eccentric bounds improve to $|\delta\hat{\varphi}_{2}|\lesssim 7\times 10^{-3}$ and $|\delta\hat{\varphi}_{2}^e|\lesssim 10^{-2}$ (measured simultaneously). The bound on $\delta\hat{\varphi}_{2}^e$ is similar ($|\delta\hat{\varphi}_{2}^e|\lesssim 10^{-2}$), when it is measured individually.

For DECIGO with $M=100\,M_{\odot}$, the circular and eccentric deviation parameters are determined with $1\sigma$ bound of $|\delta\hat{\varphi}_{2}|\lesssim 7\times 10^{-3}$ and $|\delta\hat{\varphi}_2^e|\lesssim10^{-2}$ when measured simultaneously. When $\delta\hat{\varphi}_{2}^e$ is measured individually, the errors are slightly smaller than the case where $\delta\hat{\varphi}_{2}^e$ is measured with $\delta\hat{\varphi}_{2}$. For BBHs with mass range $10^3\mbox{--}10^5\, M_{\odot}$ in the LISA band, the bounds on circular and eccentric deviations (when measured simultaneously) are $|\delta\hat{\varphi}_{2}|\lesssim 10^{-2}$ and $|\delta\hat{\varphi}_{2}^e|\lesssim 10^{-1}$. These errors become an order of magnitude larger for $M\approx 10^6\,M_{\odot}$. As the mass increases beyond $\sim 10^{5} \, M_{\odot}$, errors on the deviation parameters increase due to a sharp decrease in the number of GW cycles in the frequency band. When $\delta\hat{\varphi}_{2}^e$ is measured individually, the errors are similar to (but slightly smaller than) the simultaneous measurement case.

Apart from the mergers of supermassive BHs, LISA will also see the inspiral of heavy stellar-mass/intermediate-mass BBHs that will merge in the low-frequency band of ground-based 3G detectors. Therefore, we consider bounds on the deviation parameters from combined observations with LISA and CE. In the multiband region $(M \approx 10^2\mbox{--}10^3\,M_{\odot})$, the errors on the deviation parameters are large in the CE band. This is due to the reduced number of GW cycles and the negligible binary eccentricity near the merger. We do not show the errors on the eccentric deviation parameters, as they are not measurable by CE for this mass range. Recall that the initial eccentricity $e_0$ is defined for multiband sources relative to a reference frequency $f_0$ in the LISA band. For $M\approx 100M_{\odot}$, the errors on $\delta\hat{\varphi}_{2}$ and $\delta\hat{\varphi}_{2}^e$ are similar [$\sim\mathcal{O}(0.3)$] in the CE and LISA bands. As the mass of the binary increases, the number of inspiral cycles reduces in the CE band, while it increases in the LISA band. Therefore, for $M\approx10^3\,M_{\odot}$ in the LISA band, the errors on $\delta\hat{\varphi}_{2}$ reduce to $\sim\mathcal{O}(10^{-2})$ (one order of magnitude improvement). The errors on $\delta\hat{\varphi}_{2}^e$ are a factor $\sim 4$ larger than those for $\delta\hat{\varphi}_{2}$. Relative to LISA measurement alone, multibanding (LISA+CE) for $M=100M_{\odot}$ binaries reduces the errors on $\delta\hat{\varphi}_{2}$ by $\sim 2$ orders of magnitude and can be measured with $|\delta\hat{\varphi}_{2}| \lesssim 3\times 10^{-3}$. The bound on $\delta\hat{\varphi}^e_{2}$ improves by $\sim 1$ order of magnitude, which can be constrained to $|\delta\hat{\varphi}_{2}^e| \lesssim 3\times 10^{-2}$. For $M=10^3\,M_{\odot}$ binaries, this improvement is a factor $\sim 5$ for the error in $\delta\hat{\varphi}_{2}$ and a factor $\sim 2$ for the error in $\delta\hat{\varphi}^e_{2}$. 

\begin{figure*}
    \centering
    \begin{subfigure}{\includegraphics[width=0.235\textwidth]{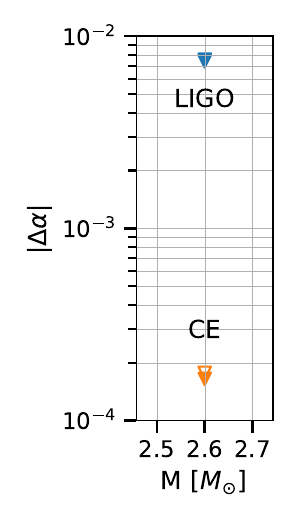}}
    \end{subfigure}
   \begin{subfigure}{\includegraphics[width=0.71\textwidth]{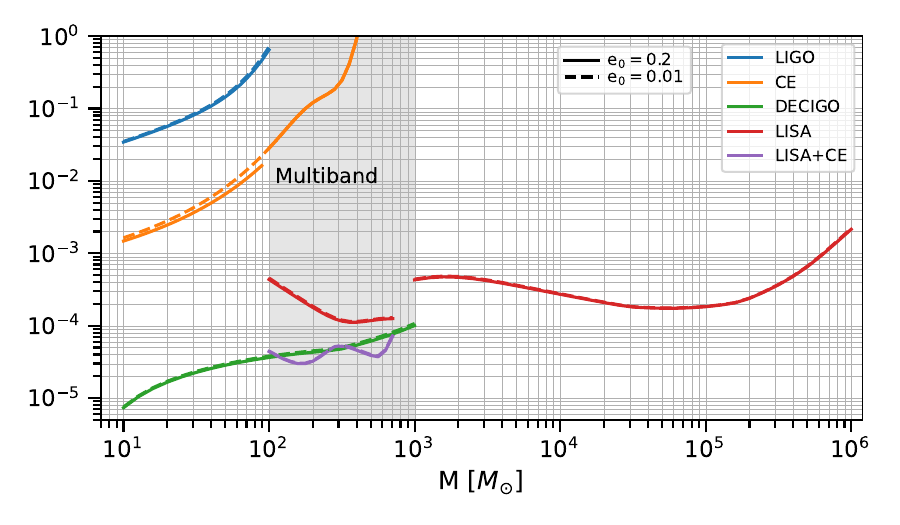}}
   \end{subfigure}
   \caption{(Color online) $1\sigma$ bound on the periastron deviation parameter $\Delta\alpha$. Different colors show the different detectors (see legend). Solid lines denote an initial eccentricity value of $e_0=0.2$ while dashed lines indicate $e_0=0.01$. The reference frequency is chosen as in Fig.~\ref{fig:first_parametrization_deviation} (see main text for details). The binary properties are the same as in Fig.~\ref{fig:first_parametrization_deviation}, except the binaries here are nonspinning. The shaded region represents the mass range where multiband observations with LISA and CE are possible (purple curves). The left panel represents errors on a BNS system with component masses $m_1=1.4 \, M_{\odot}$, $m_2=1.2 \, M_{\odot}$ located at $100$ Mpc. Filled triangles correspond to $e_0=0.2$; unfilled to $e_0=0.01$.}
  \label{fig:alpha}
\end{figure*}

\subsection{Bounds on the periastron deviation parameter $\Delta\alpha$}\label{sec:periastron-bounds}
Next, we calculate bounds on the periastron deviation parameter $\Delta\alpha$ using the parametrized phasing derived in Eq.~\eqref{eq:PsiFTecc}. Figure~\ref{fig:alpha} shows the $1\sigma$ bounds on $\Delta\alpha$ as a function of the binary total mass. Since the effect of spin on the periastron advance was not considered, and the $\Delta\alpha$ dependent terms affect both the circular and eccentric parts of the SPA phasing, we consider only nonspinning binaries in this analysis. However, the effect of spins on the $\Delta\alpha$ bounds is discussed in  Appendix~\ref{app:alpha_spin}. In Fig.~\ref{fig:alpha}, different colors represent different detectors. The solid curves denote an initial eccentricity of $e_0=0.2$, while dashed curves indicate $e_0=0.01$. Recall that $e_0$ is defined at different reference frequencies $f_0$ for ground-based and space-based detectors as discussed in Sec.~\ref{sec:results}. 

LIGO can measure the periastron deviation parameter $\Delta\alpha$ with $|\Delta\alpha| \lesssim 4\times 10^{-2}$ for $M=10M_{\odot}$. The errors on $\Delta\alpha$ increase as the binary mass increases, since the number of inspiral cycles in the detector band decreases with increasing mass. Notice that this parametrization allows us to test the deviation in periastron advance with reasonable accuracy even with LIGO sensitivity, unlike the $(\delta\hat{\varphi}_i,\delta\hat{\varphi}^e_i)$ parametrization. For binaries with total mass $10\mbox{--}10^2\, M_{\odot}$ in CE, the $\Delta\alpha$ measurement improves by an order of magnitude relative to LIGO. With CE, $\Delta\alpha$ can be measured to $|\Delta\alpha| \lesssim 2\times 10^{-3}$ for $M\approx 10\,M_{\odot}$. For the BNS system in LIGO, the $\Delta\alpha$ bounds are $|\Delta\alpha| \lesssim 8 \times 10^{-3}$, while CE can constrain $\Delta\alpha$ to within $|\Delta\alpha|\lesssim 2\times 10^{-4}$. Notice that the $e_0=0.2$ and $e_0=0.01$ bounds are very similar. This is because the periastron advance angle becomes independent of eccentricity in the small eccentricity limit [See Eq.~\eqref{eq:angular advance}]. This is further explained in Fig.~\ref{fig:alpha_circ_ecc} of Appendix~\ref{app:alpha_circ_ecc}.

The most precise single detector constraints on $\Delta\alpha$ come from binaries observed by DECIGO. This is because the sources have large SNRs [$\mathcal{O}(300\mbox{--}13775)$] in the DECIGO frequency band. The deviation parameter $\Delta\alpha$ can be constrained to $|\Delta\alpha| \lesssim 8 \times 10^{-6}$ for $M = 10\,M_{\odot}$. These constraints are better than the bounds obtained from the double pulsar observations. The current constraints on the periastron advance from double pulsar observations are $\mathcal{O}(10^{-5})$~\cite{Kramer:2021jcw}~\footnote{Reference~\cite{Kramer:2021jcw} provides the bound from double pulsar observations on the periastron advance parameter as $|k_{\rm ob}-k_{\rm GR}| /k_{\rm GR} =|\Delta k|/k_{\rm GR} = 1.5 \times 10^{-5}$, where $k_{\rm ob}$ and $k_{\rm GR}$ are the observed and GR-predicted values of $k$, respectively. In our case, the observed value of periastron advance is $k_{\rm ob} =\alpha k$ and the corresponding GR value is $k_{\rm GR}= \alpha^{\rm GR} k $, hence the double pulsar bound on $k$ is related to the bound on $\alpha$ here as $|\Delta\alpha|/\alpha_{\rm GR}=|\alpha-\alpha^{\rm GR}|/\alpha^{\rm GR} = |\Delta k| /k_{\rm GR}$ with $\alpha^{\rm GR}=1$.}. It is worth emphasizing that the extremely precise bounds obtained from GW observations represent the stronger field regime relative to the bounds from pulsar observations. It should be noted that the direct comparison of bounds obtained with different compact objects (BHs and NSs) and different dynamical regimes is associated with certain caveats~\cite{Kramer:2021jcw}. Nevertheless, the bounds obtained through double pulsar observations and GW observations can be considered complementary to one other. 

For LISA, the best measurements of $\Delta\alpha$ come from binaries with $M\approx 10^5 \,M_{\odot}$ where $\Delta\alpha$ can be measured with $|\Delta\alpha| \lesssim 2 \times 10^{-4}$. In the multiband region, LISA can constrain $\Delta\alpha$ to $|\Delta\alpha| \lesssim 4\times 10^{-4}$ for $M=100M_{\odot}$ binaries, while equivalent errors for CE are $|\Delta\alpha| \lesssim 2\times 10^{-2}$. For $e_0=0.01$, the errors in CE are large because the eccentricity ($e_0=0.01$) had almost decayed when it entered the CE band (recall that $f_0$ is defined at $4$ years before the merger in the multiband case); hence, we do not show it in the plot. As the mass of the binary increases, the errors for LISA reduce to $|\Delta\alpha| \lesssim 10^{-4}$, whereas the errors for CE increase. Combining information from LISA and CE improves the accuracy of the $\Delta\alpha$ measurement by one order of magnitude for $M=10^2\,M_{\odot}$ binaries and by a factor of $\sim 5$ (relative to LISA-only measurements) for $M \approx 10^3\,M_{\odot}$ binaries. With multibanding between LISA and CE, $\Delta\alpha$ can be measured with $|\Delta\alpha| \lesssim 3\times 10^{-5}$. This improvement is mainly due to the breaking of degeneracy among different parameters of the binary~\cite{Klein:2022rbf}. LISA provides an extremely accurate measurement of the chirp mass $\mathcal{M}$ and $e_0$. LISA measurements of $\mathcal{M}_c$ and $e_0$ are degenerate with the coalescence time $t_c$; the latter can be measured with an accuracy of $\sim\mathcal{O}(1)$ milliseconds by CE. Therefore, combining information from both detectors leads to an overall better measurement of $\Delta\alpha$. It is worth emphasizing that the periastron advance in these binaries is a few degrees per second and can be measured with a fractional precision of $\sim 10^{-5}$. This allows us to test a highly relativistic effect with exquisite precision.

\section{Summary and conclusions}\label{conclude}
The key goal of this paper was to develop GW-based parametrized tests of GR for compact binaries in eccentric orbits. We have parametrized the {\tt TaylorF2Ecc} waveform model in two different ways; either can be used to probe the GR dynamics of eccentric binaries. The first method is an extension of the standard null parametrized test of GR for quasicircular binaries~\cite{BSat95, Arun:2004hn, Arun:2006hn, Yunes:2009ke, PPE:2011ys, TIGER:2013upa, Mehta:2022pcn}. In this method, the phenomenological deviation parameters are added at each PN order in the eccentric part of the waveform phasing (assuming a low-eccentricity phase expansion). The eccentric deviation parameters ($\delta\hat{\varphi}^e$) could be measured either simultaneously with the circular deviation parameters ($\delta\hat{\varphi}$) or individually. 

For BBHs in LIGO, this method yields uninformative constraints on the deviations due to large correlations between the deviation parameters and the source parameters. However, the BNS system in the LIGO band provides informative constraints on these deviations. CE can measure both the circular and eccentric deviations for BBHs. For a BNS in CE, circular bounds improve (compared to LIGO) by $\sim 1$ order of magnitude and eccentric bounds improve by $\sim 2$ orders of magnitude. We also consider the possibility of multibanding with LISA and CE. Multibanding improves the bounds on these deviations by $\sim 1$ to $2$ orders of magnitude. 

In the second approach developed here, we build a parametrization based on the general-relativistic effect of {\it periastron advance} for nonspinning binaries. Our objective is to test if the angular advance of the periastron is consistent with the GR prediction. We parametrize the relation between the radial and azimuthal dynamical frequencies entering planar eccentric orbits, which are related via the periastron advance parameter. We introduced a non-GR parameter $\Delta \alpha$ into this relation that captures the deviation in the angular advance of the periastron relative to the GR prediction ($\Delta\alpha=0$). This relation is then used to derive the GW SPA phase in the low-eccentricity limit. When applying this new phasing, if the measured value of $\Delta\alpha$ excludes $0$, it is an indication of a GR violation or waveform systematics. The deviation parameter $\Delta\alpha$ enters at 1PN and higher orders in the phasing. This parametrization requires only one non-GR parameter to be measured; it captures a particular class of GR deviations associated with modifications to the rate of periastron advance. Even if the GR violation does not originate from periastron advance, we suspect this parametrization is sensitive to those violations despite the suboptimality of the parametrization. A separate study may be needed to quantify the ability of the second parametrization to capture more generic GR deviations appearing in various PN orders.

With the second parametrization, even LIGO can measure the periastron deviation parameter to precision $|\Delta\alpha| \lesssim 4 \times 10^{-2}$ for BBHs. These bounds improve by one order of magnitude for CE. The BNS system in the LIGO band can be constrained to $|\Delta\alpha| \lesssim 8 \times  10^{-3}$. For BNS in CE, the bounds improve by $\sim 2$ orders of magnitude (i.e $|\Delta\alpha|\lesssim 2\times 10^{-4}$). Supermassive BBH observations with LISA can provide a constraint of $|\Delta\alpha| \lesssim 2 \times 10^{-4}$. Multiband observations between LISA and CE improve the constraint to $|\Delta\alpha| \lesssim 3\times 10^{-5}$. Note that for nonspinning binaries, multibanding is possible for $M \lesssim 700\, M_{\odot}$. Sources greater that $700 \, M_{\odot}$ merge outside the CE band ($f< 5\, \text{Hz}$).  DECIGO provides the most precise constraint, $|\Delta\alpha| \lesssim 8\times 10^{-6}$, which is better than the double pulsar bounds on periastron advance set by radio observations. 

 In addition to the relativistic dynamics, orbital periastron advance can arise from spin effects~\cite{1985SvAL...11..224S}, oblateness of the compact bodies~\cite{Murray_Dermott_2000,1939MNRAS..99..451S}, and the presence of a third body~\cite{Murray_Dermott_2000,2011A&A...528A..53B,baycroft2023improving}. Our parametrization could also be used to probe for these effects, and may also lead to a non-zero value of $\Delta\alpha$. These possibilities will be explored in detail in a future study.   

Further work remains to improve the formalisms described here. Since fully analytic waveforms involving eccentricity and spin are yet to be developed, we used the {\tt TaylorF2Ecc} waveform \cite{Moore:2016qxz} as a starting point for the development of both approaches. There, spin effects are only considered in the circular part of the waveform phasing. A parametrized waveform that includes both spin and eccentricity in a fully-consistent way must begin with the development of such waveforms in the GR limit. The extension to eccentric binaries with precessing spins is more challenging (although there are some efforts in this direction, see for example Ref.~\cite{Arredondo:2024nsl}) and may require a framework beyond that currently used for parametrized GR tests. 

Additional work also remains on our periastron advance ($\Delta \alpha$) parametrization. Our formalism modifies the relation between the radial and azimuthal frequencies but assumes that the time derivatives  $(d\xi/dt)_{\rm ADM}$ and $(dk/dt)_{\rm ADM}$ are provided by their GR expressions. A more fundamental approach could begin by introducing deformation parameters at the level of the quasi-Keplerian orbital variables, and then propagating those parameters through the expressions for the energy, angular momentum, and their corresponding fluxes when expressed in terms of the orbital variables $r, \varphi, \dot{r}, \dot{\varphi}$. Consistently incorporating spins and multipole moments of the compact bodies into this approach is another challenging avenue for future work. 

Lastly, one could consider other starting points as a means to develop parametrized waveforms for eccentric binaries. For example, rather than modifying a conservative effect like the periastron advance, one could modify a dissipative effect, such as the eccentricity dependence on frequency $e_t(\xi_{\phi})$ or its rate of change $de_t/dt$.
\vspace{0.4cm}

\acknowledgments
We thank Rossella Gamba for a careful reading of the draft. K.G.A., P.S., and S.A.B.~acknowledge support from the Department of Science and Technology and the Science and Engineering Research Board (SERB) of India via  Swarnajayanti Fellowship Grant No.~DST/SJF/PSA-01/2017-18 and support from the Infosys Foundation. K.G.A also acknowledges Core Research Grant No.~CRG/2021/004565 and MATRICS grant (Mathematical Research Impact Centric Support) No.~MTR/2020/000177 of the SERB. M.F.~was supported by NSF (National Science Foundation) Grant No.~PHY-1653374. C.K.M. acknowledges the support of SERB's Core Research Grant No.~CRG/2022/007959. This paper has been assigned the LIGO Preprint No. P2400347.

\appendix
\begin{widetext}
\section{{Complete 3PN order expressions for intermediate results in Sec.~\ref{sec:periastron parametrization}}\label{app:full expressions}}
Here we provide the full expressions for the time and frequency evolution of the azimuthal frequency $\xi_\phi$ and eccentricity $e_t$. The complete expression for $d\xi_{\phi}/dt$ corresponding to Eq.~\eqref{eq:dxidt-MH-lowe} is:
\begin{multline}
\label{eq:dxidt-MH-lowe-appendix}
\frac{d\xi_{\phi}}{dt} = M \frac{d\omega_{\phi}}{dt} = \frac{96 \eta  \xi_{\phi} ^{11/3}}{5 M} \left(1+\left(\frac{1273}{336}-6 \alpha -\frac{11 \eta }{4}\right)  \xi_{\phi} ^{2/3}+4 \pi  \xi_{\phi} +\left( \frac{438887}{18144}-\frac{2365 \alpha }{42}+34 \alpha ^2-\frac{49507 \eta }{2016}+\frac{94 \alpha  \eta
   }{3}
 \right.\right.  \\  \left. \left.
+\frac{59 \eta ^2}{18} \right) \xi_{\phi}^{4/3} +\left(\frac{20033  }{672}-36   \alpha -\frac{189   \eta }{8}\right)\pi \xi_{\phi}^{5/3}+ \left[\frac{38047038863}{139708800}-\frac{1712 \gamma }{105}+\frac{16 \pi ^2}{3}-\frac{560933 \alpha
   }{1134}+\frac{179215 \alpha ^2}{336}-\frac{580\alpha^3}{3}
  \right. \right.  \\  \left. \left.
   -\frac{16554367 \eta }{31104}+\frac{287 \pi ^2 \eta
   }{24}+\frac{546365 \alpha  \eta }{1008}-\frac{41 \pi^2}{16} \alpha \eta -\frac{3215 \alpha ^2 \eta
   }{12}+\frac{617285 \eta ^2}{8064}-\frac{1367 \alpha  \eta ^2}{18}-\frac{5605 \eta ^3}{2592}-\frac{856 \log (16 \xi_{\phi}^{2/3} )}{105} \right] \xi_{\phi}^2
\right.  \\ \left.
+e_t^2 \left\{\frac{157}{24}+\left(\frac{19471}{336}-\frac{619 \alpha }{12}-\frac{673 \eta }{16}\right) \xi_{\phi} ^{2/3}
+\frac{2335 }{48} \pi  \xi_{\phi}+\left(\frac{5710421}{12096}-\frac{42457 \alpha }{48}+\frac{1491 \alpha ^2}{4}-\frac{670093 \eta }{1344}+\frac{18667
   \alpha  \eta }{36}
  \right. \right. \right.  \\  \left. \left. \left. 
   +\frac{213539 \eta ^2}{1728}\right) \xi_{\phi}
^{4/3}+\left(\frac{57311  }{96}-\frac{24713   \alpha }{48}-\frac{27645   \eta }{56}\right) \pi \xi_{\phi} ^{5/3}
+ \left[ \frac{293331877639}{46569600}-\frac{106144 \gamma }{315}+\frac{992 \pi ^2}{9}-\frac{412627363 \alpha
   }{36288}
  \right. \right. \right.    \\   \left. \left. \left.
   +\frac{1633913 \alpha ^2}{168}-\frac{48025 \alpha ^3}{18}-\frac{1167634417 \eta }{120960}+\frac{306803 \pi ^2
   \eta }{2304}+\frac{25081327 \alpha  \eta }{2016}
  -\frac{9881}{192} \pi ^2 \alpha  \eta -\frac{734999 \alpha ^2
   \eta }{144}+\frac{9062905 \eta ^2}{4032}     \right. \right. \right.   \\    \left. \left. \left. -\frac{694349 \alpha  \eta ^2}{288}
   -\frac{6874115 \eta
   ^3}{31104}+\frac{18832 \log (2)}{45}-\frac{234009 \log (3)}{560}-\frac{53072}{315} \log \left(16 \xi_\phi ^{2/3}\right) \right] \xi_{\phi} ^2\right\} \right) \, ,
\end{multline} 
where $\gamma = 0.5772156649 \cdots$ is the Euler-Mascheroni constant. The expression for $de_t/dt$ corresponding to Eq.~\eqref{eq:dedt-MH-lowe} is:
\begin{multline}
\label{eq:dedt-MH-lowe_appendix}
\frac{de_t}{dt} =  -\frac{304 e_t \eta  \xi_{\phi}^{8/3}}{15 M} \left(1+\left( \frac{14207}{2128}-8 \alpha -\frac{1021 \eta }{228} \right) \xi_{\phi}^{2/3}+\frac{985}{152}\pi  \xi_{\phi}+\left( \frac{2142631}{38304}-\frac{126363 \alpha }{1064}+60 \alpha ^2-\frac{213625 \eta }{4256}
\right. \right. \\ \left. \left.
+\frac{2411 \alpha  \eta}{38}+\frac{141 \eta ^2}{19} \right)\xi_{\phi}^{4/3}+\left( \frac{247689  }{4256}-\frac{10835   \alpha }{152}-\frac{19067 \eta }{399}\right) \pi \xi_{\phi} ^{5/3}+ \left[\frac{712414748809}{884822400}-\frac{82283 \gamma }{1995}+\frac{769 \pi ^2}{57}
  \right. \right. \\ \left. \left.
  -\frac{9736907 \alpha
   }{6384} +\frac{3088747 \alpha ^2}{2128}-\frac{1360 \alpha^3}{3}-\frac{2672079257 \eta }{2298240}+\frac{59737 \pi ^2 \eta
   }{3648}+\frac{423963 \alpha  \eta }{266} - \frac{41 \pi^2\alpha\eta}{4}-\frac{72575\alpha ^2 \eta }{228}+\frac{29374613 \eta
   ^2}{153216}
  \right. \right. \\ \left. \left.
 -\frac{72575 \alpha  \eta ^2}{342} -\frac{305005 \eta ^3}{49248}-\frac{11021 \log(2)}{285} -\frac{234009 \log (3)}{5320}-\frac{82283 \log (\xi_\phi )}{5985} \right]\xi_{\phi}
^2+e_t^2 \left\{\frac{881}{304}+\left(\frac{172835}{4256}-\frac{1185 \alpha }{38}
 \right. \right. \right.  \\  \left. \left. \left.
-\frac{51847 \eta }{1824} \right) \xi_{\phi} ^{2/3}
+\frac{21729}{608}\pi  \xi_{\phi}+\left(\frac{3216641}{7296}-\frac{1620593 \alpha }{2128}+\frac{22335 \alpha ^2}{76}-\frac{7741193 \eta
   }{17024}+\frac{133089 \alpha  \eta }{304}+\frac{274515 \eta ^2}{2432}\right)
\xi_{\phi} ^{4/3}
\right. \right.  \\  \left. \left.
+\left(\frac{1951221}{3584}-\frac{14861 \alpha }{32}-\frac{7810371 \eta }{17024} \right) 
\pi \xi_{\phi} ^{5/3} + \left[  \frac{4366680686351}{589881600}-\frac{1500461 \gamma }{3990}+\frac{14023 \pi ^2}{114}-\frac{164403703\alpha
   }{12768}   +\frac{6310213 \alpha ^2}{608}
      \right. \right. \right. \\ \left. \left. \left.  
 -\frac{152405 \alpha ^3}{57}-\frac{45208444339 \eta}{4596480}+\frac{2721703 \pi ^2 \eta }{29184} +\frac{88733171 \alpha  \eta }{6384} -\frac{76629 \pi^2\alpha\eta}{1216}-\frac{9746899 \alpha ^2 \eta
   }{1824}+\frac{96285661 \eta ^2}{38304}
 \right. \right. \right. \\ \left. \left. \left.
 -\frac{4796353 \alpha  \eta ^2}{1824}-\frac{100330729 \eta ^3}{393984}
   -\frac{3813587 \log (2)}{3990}+\frac{6318243 \log (3)}{21280}-\frac{1500461 \log \left(16 \xi \phi
   ^{2/3}\right)}{7980}   \right]\xi_{\phi} ^2\right\}\right). 
\end{multline}
Dividing Eq.~\eqref{eq:dedt-MH-lowe_appendix} with Eq.~\eqref{eq:dxidt-MH-lowe-appendix} and keeping only the terms at leading order in $e_t$ [$\mathcal{O}(e_t)$] yields:
\begin{multline}
\label{eq:dedxi-lowe_appendix}
\frac{de_t}{d\xi_{\phi}} = -\frac{19}{18}\frac{e_t}{\xi_{\phi}}\left\{1+\left(\frac{9217}{3192}-2 \alpha -\frac{197 \eta }{114}\right) \xi_{\phi} ^{2/3}+\frac{377  }{152}\pi  \xi_{\phi}+\left(\frac{200855959}{9652608}-\frac{29965 \alpha }{798}+14 \alpha ^2-\frac{23725 \eta }{2128}
\right. \right. \\ \left. \left.
+\frac{926 \alpha \eta}{57}-\frac{833 \eta ^2}{1368}\right) \xi_{\phi} ^{4/3}+\left(\frac{379951}{51072}-\frac{1885 \alpha }{152}-\frac{133157 \eta }{12768}\right)\pi \xi_{\phi} ^{5/3}+ \left[\frac{68522142014911}{178380195840}-\frac{3317 \gamma }{133}-\frac{67 \pi ^2}{38}
   \right. \right. \\ \left. \left.
  -\frac{888787231 \alpha
   }{1608768}+\frac{456417 \alpha ^2}{1064}-108 \alpha ^3-\frac{26915590901 \eta }{64350720}+\frac{5371 \pi ^2
   \eta }{1216}+\frac{1861619 \alpha  \eta }{3192}-\frac{123}{16} \pi ^2 \alpha  \eta -\frac{6561 \alpha ^2 \eta
   }{38}
  \right. \right. \\ \left. \left.
  +\frac{5350057 \eta ^2}{153216}-\frac{2625 \alpha  \eta ^2}{76}-\frac{25 \eta ^3}{608}+\frac{4601}{105} \log(2)-\frac{234009 }{5320}\log(3) -\frac{3317}{266} \log(16 \xi_{\phi} ^{2/3})\right]\xi_{\phi} ^2\right\}.
\end{multline}
The full expression for Eq.~\eqref{eq:E} is:
\begin{multline}
\label{eq:E_appendix}
{\mathcal E}(\xi_{\phi}) = \left\{1+\left(-\frac{9217}{2016}+\frac{19 \alpha }{6}+\frac{197 \eta }{72}\right) \xi_{\phi} ^{2/3}-\frac{377  }{144}\pi  \xi_{\phi}+\left(-\frac{146852651}{24385536}+\frac{26351 \alpha }{1728}-\frac{437 \alpha ^2}{72}-\frac{534599 \eta
   }{145152}
\right. \right. \\ \left. \left. 
-\frac{1813 \alpha  \eta }{432}+\frac{43807 \eta ^2}{10368}\right) \xi_{\phi} ^{4/3}
+\left(\frac{10534927}{1451520}-\frac{377 \alpha }{864}-\frac{202589 \eta }{362880}\right)\pi \xi_{\phi} ^{5/3}+ \left[-\frac{56218957069579}{386266890240}+\frac{3317 \gamma }{252}
   \right. \right. \\ \left. \left. 
 +\frac{180721 \pi ^2}{41472}+\frac{19986093559
   \alpha }{146313216}-\frac{2167165 \alpha ^2}{20736}+\frac{35245 \alpha ^3}{1296}+\frac{1439021897137 \eta
   }{8778792960}-\frac{5371 \pi ^2 \eta }{2304}-\frac{156183605 \alpha  \eta }{870912}  
   \right. \right. \\ \left. \left. 
  +\frac{779}{192} \pi ^2
   \alpha  \eta +\frac{175175 \alpha ^2 \eta }{5184}-\frac{284200123 \eta ^2}{20901888}-\frac{222731 \alpha 
   \eta ^2}{62208}+\frac{10647791 \eta ^3}{2239488}-\frac{87419}{3780}\log(2)+\frac{26001}{1120}\log(3)
   \right. \right. \\ \left. \left.
   +\frac{3317}{504}\log(16 \xi_\phi^{2/3})\right]\xi_{\phi}
^2\right\} \,.
\end{multline}
Lastly, the full expression for $d\xi_{\phi}/dt$ in the small eccentricity limit [$\mathcal{O}(e_0^2)$] is:
\begin{multline}
\label{eq:dxidtexpand2_appendix}
\frac{d\xi_{\phi}}{dt} = \frac{96 \eta  \xi_{\phi}^{11/3}}{5M} \left(1-\left(-\frac{1273}{336}+6 \alpha +\frac{11 \eta }{4}\right) \xi_{\phi} ^{2/3}+4 \pi  \xi_{\phi} +\left(\frac{438887}{18144}-\frac{2365 \alpha }{42}+34 \alpha ^2-\frac{49507 \eta }{2016}+\frac{94 \alpha  \eta }{3}
   \right. \right. \\ \left. \left.
+\frac{59
   \eta ^2}{18}\right) \xi_{\phi} ^{4/3}  - \left(-\frac{20033}{672}+36 \alpha +\frac{189 \eta }{8}\right)\pi \xi_{\phi} ^{5/3}+ \left[\frac{38047038863}{139708800}-\frac{1712 \gamma }{105}+\frac{16 \pi ^2}{3}-\frac{560933 \alpha }{1134}+\frac{179215
   \alpha ^2}{336}
   \right. \right. \\ \left. \left.
  -\frac{580 \alpha ^3}{3}-\frac{16554367 \eta }{31104}+\frac{287 \pi ^2 \eta }{24}+\frac{546365 \alpha 
   \eta }{1008}-\frac{41}{16} \pi ^2 \alpha  \eta -\frac{3215 \alpha ^2 \eta }{12}+\frac{617285 \eta ^2}{8064}-\frac{1367
   \alpha  \eta ^2}{18}-\frac{5605 \eta ^3}{2592}
   \right. \right.  \\ \left. \left. 
-\frac{856}{105} \log \left(16 \xi_{\phi} ^{2/3}\right)\right]\xi_{\phi} ^2+\frac{157}{24} e_0^2 \left(\frac{\xi_{\phi,0}}{\xi_{\phi} }\right)^{19/9}
\left\{1- \left(\frac{6451}{22608}+\frac{731 \alpha }{471}+\frac{5413 \eta }{5652}\right) \xi_{\phi} ^{2/3}+\left(\frac{9217}{1008}-\frac{19 \alpha }{3}-\frac{197 \eta }{36}\right)
\xi_{\phi,0}^{2/3}
\right. \right. \\ \left. \left.  
+\frac{24871}{11304}\pi  \xi_{\phi}+\frac{377 }{72}\pi  \xi_{\phi,0} +\left(\frac{5808587}{239283072}-\frac{2593999 \alpha }{474768}+\frac{6965 \alpha ^2}{1413}-\frac{947713 \eta
   }{712152}+\frac{73373 \alpha  \eta }{16956}-\frac{36497 \eta ^2}{101736}\right) \xi_{\phi} ^{4/3}
 \right. \right. \\ \left. \left.
   +\left(-\frac{59458867}{22788864}-\frac{1469911 \alpha }{118692}+\frac{13889 \alpha ^2}{1413}-\frac{10248923 \eta
   }{1424304}+\frac{123427 \alpha  \eta }{8478}+\frac{1066361 \eta ^2}{203472}\right) \xi_{\phi} ^{2/3} \xi_{\phi,0}^{2/3} 
+\left(\frac{227857613}{3048192}
\right. \right. \right. \\ \left. \left. \left.
-\frac{354913 \alpha }{3024}+\frac{380 \alpha ^2}{9}-\frac{614081 \eta }{9072}+\frac{6521 \alpha
    \eta }{108}+\frac{18155 \eta ^2}{1296}\right) \xi_{\phi,0}^{4/3}+\left(\frac{436799501}{28486080}-\frac{32947 \alpha }{4239}-\frac{8416733 \eta }{508680} \right)\pi \xi_{\phi} ^{5/3}
\right. \right. \\ \left. \left.
+\left(\frac{32748001}{1627776} -\frac{472549 \alpha }{33912}-\frac{4899587 \eta }{406944}\right) \pi \xi_{\phi}  \xi_{\phi,0}^{2/3} -\left(\frac{2432027}{1627776}+\frac{275587 \alpha }{33912}
+\frac{2040701 \eta }{406944}\right)
\pi \xi_{\phi} ^{2/3} \xi_{\phi,0} 
+\left(\frac{5198401}{90720}
\right. \right. \right. \\ \left. \left. \left.
-\frac{2639 \alpha }{54}-\frac{949457 \eta }{22680}\right)\pi \xi_{\phi,0}^{5/3}
+\left(\frac{53537746379}{241197336576}-\frac{71947392655 \alpha }{1435698432}+\frac{6304577 \alpha ^2}{79128}-\frac{132335
   \alpha ^3}{4239}-\frac{105965140291 \eta }{8614190592}
   \right. \right. \right. \\ \left. \left. \left.
   +\frac{2311370 \alpha  \eta }{29673}-\frac{230516 \alpha ^2 \eta
   }{4239}+\frac{2171455 \eta ^2}{542592}-\frac{483985 \alpha  \eta ^2}{22608}+\frac{7189909 \eta ^3}{3662496}\right) \xi_{\phi} ^{4/3} \xi_{\phi,0}^{2/3}
+\frac{9376367 }{813888}\pi ^2 \xi_{\phi}
\xi_{\phi,0}
\right. \right. \\ \left. \left.
+\left(-\frac{1469909461463}{68913524736}-\frac{7405218505 \alpha }{89731152}+\frac{80760581 \alpha ^2}{474768}-\frac{277780
   \alpha ^3}{4239}-\frac{900632590565 \eta }{17228381184}+\frac{40741049 \alpha  \eta }{203472}
   \right. \right. \right. \\ \left. \left. \left.  
   -\frac{252733 \alpha ^2
   \eta }{1884} +\frac{4158752159 \eta ^2}{68366592}-\frac{8094913 \alpha  \eta ^2}{101736}-\frac{98273015 \eta
   ^3}{7324992}\right) \xi_{\phi} ^{2/3} \xi_{\phi,0}^{4/3}
+ \left[\frac{9646666078845287}{66329267558400}-\frac{2491067 \gamma }{98910}
      \right. \right. \right. \\ \left. \left. \left.
   -\frac{10610699 \pi ^2}{1627776}-\frac{3848482489 \alpha }{717849216}+\frac{1095361 \alpha
   ^2}{50868}-\frac{182860 \alpha ^3}{12717}-\frac{3437742494093 \eta }{8614190592}+\frac{2838389 \pi ^2 \eta }{180864}
     \right. \right. \right. \\ \left. \left. \left.
    +\frac{80508721 \alpha  \eta
   }{4272912}+\frac{3731 \pi ^2 \alpha  \eta }{15072}-\frac{197369 \alpha ^2 \eta }{12717}+\frac{1766285201 \eta ^2}{102549888}-\frac{500063 \alpha  \eta
   ^2}{305208}-\frac{2773315 \eta ^3}{10987488}
  \right. \right. \right. \\ \left. \left. \left.
+\frac{5257873 }{296730}\log(2)-\frac{1534059 }{87920}\log(3)-\frac{2491067
}{197820}\log(16 \xi_{\phi} ^{2/3})\right]\xi_{\phi} ^2 
+\left[ \frac{141706909432211}{168991764480}-\frac{3317 \gamma }{126}+\frac{122833 \pi ^2}{10368}
   \right. \right. \right. \\ \left. \left. \left.
-\frac{14633391323 \alpha }{9144576}+\frac{11026223 \alpha
   ^2}{9072}-\frac{24035 \alpha ^3}{81}-\frac{555242763539 \eta }{548674560}+\frac{5371 \pi ^2 \eta }{1152}+\frac{21838987 \alpha  \eta }{13608}-\frac{779}{96} \pi ^2
   \alpha  \eta 
     \right. \right. \right. \\ \left. \left. \left.
   -\frac{186691 \alpha ^2 \eta }{324}+\frac{341678423 \eta ^2}{1306368}-\frac{1033961 \alpha  \eta ^2}{3888}-\frac{3090307 \eta ^3}{139968}+\frac{87419
}{1890}\log(2) -\frac{26001 }{560}\log(3)
 \right. \right. \right. \\ \left. \left. \left.
-\frac{3317}{252} \log(16 \xi_{\phi,0}^{2/3}) \right] \xi_{\phi,0}^2 \right\} \right).
\end{multline}

\section{Alternate method for deriving the $\Delta\alpha$-dependent SPA phasing}\label{app:alternate method}
Here we provide an alternative approach for deriving the SPA phasing in the $\Delta \alpha$ parametrization. This approach relies on low-eccentricity expansions of the orbital energy and GW luminosity. The expressions here are written in terms of $\alpha$ and can be converted to $\Delta\alpha$ via $\alpha=1+\Delta\alpha$.

In the quasicircular limit and the \emph{stationary phase approximation}, the phase evolution is governed by the following differential equations:
\begin{subequations}
\label{eq:approx_timedom}
\begin{align}
\label{eq:approx_timedom_phi}
 \frac{d\phi}{dt} &=\frac{\xi_{\phi}}{M}=\frac{v^3}{M},\\
\label{eq:approx_timedom_v}
 \frac{dv}{dt} &=-\frac{{\mathcal F}(v)}{dE(v)/dv} \, ,
 \end{align}
 \end{subequations}
where $v=\xi_{\phi}^{1/3}=(\pi M f)^{1/3}$, $v_0=\xi_{\phi,0}^{1/3}=(\pi M f_0)^{1/3}$, $\mathcal{F}(v)$ is the energy flux, and $E(v)$ is the orbital energy. Expressions for $\mathcal{F}(v)$ and $E(v)$ are given in terms of $x$ and $e_t$ in ADM gauge in Eqs.~(6.5a) and (7.4a) of Ref.~\cite{ABIS09}. We reexpress these two equations in terms of $\xi$ by using Eq.~(4.17) of Ref.~\cite{ABIS09} and then substitute for $\xi$ in terms of $\xi_\phi$, $\alpha$ and $e_{t}$ in ADM gauge given by Eq.~\eqref{eq:zeeADM}. Now we have obtained $E(v)$ and $\mathcal{F}(v)$ as a power series in $\xi_\phi$ containing $\alpha$ explicitly as well as $e_t$ in ADM gauge. We can convert these equations into MH gauge via Eq.~\eqref{eq:ADMtoMH} to obtain the final required equations for the energy and flux. Finally, we use the eccentricity evolution Eq.~\eqref{eq:etofxi} and simultaneously expand the energy and flux in $v$ and $v_0$, yielding the low-eccentricity limit of $E(v)$ and $\mathcal{F}(v)$ in terms of $v$ and $\alpha$:

\begin{multline}
\label{eq:energy_v_appendix}
E(v)=-\frac{1}{2}\eta M v^2 \left(1+\left(\frac{5}{4}-2 \alpha -\frac{\eta }{12}\right)v^2 + \left(\frac{45}{8}-18 \alpha +9 \alpha ^2-\frac{21 \eta }{8} + 5 \alpha  \eta -\frac{\eta ^2}{24}\right)v^4 +
\left[\frac{7975}{192}-\frac{685 \alpha }{4}
\right. \right. \\ \left. \left.
+\frac{341 \alpha ^2}{2}-\frac{154 \alpha ^3}{3}-\frac{30403 \eta
   }{576}+\frac{41 \pi ^2 \eta }{96}+\frac{663 \alpha  \eta }{4}-\frac{41}{16} \pi ^2 \alpha  \eta -\frac{319
   \alpha ^2 \eta }{6}+\frac{1031 \eta ^2}{288}-\frac{187 \alpha  \eta ^2}{36}-\frac{35 \eta ^3}{5184}\right]v^6
   \right. \\ \left.
  +e_0^2\left(\frac{v_0}{v}\right)^{19/3} \left \{ -2 \alpha v^2 
 + \left(\frac{5}{2}-\frac{10691 \alpha }{504}+\frac{16 \alpha ^2}{3}-\eta +\frac{55 \alpha  \eta }{18} \right)v^4+ \left(-\frac{9217 \alpha}{504} +\frac{38 \alpha ^2}{3}+\frac{197 \alpha  \eta }{18} \right)v^2 v_0^2
	\right. \right. \\ \left. \left.
 +\frac{377 }{36} \alpha \pi v^5-\frac{377}{36} \alpha \pi  v^2 v_0^3+\left(-\frac{227857613 \alpha }{1524096}+\frac{354913 \alpha ^2}{1512}-\frac{760 \alpha ^3}{9}+\frac{614081 \alpha 
   \eta }{4536}-\frac{6521 \alpha ^2 \eta }{54}-\frac{18155 \alpha  \eta ^2}{648}\right)v^2 v_0^4
   \right. \right. \\ \left. \left.
  + \left(\frac{46085}{2016}-\frac{106582787 \alpha }{508032}
   +\frac{276865 \alpha ^2}{1512}-\frac{304 \alpha
   ^3}{9}-\frac{7669 \eta }{336}+\frac{1363987 \alpha  \eta }{9072}-\frac{2621 \alpha ^2 \eta }{54}+\frac{197
   \eta ^2}{36}-\frac{10835 \alpha  \eta ^2}{648}\right)v^4 v_0^2
 \right.  \right. \\  \left. \left. 
 +\left[\frac{77395}{2016}-\frac{192189365 \alpha }{762048}
+\frac{17243 \alpha ^2}{108}-\frac{322 \alpha
   ^3}{9}-\frac{19631 \eta }{336}+\frac{41 \pi ^2 \eta }{64}+\frac{434725 \alpha  \eta }{1296}-\frac{533}{64}
   \pi ^2 \alpha  \eta -\frac{973 \alpha ^2 \eta }{27}
   \right. \right. \right. \\ \left. \left. \left.
 +\frac{121 \eta ^2}{36}  +\frac{331 \alpha  \eta ^2}{162}\right]v^6\right \} + \mathcal{O}(v^8) \right) \,,
 \end{multline}

 \begin{multline}
\label{eq:energyflux_v_appendix}
{\cal F}(v)=\frac{32}{5} \eta^2v^{10}\left(1+ \left(\frac{2113}{336}-10 \alpha -\frac{35 \eta }{12}\right)v^2+4 \pi  v^3+ \left(\frac{458461}{9072}-\frac{3933 \alpha }{28}+85 \alpha ^2-\frac{20129 \eta }{504}+\frac{175 \alpha  \eta
   }{3}
  \right. \right.  \\  \left. \left.
  +\frac{65 \eta ^2}{18}\right)v^4 + \left(\frac{26753}{672}-52 \alpha -\frac{583 \eta }{24}\right)\pi  v^5
+ \left[\frac{13106635373}{23284800}-\frac{1712 \gamma }{105}+\frac{16 \pi ^2}{3}-\frac{976952 \alpha
   }{567}+\frac{109307 \alpha ^2}{56}-\frac{2090 \alpha ^3}{3}
 \right. \right. \\ \left. \left.
  -\frac{6881951 \eta }{7776}+\frac{41 \pi ^2 \eta}{3} +\frac{14792 \alpha  \eta }{9}
   -\frac{205}{16} \pi ^2 \alpha  \eta -\frac{4655 \alpha ^2 \eta
   }{6}+\frac{375997 \eta ^2}{3024}-\frac{1400 \alpha  \eta ^2}{9}-\frac{775 \eta ^3}{324}-\frac{856}{105}
\log(16 v^2)\right]v^6 
\right. \\  \left.
+\frac{157}{24} e_{0}^2 \left(\frac{v_0}{v}\right)^{19/3} \left\{1+ \left(\frac{50069}{22608}-\frac{2447 \alpha }{471}-\frac{6355 \eta }{5652}\right)v^2+  \left(\frac{9217}{1008}-\frac{19 \alpha }{3}-\frac{197 \eta }{36} \right)v_0^2 
+\frac{24871}{11304} \pi  v^3
+\frac{377 }{72}\pi
v_0^3 
 \right. \right.  \\  \left. \left.
+ \left(\frac{3857776007}{239283072}-\frac{28633531 \alpha }{474768}+\frac{50672 \alpha ^2}{1413}-\frac{32835973 \eta
   }{2848608}+\frac{379103 \alpha  \eta }{16956}-\frac{32975 \eta ^2}{101736}\right)v^4+
\left(\frac{461485973}{22788864}-\frac{3651647 \alpha }{59346}
\right. \right. \right. \\ \left. \left. \left.
+\frac{46493 \alpha ^2}{1413}
-\frac{63809593 \eta}{2848608}+\frac{150701 \alpha  \eta }{4239}+\frac{1251935 \eta ^2}{203472}\right)  v^2 v_0^2
+  \left(\frac{227857613}{3048192}-\frac{354913 \alpha}{3024}+\frac{380 \alpha ^2}{9}-\frac{614081 \eta
   }{9072}
  \right. \right. \right. \\ \left. \left. \left.
  +\frac{6521 \alpha  \eta }{108}+\frac{18155 \eta ^2}{1296}\right)v_0^4
 + \left(\frac{593486801}{28486080}-\frac{123887 \alpha }{8478} -\frac{17206531 \eta }{1017360} \right)\pi v^5+  \left(\frac{32748001}{1627776}-\frac{472549 \alpha }{33912}
 \right. \right. \right. \\  \left. \left. \left.
-\frac{4899587 \eta }{406944} \right)\pi v^3 v_0^2
+ \left(\frac{18876013}{1627776}
-\frac{922519 \alpha }{33912}
-\frac{2395835 \eta }{406944} \right)\pi v^2  v_0^3+ \left(\frac{5198401 }{90720}-\frac{2639  \alpha }{54}-\frac{949457  \eta }{22680} \right)\pi v_0^5 
\right.  \right. \\  \left. \left.
+  \left[\frac{20660014769574287}{66329267558400}-\frac{2491067 \gamma }{98910}-\frac{10610699 \pi
   ^2}{1627776}-\frac{528314926957 \alpha }{717849216}+\frac{1154130889 \alpha ^2}{1424304}-\frac{3471091 \alpha
   ^3}{12717}
\right. \right. \right. \\ \left. \left. \left.
  -\frac{5466800871683 \eta }{8614190592}+\frac{3162125 \pi ^2 \eta }{180864}+\frac{6756239213 \alpha
    \eta }{8545824}-\frac{166747 \pi ^2 \alpha  \eta }{15072}-\frac{14651489 \alpha ^2 \eta
   }{50868}+\frac{284237683 \eta ^2}{7324992}-\frac{10447469 \alpha  \eta ^2}{305208}
\right. \right. \right. \\  \left. \left. \left.
-\frac{274435 \eta
   ^3}{2746872}+\frac{5257873 }{296730}\log(2) 
-\frac{1534059 }{87920}\log(3) -\frac{2491067
}{197820}	\log(16 v^2)\right]v^6+ \left(\frac{35557121456519}{241197336576}-\frac{938341253947 \alpha }{1435698432}
   \right. \right. \right. \\ \left. \left. \left.
  +\frac{337026971 \alpha
   ^2}{474768}
 -\frac{962768 \alpha ^3}{4239}-\frac{833964681401 \eta }{4307095296}+\frac{288410137 \alpha  \eta
   }{474768} -\frac{5728447 \alpha ^2 \eta }{16956}+\frac{1027459351 \eta ^2}{17091648}-\frac{24476747 \alpha 
   \eta ^2}{203472}
\right. \right. \right. \\ \left. \left. \left.
+\frac{6496075 \eta ^3}{3662496}\right)v^4 v_0^2 
+\frac{9376367 }{813888}\pi ^2 v^3 v_0^3+ \left(\frac{11408602825297}{68913524736}-\frac{232685124487 \alpha }{358924608}+\frac{333885217 \alpha
   ^2}{474768}
   -\frac{929860 \alpha ^3}{4239}
 \right. \right. \right.  \\  \left. \left. \left.
 -\frac{4030734544091 \eta }{17228381184}+\frac{1758599431 \alpha 
   \eta }{2848608}-\frac{6123929 \alpha ^2 \eta }{16956}+\frac{90423695 \eta ^2}{844032}-\frac{1788880 \alpha 
   \eta ^2}{12717}
   -\frac{115375025 \eta ^3}{7324992}\right)v^2  v_0^4
        \right. \right. \\  \left. \left.
  +  \left[\frac{141706909432211}{168991764480}-\frac{3317 \gamma }{126}+\frac{122833 \pi ^2}{10368}-\frac{14633391323
   \alpha }{9144576}+\frac{11026223 \alpha ^2}{9072}-\frac{24035 \alpha ^3}{81}-\frac{555242763539 \eta
   }{548674560}
\right. \right. \right. \\ \left. \left. \left.
+\frac{5371 \pi ^2 \eta }{1152}+\frac{21838987 \alpha  \eta }{13608}-\frac{779}{96} \pi ^2 \alpha
    \eta -\frac{186691 \alpha ^2 \eta }{324}+\frac{341678423 \eta ^2}{1306368}-\frac{1033961 \alpha  \eta
   ^2}{3888}-\frac{3090307 \eta ^3}{139968} 
+\frac{87419}{1890} \log(2)
  \right. \right. \right. \\ \left. \left. \left.
-\frac{26001 }{560}\log(3)-\frac{3317}{252} \log(16
v_0^2)\right]v_0^6 \right\}\right).
\end{multline} 

The TaylorT2 approximant is obtained by series expanding the ratio in Eq.~\eqref{eq:approx_timedom_v} to the appropriate PN order. One then analytically obtains a parametric solution for the phase $[\langle\phi\rangle(v), t(v)]$ by integrating: 
\begin{subequations}
\label{eq:approx_freqdom_c}
\begin{align}
\label{eq:approx_freqdom_phi_c}
\frac{d\langle\phi\rangle}{dv} &= \frac{d\langle\phi\rangle}{dt}\frac{dt}{dv}=-\frac{v^3}{M} \frac{dE(v)/dv}{{\mathcal F}(v)} \,, \\
\label{eq:appox_freqdom_t}
\frac{dt}{dv} &= -\frac{dE(v)/dv}{{\mathcal F}(v)} \, .
\end{align}
\end{subequations}
We need to calculate $dE/dv$ to obtain the TaylorT2 approximant. This can be obtained by differentiating Eq.~\eqref{eq:energy_v_appendix} with respect to $v$; the result is:
\begin{multline}
\label{eq:denergy_dv}
\frac{dE}{dv}= -\eta M v\left(1+\left(\frac{5}{2}-4 \alpha -\frac{\eta }{6} \right)v^2+ \left[\frac{135}{8}-54 \alpha +27 \alpha ^2-\frac{63 \eta }{8}+15 \alpha  \eta -\frac{\eta ^2}{8} \right]v^4+ \left[\frac{7975}{48}-685 \alpha +682 \alpha ^2
\right. \right. \\ \left. \left.
-\frac{616 \alpha ^3}{3}
-\frac{30403 \eta }{144}+\frac{41 \pi ^2 \eta
   }{24}+663 \alpha  \eta -\frac{41}{4} \pi ^2 \alpha  \eta -\frac{638 \alpha ^2 \eta }{3}+\frac{1031 \eta
   ^2}{72}-\frac{187 \alpha  \eta ^2}{9}-\frac{35 \eta ^3}{1296} \right]v^6
+e_{0}^2 \left(\frac{v_0}{v}\right)^{19/3} \left\{\frac{7}{3} \alpha v^2 
\right. \right. \\ \left. \left.
+ \left(\frac{9217 \alpha }{432}-\frac{133 \alpha ^2}{9}-\frac{1379 \alpha  \eta }{108}\right)v^2
v_0^2+ \left(-\frac{5}{12}+\frac{10691 \alpha }{3024}-\frac{8 \alpha ^2}{9}+\frac{\eta }{6}-\frac{55 \alpha  \eta }{108}\right)v^4+\frac{377}{108} \alpha \pi  v^5+\frac{2639}{216} \alpha \pi  v^2 v_0^3
\right.  \right. \\ \left. \left.
+\left(\frac{227857613 \alpha }{1306368}-\frac{354913 \alpha ^2}{1296}+\frac{2660 \alpha ^3}{27}-\frac{614081 \alpha 
   \eta }{3888}+\frac{45647 \alpha ^2 \eta }{324}+\frac{127085 \alpha  \eta ^2}{3888}\right)v^2 v_0^4
 + \left[\frac{386975}{12096}-\frac{960946825 \alpha }{4572288}
\right. \right. \right. \\ \left. \left. \left.
 +\frac{86215 \alpha ^2}{648}-\frac{805 \alpha
   ^3}{27}-\frac{98155 \eta }{2016}+\frac{205 \pi ^2 \eta }{384}+\frac{2173625 \alpha  \eta
   }{7776}-\frac{2665}{384} \pi ^2 \alpha  \eta -\frac{4865 \alpha ^2 \eta }{162}+\frac{605 \eta
   ^2}{216}+\frac{1655 \alpha  \eta ^2}{972}\right]v^6
\right.  \right. \\ \left. \left.
 + \left(-\frac{46085}{12096}+\frac{106582787 \alpha }{3048192}-\frac{276865 \alpha ^2}{9072}+\frac{152 \alpha
   ^3}{27}+\frac{7669 \eta }{2016}-\frac{1363987 \alpha  \eta }{54432}+\frac{2621 \alpha ^2 \eta
   }{324}-\frac{197 \eta ^2}{216}+\frac{10835 \alpha  \eta ^2}{3888}\right)v^4 v_0^2
   \right\}\right)\,.
\end{multline}

The resulting equation for $\langle\phi\rangle(v)$ is:
\begin{subequations}
\label{eq:taylort2_phi}
\begin{equation}
\label{eq:taylort2_phi_b}
\langle\phi\rangle-\phi_{c}=-\frac{1}{32 v^5 \eta } \Lambda_f(v,v_0,e_0) \,,
\end{equation}
\begin{multline}
\label{eq:Lambdaf_PNapprox_b}
\Lambda_f(v,v_0,e_0) = \left(1+ \left(-\frac{6365}{1008}+10 \alpha +\frac{55 \eta }{12}\right)v^2-10 \pi  v^3+ \left(-\frac{49964555}{1016064}+\frac{4555 \alpha }{84}+10 \alpha ^2+\frac{18745 \eta }{1008}+\frac{25 \alpha  \eta
   }{3}
   \right. \right. \\  \left. \left.
   +\frac{3085 \eta ^2}{144}\right)v^4+\left(-\frac{1675}{2016}+20 \alpha -\frac{65 \eta }{24}  \right)\pi \log(v^3) v^5 + \left[\frac{14996074652051}{18776862720}-\frac{1712 \gamma }{21}-\frac{160 \pi ^2}{3}-\frac{91641805 \alpha
   }{508032}
   \right. \right. \\ \left. \left.
   +\frac{15475 \alpha ^2}{336}-\frac{20 \alpha ^3}{3}-\frac{20211350275 \eta }{12192768}+\frac{1435
   \pi ^2 \eta }{24}+\frac{31625 \alpha  \eta }{84}-\frac{205}{16} \pi ^2 \alpha  \eta -\frac{115 \alpha ^2 \eta
   }{12}+\frac{89975 \eta ^2}{6912}-\frac{145 \alpha  \eta ^2}{72}
 \right. \right. \\ \left. \left.
 -\frac{127825 \eta ^3}{5184}-\frac{856}{21} \log(16 v^2)\right]v^6
-\frac{785}{272} e_0^2 \left(\frac{v_0}{v}\right)^{19/3}
\left\{1
+\left(-\frac{21153491}{2215584}+\frac{11951 \alpha }{942}+\frac{436441 \eta }{79128}\right)v^2
+ \left(\frac{9217}{1008}-\frac{19 \alpha }{3}
\right. \right. \right. \\ \left. \left. \left.
-\frac{197 \eta }{36} \right)v_0^2 -\frac{1114537 }{141300}\pi  v^3
+\frac{377 }{72}\pi  v_0^3+ \left(-\frac{330629039}{68366592}-\frac{168696389 \alpha }{5222448}+\frac{57443 \alpha ^2}{1413}-\frac{437894789 \eta
   }{31334688}+\frac{5947195 \alpha  \eta }{186516}
    \right. \right. \right. \\  \left. \left. \left.
 +\frac{36339727 \eta ^2}{2238192}\right)v^4 + \left(-\frac{194971726547}{2233308672}+\frac{586491449 \alpha }{3323376} -\frac{227069 \alpha
   ^2}{2826}+\frac{1023739303 \eta }{9970128}-\frac{3096601 \alpha  \eta }{29673}
   \right. \right. \right. \\ \left. \left. \left.
   -\frac{85978877 \eta
   ^2}{2848608}\right)v^2 v_0^2 +  \left(\frac{227857613}{3048192}-\frac{354913 \alpha }{3024}+\frac{380 \alpha ^2}{9}-\frac{614081 \eta
   }{9072}+\frac{6521 \alpha  \eta }{108}+\frac{18155 \eta ^2}{1296}\right)v_0^4 + 
\left(\frac{488067161}{8456805}
 \right. \right. \right. \\ \left. \left. \left.
-\frac{5902519 \alpha }{80541}-\frac{268652717 \eta }{9664920}\right)\pi v^5+ \left(-\frac{10272687529}{142430400}+\frac{21176203 \alpha }{423900}+\frac{219563789 \eta }{5086800} \right)\pi v^3 v_0^2 +  \left(-\frac{7974866107}{159522048}
\right. \right. \right. \\  \left. \left. \left.
+\frac{4505527 \alpha }{67824}+\frac{164538257 \eta }{5697216}\right)\pi v^2 v_0^3 +  \left(\frac{5198401}{90720}-\frac{2639 \alpha }{54}-\frac{949457 \eta }{22680} \right) \pi v_0^5
+ \left[-\frac{144704938285200649}{1061268280934400}+\frac{12483797 \gamma }{791280}
   \right. \right. \right. \\ \left. \left. \left.
  +\frac{365639621 \pi
   ^2}{13022208}-\frac{1871487375475 \alpha }{11485587456}+\frac{123085525 \alpha ^2}{1424304}+\frac{252875
   \alpha ^3}{12717}+\frac{58822114372195 \eta }{137827049472}-\frac{25283675 \pi ^2 \eta
   }{1446912}
   \right. \right. \right. \\ \left. \left. \left.
 -\frac{1778936525 \alpha  \eta }{9766656}+\frac{1376575 \pi ^2 \alpha  \eta }{120576}+\frac{394825
   \alpha ^2 \eta }{50868}+\frac{12671787437 \eta ^2}{1640798208}+\frac{201434275 \alpha  \eta
   ^2}{4883328}+\frac{5885194385 \eta ^3}{175799808}
  \right. \right. \right.  \\ \left. \left. \left.
  +\frac{89383841}{2373840} \log(2) -\frac{26079003 }{703360}\log(3)
+\frac{12483797}{1582560}\log(16 v^2)\right]v^6+ \left(-\frac{3047407852463}{68913524736}-\frac{2090456784091 \alpha }{7896341376}+\frac{188108434 \alpha
   ^2}{326403}
\right. \right. \right.\\  \left. \left. \left.
-\frac{1091417 \alpha ^3}{4239} -\frac{19201145728687 \eta }{189512193024}+\frac{17448081155 \alpha
    \eta }{31334688}-\frac{39579281 \alpha ^2 \eta }{93258}+\frac{169157936875 \eta
   ^2}{752032512}-\frac{51723673 \alpha  \eta ^2}{186516}
   \right. \right. \right. \\ \left. \left. \left.
  -\frac{7158926219 \eta ^3}{80574912}\right)v^4  v_0^2-\frac{420180449 }{10173600}\pi ^2 v^3 v_0^3+
\left(-\frac{4819983965876983}{6753525424128}+\frac{20792415592295 \alpha }{10049889024}-\frac{12576427807 \alpha
   ^2}{6646752}
\right. \right. \right. \\ \left. \left. \left.
+\frac{2270690 \alpha ^3}{4239}+\frac{255325887356585 \eta }{241197336576} -\frac{20763716513
   \alpha  \eta }{9970128}+\frac{79041653 \alpha ^2 \eta }{79128}-\frac{485360710663 \eta
   ^2}{957132288}+\frac{363735383 \alpha  \eta ^2}{712152}
    \right. \right. \right. \\ \left. \left. \left.
  +\frac{7923586355 \eta ^3}{102549888}\right)v^2 v_0^4
+ \left[\frac{141706909432211}{168991764480}-\frac{3317 \gamma }{126}+\frac{122833 \pi ^2}{10368}-\frac{14633391323
   \alpha }{9144576}+\frac{11026223 \alpha ^2}{9072}-\frac{24035 \alpha ^3}{81}
    \right. \right. \right. \\ \left. \left. \left.
   -\frac{555242763539 \eta
   }{548674560}+\frac{5371 \pi ^2 \eta }{1152}+\frac{21838987 \alpha  \eta }{13608}-\frac{779}{96} \pi ^2 \alpha
    \eta -\frac{186691 \alpha ^2 \eta }{324}+\frac{341678423 \eta ^2}{1306368}-\frac{1033961 \alpha  \eta
   ^2}{3888}
   \right. \right. \right. \\ \left. \left. \left.
   -\frac{3090307 \eta ^3}{139968}
+\frac{87419 }{1890}\log(2)-\frac{26001 }{560}\log(3)-\frac{3317}{252} \log(16 v_0^2)\right]v_0^6\right\} \right)\,.
\end{multline}
\end{subequations} 
Similarly, the resulting equation for $t(v)$ is:
\begin{subequations}
\label{eq:taylort2_t}
\begin{equation}
\label{eq:taylort2_t_a}
t_{c}-t=\frac{5}{256 } \frac{M}{\eta} \frac{1}{v^8} {\mathcal T}(v,v_0,e_0) \,,
\end{equation}
\begin{multline}
\label{eq:T_PNapprox}
{\mathcal T}(v,v_0,e_0) = \left(1+ \left(-\frac{1273}{252}+8 \alpha +\frac{11 \eta }{3}\right)v^2-\frac{32 }{5}\pi  v^3+ \left(-\frac{9992911}{508032}+\frac{911 \alpha }{42}+4 \alpha ^2+\frac{3749 \eta }{504}+\frac{10 \alpha  \eta
   }{3}
   \right. \right. \\ \left. \left.
   +\frac{617 \eta ^2}{72}\right)v^4+ \left(\frac{335}{252}-32 \alpha +\frac{13 \eta }{3}\right)\pi v^5
+ \left[-\frac{12699932582291}{23471078400}+\frac{6848 \gamma }{105}+\frac{128 \pi ^2}{3}+\frac{18328361 \alpha
   }{127008}-\frac{3095 \alpha ^2}{84}
   \right. \right. \\ \left. \left.
  +\frac{16 \alpha ^3}{3}+\frac{4042270055 \eta }{3048192}-\frac{287 \pi ^2\eta }{6} -\frac{6325 \alpha  \eta }{21}+\frac{41}{4} \pi ^2 \alpha  \eta +\frac{23 \alpha ^2 \eta
   }{3}-\frac{17995 \eta ^2}{1728}+\frac{29 \alpha  \eta ^2}{18}+\frac{25565 \eta ^3}{1296}  
\right. \right.  \\  \left. \left.
+\frac{3424}{105} \log(16 v^2)\right]v^6
-\frac{157}{43} e_0^2 \left(\frac{v_0}{v}\right)^{19/3} \left\{1+\left(-\frac{53505889}{5855472}+\frac{5719 \alpha }{471}+\frac{1103939 \eta }{209124}\right)v^2
+\left(\frac{9217}{1008}-\frac{19 \alpha }{3}
\right. \right. \right.  \\  \left. \left. \left.
-\frac{197 \eta }{36} \right) v_0^2-\frac{2819123 }{384336}\pi  v^3+\frac{377 }{72}\pi  v_0^3
+ \left(-\frac{9199266791}{2119364352}-\frac{13764601 \alpha }{474768}+\frac{51557 \alpha ^2}{1413}-\frac{1107616231
   \eta }{88306848}+\frac{485255 \alpha  \eta }{16956}
  \right. \right. \right. \\  \left. \left. \left.
   +\frac{91918133 \eta ^2}{6307632}\right)v^4+ \left(-\frac{493163778913}{5902315776}+\frac{1483478371 \alpha }{8783208}-\frac{108661 \alpha
   ^2}{1413}+\frac{2589458237 \eta }{26349624}-\frac{15665158 \alpha  \eta }{156843}
   \right. \right. \right.  \\  \left. \left. \left.
  -\frac{217475983 \eta
   ^2}{7528464}\right) v^2 v_0^2+ \left(\frac{227857613}{3048192}-\frac{354913 \alpha }{3024}+\frac{380 \alpha ^2}{9}-\frac{614081 \eta
   }{9072} +\frac{6521 \alpha  \eta }{108}+\frac{18155 \eta ^2}{1296}\right)v_0^4 + \left(\frac{1234522819}{24925320}
  \right. \right. \right. \\ \left. \left. \left.
  -\frac{2132843 \alpha }{33912}-\frac{679533343 \eta }{28486080}\right)\pi v^5+ \left(-\frac{25983856691}{387410688}+\frac{53563337 \alpha }{1153008} +\frac{555367231 \eta }{13836096}\right)\pi v^3 v_0^2 + \left(-\frac{20171720153}{421593984}
  \right. \right. \right.  \\  \left. \left. \left.
   +\frac{2156063 \alpha }{33912}+\frac{416185003 \eta }{15056928}\right)\pi v^2 v_0^3 + \left(\frac{5198401}{90720}-\frac{2639 \alpha }{54}-\frac{949457 \eta }{22680} \right) \pi v_0^5+ \left[-\frac{1844385248682305311}{16582316889600000}
   \right. \right. \right. \\ \left. \left. \left.
   +\frac{31576663 \gamma }{2472750}+\frac{924853159 \pi
   ^2}{40694400}-\frac{189350487401 \alpha }{1435698432}+\frac{12453359 \alpha ^2}{178038}+\frac{204680 \alpha
   ^3}{12717}+\frac{29757069623581 \eta }{86141905920}
   \right. \right. \right. \\ \left. \left. \left.
  -\frac{2558113 \pi ^2 \eta }{180864}-\frac{179986519
   \alpha  \eta }{1220832}+\frac{139277 \pi ^2 \alpha  \eta }{15072}+\frac{79894 \alpha ^2 \eta
   }{12717}+\frac{32052168223 \eta ^2}{5127494400}+\frac{20380409 \alpha  \eta ^2}{610416}
  \right. \right. \right. \\ \left. \left. \left.
  +\frac{2977215983 \eta
   ^3}{109874880}+\frac{226088539 }{7418250}\log(2)-\frac{65964537 }{2198000}\log(3) +\frac{31576663
}{4945500}\log(16 v^2)\right]v^6+ \left(-\frac{84789642012647}{2136319266816}
   \right. \right. \right. \\ \left. \left. \left.
   -\frac{5287625983289 \alpha }{22253325696}+\frac{15348506 \alpha
   ^2}{29673}-\frac{979583 \alpha ^3}{4239}-\frac{48567603901973 \eta }{534079816704} +\frac{44133381745 \alpha 
   \eta }{88306848}-\frac{3229429 \alpha ^2 \eta }{8478}
   \right. \right. \right. \\ \left. \left. \left.
   +\frac{427870075625 \eta ^2}{2119364352}
   -\frac{130830467
   \alpha  \eta ^2}{525636}-\frac{18107872201 \eta ^3}{227074752}\right)v^4 v_0^2 -\frac{1062809371 }{27672192}\pi ^2 v^3 v_0^3
   +\left(-\frac{12191724148982957}{17848602906624}
   \right. \right.  \right. \\  \left. \left. \left.
   +\frac{52592580615805 \alpha }{26560420992}-\frac{31810964453 \alpha
   ^2}{17566416}+\frac{2173220 \alpha ^3}{4239}+\frac{645824303313715 \eta }{637450103808}-\frac{52519988827
   \alpha  \eta }{26349624}
    \right. \right. \right. \\ \left. \left. \left.
   +\frac{199928887 \alpha ^2 \eta }{209124}-\frac{1227677091677 \eta
   ^2}{2529563904}+\frac{920036557 \alpha  \eta ^2}{1882116}+\frac{20042012545 \eta ^3}{271024704}\right)v^2v_0^4 +  \left[\frac{141706909432211}{168991764480}-\frac{3317 \gamma }{126}
    \right. \right. \right.\\ \left. \left. \left.
  +\frac{122833 \pi ^2}{10368}-\frac{14633391323
   \alpha }{9144576}+\frac{11026223 \alpha ^2}{9072}-\frac{24035 \alpha ^3}{81}-\frac{555242763539 \eta
   }{548674560}+\frac{5371 \pi ^2 \eta }{1152}+\frac{21838987 \alpha  \eta }{13608} 
    \right. \right. \right. \\ \left. \left. \left.
  -\frac{779}{96} \pi ^2 \alpha
    \eta -\frac{186691 \alpha ^2 \eta }{324}+\frac{341678423 \eta ^2}{1306368}-\frac{1033961 \alpha  \eta
   ^2}{3888}-\frac{3090307 \eta ^3}{139968}+\frac{87419 }{1890}\log(2)
  \right. \right. \right. \\ \left. \left. \left.
   -\frac{26001 }{560}\log(3)-\frac{3317}{252} \log(16 v_0^2) \right] v_0^6 \right\} \right).
\end{multline}
\end{subequations}
The SPA phase can now be obtained using Eq.~\eqref{eq:PsiFT}. The complete expression is given by Eq.~\eqref{eq:PsiFTecc}. The result obtained via the approach here agrees with that found in the main text.
\end{widetext}

\section{\label{app:eccentricity_measurement}Constraints on the eccentricity parameter $e_0$}
\begin{figure*}
    \centering
    \begin{subfigure}{\includegraphics[width=0.235\textwidth]{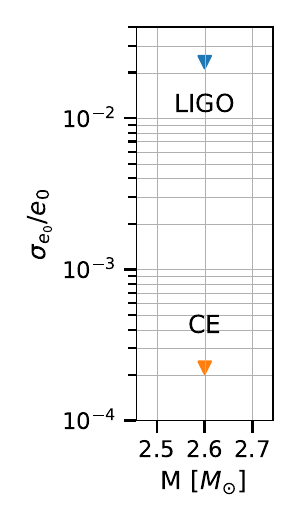}}
    \end{subfigure}
   \begin{subfigure}{\includegraphics[width=0.71\textwidth]{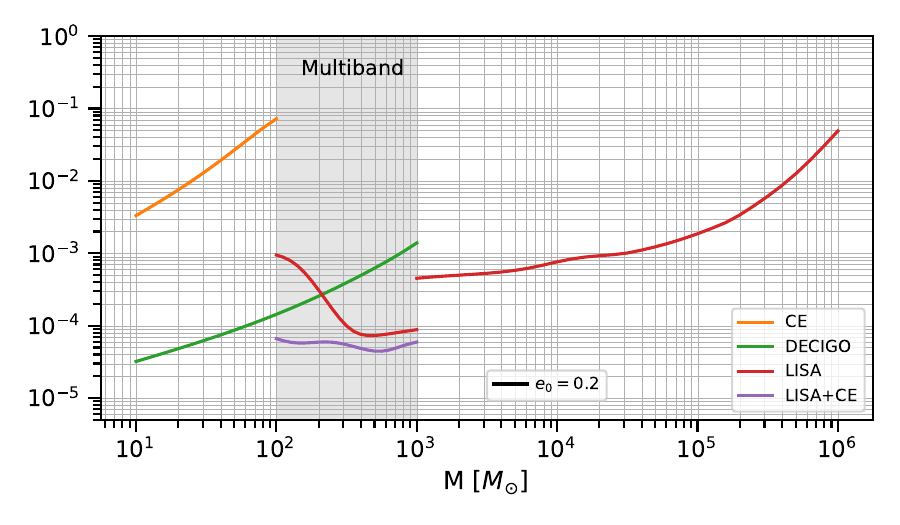}}
   \end{subfigure}
  \caption{(Color online) $1\sigma$ fractional error on the initial eccentricity $e_0$ measured along with $(\delta\phi_2$, $\delta\phi_2^e)$ and the other binary parameters as in Fig.~\ref{fig:first_parametrization_deviation}.  The different colored curves represent different detectors. The binaries properties are the same as in Fig.~\ref{fig:first_parametrization_deviation}. The left panel shows errors on a BNS system with component masses $m_1=1.4 \, M_{\odot}$, $m_1=1.2 \, M_{\odot}$, spins $\chi_{1,2}=0.05$. The luminosity distance is $100$ Mpc. In all cases we assume $e_0=0.2$ at the reference frequency.}
  \label{fig:eccentricity_dphi}
\end{figure*}

\begin{figure*}
    \centering
    \begin{subfigure}{\includegraphics[width=0.234\textwidth]{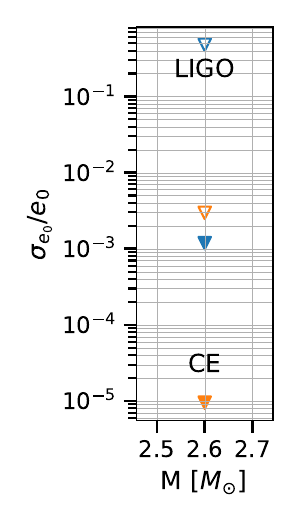}}
    \end{subfigure}
   \begin{subfigure}{\includegraphics[width=0.71\textwidth]{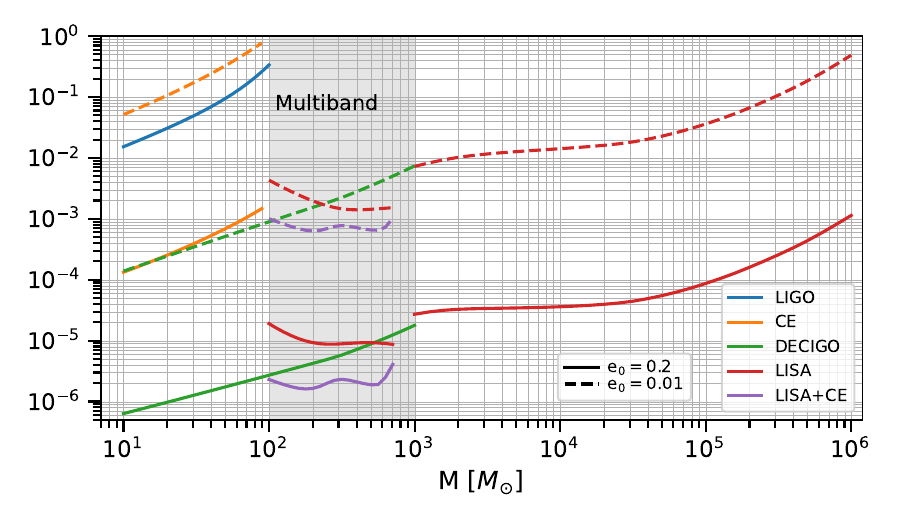}}
   \end{subfigure}
\caption{(Color online) $1\sigma$ errors on the initial eccentricity $e_0$ measured along with the periastron advance parameter $\Delta\alpha$ and the binary source parameters. The different colored curves represent different detectors. The binaries properties are the same as in Fig.~\ref{fig:alpha}. The left panel shows errors on a nonspinning BNS system with component masses $m_1=1.4 \, M_{\odot}$, $m_2=1.2 \, M_{\odot}$. The luminosity distance is $100$ Mpc. Orange triangles are for a CE detector, blue triangles for LIGO. The filled triangles and solid curves assume $e_0=0.2$ at the reference frequency; the unfilled triangles and dashed curves assume $e_0=0.01$.}
 \label{fig:eccentricity_alpha}
\end{figure*}

\begin{figure*}[th]
    \centering
    \begin{subfigure}{\includegraphics[width=0.255\textwidth]{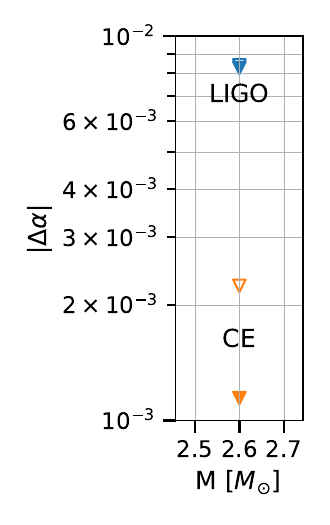}}
    \end{subfigure}
   \begin{subfigure}{\includegraphics[width=0.71\textwidth]{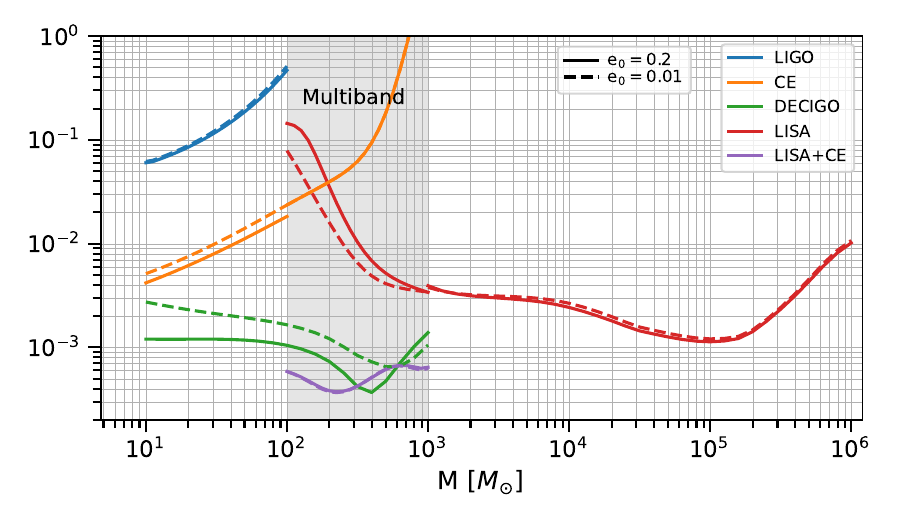}}
   \end{subfigure}
\caption{(Color online) Same as Fig.~\ref{fig:alpha} but for spinning binaries, with spin terms included as described in Appendix~\ref{app:alpha_spin}. The dimensionless spins are $\chi_{1}=0.5$ and $\chi_2 = 0.4$. Other binary parameters are the same as in Fig.~\ref{fig:alpha}.}
    \label{fig:alpha_spin}
\end{figure*}

Here we examine the parameter errors on the eccentricity parameter $e_0$ for our two waveform parametrizations. In the main text, the eccentricity is treated as a free parameter in our Fisher matrix calculations, along with the other source parameters and the beyond GR parameters. While there we focused on the parameter bounds on either $(\delta \hat{\varphi}_i,\delta \hat{\varphi}_i^{e})$ or $\Delta\alpha$, here we report the $1\sigma$ constraints on $e_0$. Note that the eccentricity is defined at a reference frequency $f_0$ that varies depending on the detector configuration considered; see the main text for details.

Figure~\ref{fig:eccentricity_dphi} shows the fractional errors on the binary eccentricity $(\sigma_{e_0}/e_0)$, when $e_0$ is measured along with the 1PN deviation parameters $(\delta\hat{\varphi}_2, \delta\hat{\varphi}_2^e)$ and the other binary parameters. The initial eccentricity is chosen to be $e_0=0.2$ for all binaries. Other binary parameters are the same as in Fig.~\ref{fig:first_parametrization_deviation}. In each detector, the eccentricity is better measured for low-mass binaries. This is due to the longer inspiral time in the detector band for low-mass binaries. The measurement of eccentricity is inversely proportional to $e_0^2$ and directly proportional to $M^{5/3}$~\cite{Favata:2021vhw}:
\begin{equation}
    \frac{\sigma_{e_0}}{e_0} \propto \frac{1}{e_0^2}\frac{\eta}{\rho} (M f_{\rm low})^{5/3} \bigg(\frac{f_{\rm low}}{f_0}\bigg)^{19/9} \,.
\end{equation}
Therefore, low-mass binaries with higher values of eccentricity have better eccentricity measurements. The errors increase as the binary mass increases.
The LIGO curve for BBHs lies outside the plot range since $\sigma_{e_0}/e_0 >1$ in that case. However, LIGO can measure the eccentricity of the BNS system with $\sigma_{e_0}/e_0 \approx 2 \times 10^{-2}$ (left panel). In CE, the error on $e_0$ for the BNS decreases by $2$ orders of magnitude $(\sigma_{e_0}/e_0 \approx 2 \times 10^{-4})$. This is due to the better overall sensitivity of CE and its enhanced sensitivity at low frequencies (where the instantaneous binary eccentricity is larger). The better the measurement of eccentricity in Fig.~\ref{fig:eccentricity_dphi}, the better the measurement of deviation parameters in Fig.~\ref{fig:first_parametrization_deviation}. Since the eccentricity is defined at $4$ years before the merger for multiband sources, all the eccentricity has decayed by the time the binary enters the CE band; therefore it is not measurable by CE. Hence, we do not show CE-only curves in the multiband region of the plot. However, when LISA and CE observations are combined, the measurement accuracy improves by one order of magnitude for $M=100\, M_{\odot}$. The best measurement of eccentricity comes from DECIGO and multiband observations between LISA and CE $(\sigma_{e_0}/e_0 \approx 4 \times 10^{-5})$.

Figure~\ref{fig:eccentricity_alpha} shows the fractional errors on the initial binary eccentricity $e_0$, assuming binaries with either $e_0=0.2$ or $0.01$. There, the eccentricity is jointly measured with the periastron deviation parameter $\Delta\alpha$ and the other binary source parameters. The same binary properties as in Fig.~\ref{fig:alpha} are assumed. As expected, we see that the fractional errors on $e_0$ decrease as the binary eccentricity increases from $0.01$ to $0.2$. Low-mass binaries generally have better measurements of eccentricity compared to high-mass binaries (except for multiband LISA sources). A binary with $M=10 \,M_{\odot}$ and $e_0=0.2$ can be measured with fractional precision $\sigma_{e_0}/e_0 \sim 10^{-2}$ in LIGO. For the same system, CE can measure the eccentricity with $\sigma_{e_0}/e_0\sim 2\times 10^{-4}$. For an eccentricity of $e_0=0.01$, CE can set a constraint of $\sigma_{e_0}/e_0\approx 0.1$ for $M=10\,M_{\odot}$. Such a small eccentricity is not measurable with LIGO and is not shown on the plot. For the BNS system in LIGO, $e_0 = 0.2$ can be measured with $\sigma_{e_0}/e_0\approx 10^{-3}$, while in CE it can be measured with $\sigma_{e_0}/e_0\approx 10^{-5}$. For a BNS system with $e_0=0.01$, these errors are $\sim 2$ orders of magnitude larger compared to $e_0=0.2$.

DECIGO can measure an eccentricity of $e_0=0.2$ with $\sigma_{e_0}/e_0 \sim 10^{-6}$ for lower mass BBHs. The errors increase by two orders of magnitude for $e_0=0.01$. In LISA, an eccentricity $e_0=0.2$ can be measured with $\sigma_{e_0}/e_0 \sim 3\times 10^{-5}$ for a $M=10^3\,M_{\odot}$ binary. For $M=10^6\,M_{\odot}$, the errors increase to $\sigma_{e_0}/e_0 \sim 10^{-3}$. LISA can also constrain the eccentricity to $\sigma_{e_0}/e_0 \sim 10^{-5}$ for multiband binaries. For $e_0=0.01$, the errors increase by nearly two orders of magnitude.

\section{\label{app_addionalboundresults}Additional results on the $\Delta\alpha$ bounds}
In this appendix we explore additional dependencies related to projected constraints on $\Delta\alpha$.
\subsection{Spin dependence on the $\Delta\alpha$ bounds}\label{app:alpha_spin}
When deriving the SPA phase for the periastron-based parametrization in Sec.~\ref{sec:periastron parametrization}, we ignored spin terms in all parts of the waveform. This was done because all of our starting expressions in that calculation account for eccentricity but not spin. Because the $\Delta\alpha$ parametrization modifies both the circular and eccentric parts of the phasing, inclusion of spin effects in the eccentric dynamics is needed to consistently include spin effects in the circular piece of the phasing. Because spin parameters were not included in the analysis in Sec.~\ref{sec:periastron-bounds}, the bounds there are overly optimistic. Here, we attempt to correct for this by including spin terms ``by hand'' in a straightforward (if not entirely consistent) way---with the goal of introducing the additional spin parameters and making the resulting bounds on $\Delta\alpha$ more realistic. To do this, we simply add the spin terms to the circular piece of the $\Psi_{\rm GR}$ part of the phasing, as in the TaylorF2 waveform up to 3.5PN order.  (In principle, there should also be spin corrections to both $\delta\Psi_{\rm circ.}$ and $\delta\Psi_{\rm ecc.}$, but they cannot be computed given the starting points of our formalism. In the $\Psi_{\rm GR}$ piece, there are likewise cross terms involving both the eccentricity and spin, which are unknown.)  

In Fig.~\ref{fig:alpha} of Sec.~\ref{sec:periastron-bounds} we showed the bounds on $\Delta\alpha$ for nonspinning binaries.  In Fig.~\ref{fig:alpha_spin} we reconsider those bounds after adding $\chi_1$ and $\chi_2$ to the parameter space as described above. The total parameter space is now: $\{t_{c}, \, \phi_{c}, \, {\log \mathcal{M}}, {\log \eta},  \, \log e_0, \, \chi_1,\, \chi_2, \, \Delta\alpha \}$. 
Including spins has the effect of degrading the bounds on $\Delta\alpha$. In LIGO, $\Delta\alpha$ can still be measured with good accuracy: $|\Delta\alpha| \lesssim 6 \times 10^{-2}$ for $10 \, M_{\odot}$. CE can measure $\Delta\alpha$ with $|\Delta\alpha|\lesssim 4 \times 10^{-3}$. For the considered BNS system, LIGO can constrain $|\Delta\alpha|\lesssim 9 \times  10^{-3}$, while the corresponding CE constraint is $|\Delta\alpha|\lesssim 2 \times 10^{-3}$ ($|\Delta\alpha|\lesssim 10^{-3}$) for $e_0=0.2$ ($0.01$). LISA can constrain $|\Delta\alpha|$ to $\lesssim 10^{-3}$ for $10^{5}\, M_{\odot}$. With DECIGO, $\Delta\alpha$ can be measured to $|\Delta\alpha|\lesssim 4 \times 10^{-4}$ for $M\approx 400 \,M_{\odot}$ and $e_0=0.2$. The best multiband bounds obtained are $|\Delta\alpha| \lesssim 4 \times 10^{-4}$.

\subsection{\label{app:alpha_circ_ecc}{Contributions to the $\Delta\alpha$ constraint from $\delta\Psi_{\rm circ.}$ and $\delta\Psi_{\rm ecc.}$}}

\begin{figure}[t]
    \centering
    \includegraphics[width=0.47\textwidth]{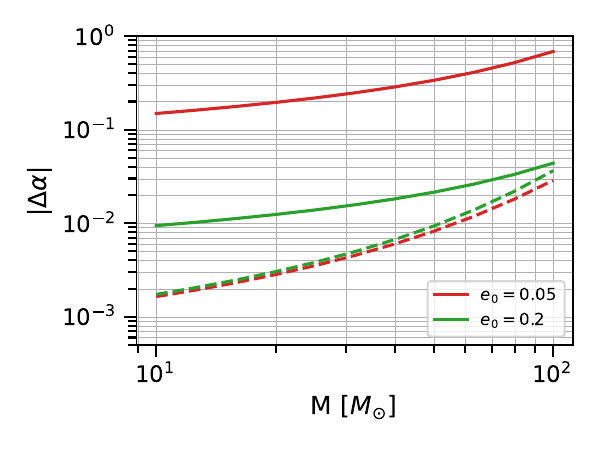}
\caption{\label{fig:alpha_circ_ecc} (Color online) Contributions to the $1\sigma$ bounds on $\Delta\alpha$ from $\delta\Psi_{\rm circ.}$ or $\delta\Psi_{\rm ecc.}$ alone for a CE detector. Different colors represent the two indicated values of eccentricity. Dashed lines include only $\Psi_{\rm GR}$ and the circular piece $\delta\Psi_{\rm circ.}$ of the $\Delta\alpha$ SPA phasing; solid lines include only $\Psi_{\rm GR}$ and the eccentric piece $\delta\Psi_{\rm ecc.}$. The bounds on $\Delta\alpha$ in the solid lines improve by one order of magnitude when $e_0$ increases from $0.05$ to $0.2$ (defined at $10\, \text{Hz}$.) However, the strongest constraints on $\Delta\alpha$ come from the $\delta\Psi_{\rm circ.}$ contribution (dashed lines). The binary parameters are the same as in Fig.~\ref{fig:alpha}.}
\end{figure}

As discussed in Sec.~\ref{subsec:SPAphase}, the dependence on $\Delta\alpha$ in the SPA phasing enters via two terms, one independent of $e_0$ ($\delta \Psi_{\rm circ.}$) and one proportional to $e_0^2$ ($\delta\Psi_{\rm ecc.}$). Here we examine the relative contribution of these two terms to the $\Delta\alpha$ constraint. This is shown in Fig.~\ref{fig:alpha_circ_ecc}  for CE, with only the circular contribution shown in dashed lines and only the eccentric contribution shown in solid. The contribution of $\delta\Psi_{\rm ecc.}$ to the overall $\Delta\alpha$ bound is small in the low eccentricity limit $(e_0 \lesssim 0.2)$. The contribution from $\delta\Psi_{\rm circ.}$ to the $\Delta\alpha$ bound is $\sim \mathcal{O}(10^{-3})$ for $10 \,M_{\odot}$. When only $\delta\Psi_{\rm circ.}$ is considered (but eccentricity still enters as a parameter through $\Psi_{\rm GR}$), there is negligible eccentricity dependence on the $\Delta\alpha$ bound due to correlations between $\Delta\alpha$, $e_0$, and the other source parameters (cf., two dashed curves in Fig.~\ref{fig:alpha_circ_ecc}). The bounds on $\Delta\alpha$ from only the eccentric part of the phasing ($\delta\Psi_{\rm ecc.}$) is $\sim \mathcal{O}(10^{-2})$ for $e_0=0.2$. Therefore, the overall bounds on $\Delta\alpha$ have negligible eccentricity dependence as seen in Fig.~\ref{fig:alpha}. However, when eccentricity increases (e.g., from $0.05$ to $0.2$ as shown in the figure), the overall contribution to the $\Delta\alpha$ bound from $\delta\Psi_{\rm ecc.}$ increases and can become comparable to the bound from the $\delta\Psi_{\rm circ.}$ term.

\bibliographystyle{apsrev}
\bibliography{refs-list}
\end{document}